\documentclass[aps,prd,twocolumn,superscriptaddress,nofootinbib,preprintnumbers]{revtex4-2}

\usepackage{amsmath}
\usepackage{mathtools}
\usepackage{amsfonts}
\usepackage{amssymb}
\usepackage{bm}
\usepackage{slashed}

\usepackage{graphicx}
\graphicspath{{figures/}}
\usepackage{tikz}
\usepackage{tikz-feynman}

\usepackage[caption=false]{subfig} 

\usepackage[usenames,dvipsnames,svgnames,table]{xcolor}
\usepackage{dcolumn}
\usepackage{float}
\usepackage{placeins}
\usepackage{comment}
\usepackage{changepage}
\usepackage{listings}
\usepackage{xpatch}
\usepackage{orcidlink}

\usepackage{hyperref}
\hypersetup{
    pdfnewwindow=true,
    colorlinks=true,
    allcolors=[RGB]{31 119 180},
    bookmarks=true
}
\usepackage{nameref}
\usepackage[capitalise]{cleveref} 

\DeclareMathOperator{\Ima}{Im}
\DeclareMathOperator{\Tr}{Tr}
\newcommand{\dd}{\mathrm{d}}
\newcommand{\ii}{\mathrm{i}}

\newcommand{\TT}{\mathrm{T}}
\newcommand{\LL}{\mathrm{L}}
\newcommand{\diff}{\ensuremath{\mathrm{d}}}
\newcommand{\eps}{\varepsilon}
\newcommand{\refl}{\epsilon}
\newcommand\bsub{\begin{subequations}}
\newcommand\esub{\end{subequations}}

\makeatletter
\xpatchcmd{\@ssect@ltx}{\@xsect}{\protected@edef\@currentlabelname{#8}\@xsect}{}{}
\xpatchcmd{\@sect@ltx}{\@xsect}{\protected@edef\@currentlabelname{#8}\@xsect}{}{}
\makeatother

\newcommand{\ub}{Departament de F\'isica Qu\`antica i Astrof\'isica and Institut de Ci\`encies del Cosmos, Universitat de Barcelona, E-08028 Barcelona, Spain}
\newcommand{\glasgow}{School of Physics and Astronomy, University of Glasgow, Glasgow, G12 8QQ, UK}
\newcommand{\jlab}{Thomas Jefferson National Accelerator Facility, Newport News, Virginia 23606, USA}
\newcommand{\messina}{Dipartimento di Scienze Matematiche e Informatiche, Scienze Fisiche e Scienze della Terra,
Universit\`a degli Studi di Messina, I-98166 Messina, Italy}
\newcommand{\catania}{INFN Sezione di Catania, I-95123 Catania, Italy}
\newcommand{\icn}{Instituto de Ciencias Nucleares,
    Universidad Nacional Aut\'onoma de M\'exico, Ciudad de M\'exico 04510, Mexico}
\newcommand{\york}{School of Physics, Engineering and Technology, University of York, York, YO10 5DD, United Kingdom}

\begin{document}

\allowdisplaybreaks
\pagenumbering{arabic}

\title{A Partial Wave Formalism for High-Energy Meson Electroproduction}

\author{D.~\surname{Leahy}\orcidlink{0009-0002-0821-5842}}

\affiliation{\glasgow}

\author{D.~I.~\surname{Glazier}\orcidlink{0000-0002-8929-6332}}
\affiliation{\glasgow}

\author{V.~\surname{Mathieu}\orcidlink{0000-0003-4955-3311}}
\affiliation{\ub}

\author{M.~\surname{Filippini}\orcidlink{0009-0000-6347-1123}}
\affiliation{\messina}
\affiliation{\catania}

\author{D.~G.~\surname{Ireland}\orcidlink{0000-0001-7713-7011}}
\affiliation{\glasgow}

\author{G.~\surname{Monta\~na}\orcidlink{0000-0001-8093-6682}}
\affiliation{\ub}

\author{A.~\surname{Pilloni}\orcidlink{0000-0003-4257-0928}}
\affiliation{\messina}
\affiliation{\catania}

\author{B.~\surname{Singh}\orcidlink{0000-0001-8997-0019}}
\affiliation{\jlab}

\author{C.~\surname{Velasquez}\orcidlink{0009-0000-5632-8038}}
\affiliation{\york}

\author{D.~Winney\orcidlink{0000-0002-8076-243X}}
\affiliation{\icn}

\begin{abstract}
We present a comprehensive partial-wave analysis formalism for meson electroproduction based on reflectivity amplitudes. By applying parity-based symmetries, we construct a fully diagonalized framework that relates complex production amplitudes to measurable spherical harmonic moments and Spin Density Matrix Elements. We demonstrate that, under the assumption of high-energy Regge factorization, longitudinal and transverse cross sections can be natively separated without requiring beam-energy-varying Rosenbluth separations. We validate this inversion methodology through numerical closure tests and apply it to existing unpolarized $\rho^0$ and $\omega$ electroproduction data from HERMES and COMPASS. The extracted amplitudes align with $s$-channel helicity conservation expectations for the $\rho^0$, while clearly isolating the large unnatural-parity pion exchange contribution in $\omega$ production without model-dependent background approximations. Finally, we systematically generalize the formalism to include initial target and recoil baryon polarization, establishing the observables required for complete amplitude extraction. This unified framework directly supports upcoming spectroscopy and nucleon structure programs at CLAS12, GlueX, and the future Electron-Ion Collider.
\end{abstract}
\maketitle

\section{Introduction}
The goal of identifying the presence of exotic mesons in partial waves forbidden by pure quark-model states remains a primary motivation in hadron spectroscopy. The number of exotic hadron candidates with heavy quarks has increased greatly~\cite{Johnson:2024omq,MEZZADRI2022100070,Brambilla:2019esw}, while the light-quark sector still poses many questions~\cite{KLEMPT20071}. Therefore, additional experimental evidence that would shed light on these issues is still critical.

One of the key reactions to study such states is photoproduction, driving facilities such as the GlueX experiment~\cite{PhysRevLett.133.261903}. 
The mass energy dependence of the cross sections may be used to search for resonances, and to understand the details of their production mechanisms in terms of exchanged Reggeons. Furthermore, extending to a dependence on $Q^2$ in virtual photoproduction  can be used to determine quantities directly related to the resonance structure and spatial extension, such as transition form factors. This probe of different distance scales within the resonance has previously been shown to provide clear evidence for the nature of the Roper resonance in the baryon sector~\cite{burkert2017roperresonancesolution}. 

Previously, Diehl~\cite{DIEHL200341,Diehl:2007jy} proposed a formalism relating the helicity amplitudes and exchange naturality to measurable Spin Density Matrix Elements (SDMEs) for the specific case of vector meson electroproduction on polarized initial nucleons, building on the standard framework of Schilling and Wolf~\cite{schilling_eprod}. 
In contrast, here we propose making the partial wave amplitudes the experimental quantities we wish to extract and explicitly extend to any produced spin decaying to a two pseudoscalar mesons final state, and include polarized recoil nucleons. 
We adopt the reflectivity basis, which projects onto exchange particle naturality, and has seen wide adoption and success for real photoproduction at GlueX~\cite{jfzb-rfl4,Mathieu2019}.
To extend to general spin, we consider moments of spherical harmonic distributions $H(LM)$ that can be determined from a Fourier analysis of the decay distributions in $Y^M_L\left(\theta,\phi\right)$, in place of straight SDMEs. These moments become linear combinations of the contributing SDMEs from different spin states.

In the spectroscopy context, these partial waves are readily identified with production of states of particular total spin ($J=\ell$), produced spin projection ($m$) and exchange naturality ($\epsilon$) with a related invariant mass. One may then use a suitable mass dependent model to extract the underlying resonance mass and width. In photoproduction, we may also measure the partial waves as a function of the reaction momentum transfer squared $t$, which leads to different strengths of exchanges. 

Measurements of these spherical harmonic moments with the CLAS detector at Jefferson Lab allowed the separation and determination of differential cross sections for both S and P waves in $\pi^+\pi^-$ photoproduction, dominated by the $f_0$ and $\rho$ mesons, respectively~\cite{CLAS:2009ngd}. Subsequent work on $K^+K^-$ photoproduction showed similar interference for production of the dominant $\phi$ meson with the S wave~\cite{PhysRevD.98.052009}. Such measurements show clearly the need to consider the full partial wave contribution to photo or electro production of vector, or other, mesons. The $Q^2$ dependence may then be modeled in terms of form factors providing a route to measuring the spatial extent of these states. For example, the charged pion electromagnetic form factor has been measured at Jefferson Lab Hall~C~\cite{PhysRevLett.86.1713,PhysRevC.75.055205}. This required a longitudinal to transverse cross section separation, which was performed at different beam energy settings. The key to this is that the longitudinal cross section is dominated by unnatural pion exchange at low $t$.

The same principle may be applied to the production and exchange of other states to determine the transition form factors, where the photon couples to two different particles. Goloskokov and Kroll~\cite{Goloskokov2014FF} performed such an analysis on combinations of HERMES $\omega$ SDMEs~\cite{HERMESomega} to determine the $\pi-\omega$ transition form factor using a GPD-based model. Such an analysis could be made less model dependent by directly using partial waves as derived here. 

Deeply Virtual Meson Production (DVMP) is a powerful mechanism for probing the 3D structure of the nucleon via Generalized Parton Distributions (GPDs). However, the theoretical bridge between experimental observables and GPDs is highly sensitive to the polarization of the exchanged virtual photon and the target baryon. The partial-wave analysis and reflectivity formalism presented in this work offer a distinct advantage for GPD extraction, as the underlying Compton form factors are directly related to our isolated reflectivity partial-wave amplitudes. In addition, this framework provides great synergetic potential for studies in conventional meson structure experiments and femtography. The extracted reflectivity partial waves can be mapped to the Compton form factors of the Generalized Parton Distribution framework via standard kinematic Wigner rotations, allowing the full magnitude and phase of these objects to be constrained by experimental data. However, while unpolarized and singly-polarized measurements leave the extracted amplitudes subject to continuous rotational ambiguities, we demonstrate that the inclusion of simultaneous target and recoil polarimetry strictly breaks this degeneracy. Measurement of the fully polarized initial and final nucleon states is therefore mathematically required for a completely model-independent extraction of the full amplitude space.
The electroproduction scheme outlined here can be applied to the current CLAS12 experiment~\cite{BURKERT2020163419} and could also facilitate a spectroscopy program in a proposed EIC~\cite{ABDULKHALEK2022122447}, where charmonium-like exotics would be within range.

In summary, we provide a unified analysis framework for meson electroproduction experiments. The paper is organized as follows: Sec.~\ref{sec:formalism} establishes the core theoretical framework for unpolarized nucleons, detailing the leptoproduction intensity, the spherical-harmonic moment expansion, and the construction of the reflectivity basis. Sec.~\ref{sec:results} presents the numerical inversion procedure, validating it via closure tests before applying it to published $\rho^0$ and $\omega$ electroproduction data to extract underlying amplitudes and the longitudinal-to-transverse ratio $R$. In Sec.~\ref{sec:polarized_nucleons_main}, we generalize the framework to incorporate polarized targets and recoil baryons, specifically addressing the required frame transformations and the resolution of continuous ambiguities within Sec.~\ref{sec:symmetry}. Section~\ref{sec:macro_cross_sections} then directly maps our moments to measurable experimental cross sections. Finally, Sec.~\ref{sec:conclusion} offers a discussion of future experimental applications and concluding remarks. 
For clarity of presentation, all technical details are collated in the appendices. In particular, the field-theoretic derivations and calculations of the kinematics are contained within Appendix~\ref{app:electroformalism}. Within Appendix~\ref{app:angular_ints_moments} the definition of electroproduction moments are given, with Appendix~\ref{app:reflectivity} extending the reflectivity basis to reactions mediated by a virtual-photon. Appendix~\ref{app:factorization_coherence} gives a physical basis for the single reggeon exchange approximation and details where it breaks down. The full calculation for the generalization of this formalism to polarised nucleon states is given within Appendix~\ref{app:fullpol}, with nucleon polarization frame transformations given within Appendix~\ref{app:nuc_pol_trans}. Underlying symmetries at the amplitude level and subsequent parameter redundancies are discussed within Appendix~\ref{app:symmetries} with the total number of observables and parameters (for unpolarized and polarized nucleons) for different angular momenta listed within Appendix~\ref{app:counting}. Finally, supplementary plots are collected within Appendix~\ref{app:supplementary}.

\section{Unpolarized nucleons}
\label{sec:formalism}
\subsection{Leptoproduction intensity}
\label{subsec:lepto_main}

\begin{figure}[h]
    \centering
    \includegraphics[width=0.9\linewidth]{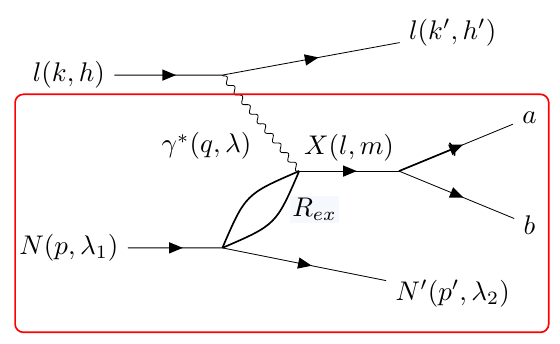}
    \caption{One-photon-exchange representation of Eq.~\eqref{eq:full_reaction}. The red box highlights the virtual-photon subprocess whose spin structure is analysed below.}
    \label{fig:feyn_diag}
\end{figure}

The exclusive reaction considered throughout is
\begin{align}
\label{eq:full_reaction}
\ell(k,h)+N(p,\lambda_1)&\to \ell'(k',h')+N'(p',\lambda_2)+a+b,
\end{align}
where $h,h'$ are the lepton helicities, $\lambda_1,\lambda_2$ are the nucleon helicities and $a,b$ are the two spin-zero mesons.
The reaction \eqref{eq:full_reaction} is approximated in the one-photon-exchange as depicted in Fig.~\ref{fig:feyn_diag}.

In this approximation the hadronic subprocess is 
\begin{equation}
\begin{aligned}
\gamma^*(q,\lambda)+N(p,\lambda_1)&\to N'(p',\lambda_2) + a(p_a) + b(p_b),
\end{aligned}
\label{eq:hadronic_subprocess_main}
\end{equation}
with virtual-photon helicity $\lambda=+1,0,-1$. In addition to the photon virtuality $Q^2\equiv -q^2>0$, with $q=k-k'$, there are five independent variables describing the process~\eqref{eq:hadronic_subprocess_main}, we choose to use: $s = W^2  = (q+p)^2$ is the total energy squared of the $\gamma^* N$ system, $t = (p-p')^2$ the momentum transferred between the nucleons, $m^2 = (p_a+p_b)^2$ the invariant mass squared of the $ab$ system, and $\Omega = (\theta,\phi)$ the angles of particle $a$ in the $ab$ rest frame. The angles depend on the choice of the axes. The $xz$ plane is always chosen as the production plane containing the photon and nucleon momenta. The two common choices for the $z$-axis are the Gottfried-Jackson frame, in which the $z$-axis is parallel to the virtual photon momentum, and the helicity frame, in which the $z$-axis is opposite to the recoiling nucleon's momentum. 
Although the formalism is identical in both frames, in practice, a $z$-axis needs to be chosen. 

\begin{figure}
    \centering
    \includegraphics[width=1.05\linewidth]{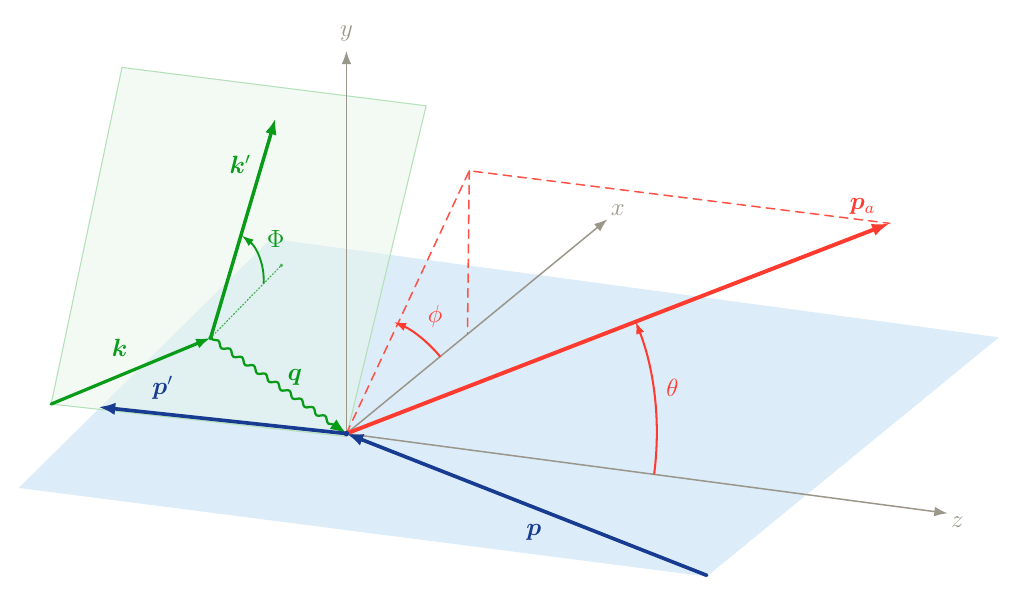}
        \caption{Helicity frame used for the decay analysis. The $z$ axis is defined along the recoiling nucleon direction, the hadronic plane defines the $xz$ plane, and $\Phi$ is the azimuthal angle between the lepton and hadron planes.}
    \label{fig:angle_diagram}
\end{figure}

The reduced angular distribution is obtained after factorizing the virtual-photon flux, 
\begin{equation}
\begin{aligned}
\frac{\dd^7\sigma}{\dd W\,\dd Q^2\,\dd t\,\dd m\,\dd\Omega\,\dd\Phi}
&=\Gamma(Q^2,W)\,\frac{\dd^5\sigma^{*}}{\dd t\,\dd m\,\dd\Omega\,\dd\Phi}\\
&=\Gamma(Q^2,W)\,I(\Omega,\Phi),
\end{aligned}
\label{eq:masterxs_main}
\end{equation}

where $\Phi$ is the azimuth of the lepton plane with respect to the hadronic production plane. The helicity frame and the corresponding angles are shown in Fig.~\ref{fig:angle_diagram}. 
The virtual-photon flux factor in the Hand convention~\cite{Hand:1963bb} is written as 
\begin{equation}
\Gamma(Q^2,W) = \frac{\alpha}{2\pi} \frac{W\left(W^2 - m^2_N\right)}{2m_N^2 E^2 Q^2} \frac{1}{1-\varepsilon}.
\label{eq:GammaV_main}
\end{equation}
For negligible lepton mass the virtual photon polarization parameter is
\begin{equation}
    \varepsilon = \frac{1-y-Q^2/4E^2}{1-y+y^2/2+Q^2/4E^2},
\label{eq:epsilon_main}
\end{equation}
where $E$ is the incident lepton energy in the target rest frame, and $y=(W^2+Q^2-m_N^2)/2m_N E$ is the fraction of energy lost by the lepton.
We denote the amplitude of the subprocess in Eq.~\eqref{eq:hadronic_subprocess_main} by $A_{\lambda;\lambda_1\lambda_2}(\Omega)$. With target and recoil spins unobserved, the intensity is
\begin{equation}
\begin{aligned}
I(\Omega,\Phi)={}&\kappa
\sum_{\lambda\lambda'\lambda_1\lambda_2}
A_{\lambda;\lambda_1\lambda_2}(\Omega)
\rho^{\gamma^*}_{\lambda\lambda'}(\Phi)
A^{*}_{\lambda';\lambda_1\lambda_2}(\Omega),
\end{aligned}
\label{eq:leptoproduction_intensity_main}
\end{equation}
where $\kappa$ (\textit{cf.} Eq.~\eqref{eq:app_phase_factor}) contains the hadronic phase space and flux factors, and $\rho^{\gamma^*}$ is the virtual-photon spin-density matrix generated by the lepton tensor, which can be projected into the nine Hermitian Schilling-Wolf response matrices~\cite{schilling_eprod},
\begin{equation}
\rho^{\gamma^*}_{\lambda\lambda'}(\Phi)=\frac{1}{2}
\sum_{\alpha=0}^{8}\Pi_\alpha(\Phi)\Sigma^\alpha_{\lambda\lambda'}.
\label{eq:photon_density_decomp_main}
\end{equation}
 The first four responses $\alpha=0,1,2,3$ are purely transverse and reduce to the photoproduction basis, $\alpha=4$ is the purely longitudinal response, and $\alpha=5,6,7,8$ are longitudinal-transverse interference responses. The lepton kinematics and the matrices $\Sigma^\alpha$ derived in Appendix~\ref{app:electroformalism} enter through
\begin{align} \nonumber
    \Pi_{0-3}(\Phi) & = \left[1,-\eps\cos2\Phi, -\eps\sin2\Phi, \sqrt{1-\eps^2}\,P_l\right],
    \\ \nonumber
    \Pi_{4-6}(\Phi) & = \left[\eps, \sqrt{2\eps(1+\eps)}\cos\Phi, \sqrt{2\eps(1+\eps)}\sin\Phi\right],
    \\
    \Pi_{7-8}(\Phi) & =\sqrt{2\eps(1-\eps)}\,P_l \left[\cos\Phi, \sin\Phi\right].
\end{align}
This paper uses the same polarization coefficients of Ref.~\cite{glazier_elliptical_2025}, in which the $E^2\gg m_l^2$ (the mass of the lepton) limit is taken for the polarization vector. Thus, the angular distribution may be written in the beam-response form
\begin{align}
    I(\Omega,\Phi) & = \kappa\sum_{\alpha=0}^8 \Pi_\alpha(\Phi) I^\alpha(\Omega),
    \label{eq:intensities_schilling}
\end{align}
defining the intensity functions \begin{equation}
    I^\alpha(\Omega) = \frac{1}{2}\sum_{\lambda\lambda'\lambda_1\lambda_2} A_{\lambda;\lambda_1\lambda_2}(\Omega)\Sigma^\alpha_{\lambda\lambda'}A^*_{\lambda';\lambda_1\lambda_2}(\Omega).
\end{equation}
Following the approach of Ref.~\cite{Mathieu2019}, the hadronic amplitude is expanded into partial waves as
\begin{equation}
A_{\lambda;\lambda_1\lambda_2}(\Omega) = \sum_{\ell m}T^\ell_{\lambda m; \lambda_1 \lambda_2} Y^m_\ell(\Omega),
\label{eq:partial_wave_main}
\end{equation}
where $\ell$ and $m$ are the total angular momentum and spin-projection labels of the two-body system.

\subsection{Moments, SDMEs and reflectivity}
\label{subsec:moments_main}
Each beam response is expanded in spherical-harmonic moments (more detail on the construction within Appendix~\ref{app:angular_ints_moments}),
\begin{equation}
I^\alpha(\Omega)=
\sum_{LM}\left(\frac{2L+1}{4\pi}\right)\, H^\alpha(LM)Y_L^M(\Omega).
\label{eq:Ialpha_main}
\end{equation}

The $\alpha=0$ and $\alpha=4$ responses have the same external lepton polarization and $\Phi$ dependence [see Eqs.~\eqref{eq:app_moments_fourier}~and~\eqref{eq:app_moments_fourier_norms}], so only their weighted combination is directly accessible in a single-energy measurement unless $R = \sigma_\LL/\sigma_\TT$ is fixed independently or determined self-consistently from the amplitude solution. The fundamental hadronic moments are bilinear in partial-wave amplitudes through the response SDMEs,
\begin{equation}
\begin{aligned}
H^{\alpha}(LM)={}&
\sum_{\substack{\ell,\ell'\\m,m'}}
\left(\frac{2\ell'+1}{2\ell+1}\right)^{1/2}
C^{\ell'0}_{\ell0\,L0}
C^{\ell'm'}_{\ell m\,LM}
\rho^{\alpha,\ell\ell'}_{mm'}.
\end{aligned}
\label{eq:momentformula_main}
\end{equation}
The Clebsch-Gordan coefficients above enforce the usual angular-momentum selection rules and fix the relation between the spherical-harmonic moments and the spin-density matrices. The $H^0(00)$ moments are proportional to the transverse cross section, while the $H^4(00)$ moments are proportional to the longitudinal cross section (see Eqs.~\eqref{eq:app_integrated_cross_sections_moments}).
Experimentally with a single beam energy, one may only directly measure the photon polarization weighted moment sum of $H^0$ and $H^4$ moments.

The terms $\rho^{\alpha,\ell\ell'}_{mm'}$ entering Eq.\eqref{eq:momentformula_main} represent the unnormalized SDMEs of the resonance, consistent with the Schilling and Wolf formalism~\cite{schilling_eprod} prior to trace normalization. We choose to normalize directly at the moment level (see Sec.~\ref{sec:macro_cross_sections}), which leaves the final observables invariant. The SDMEs are defined as:
\begin{equation}
    \rho^{\alpha,\ell\ell'}_{mm'} = \frac{1}{2}\sum_{\lambda\lambda'\lambda_1\lambda_2}T^{\ell}_{\lambda m; \lambda_1 \lambda_2} \Sigma^\alpha_{\lambda\lambda'} \left(T^{\ell'}_{\lambda' m'; \lambda_1 \lambda_2}\right)^*,
    \label{eq:Moments_SDME}
\end{equation}
 where the full expansion for all response classes is provided in Appendix~\ref{app:angular_ints_moments}.
 
It is advantageous to reorganize the production amplitudes into transverse
($\TT$) and longitudinal ($\LL$) photon reflectivity states, labelled by the
eigenvalue $\refl=\pm1$:
\begin{align}
[\ell]^\refl_{\TT m;\lambda_1\lambda_2}
&=
\frac{1}{2}
\left(
T^\ell_{+1,m;\lambda_1\lambda_2}
-\refl(-1)^m
T^\ell_{-1,-m;\lambda_1\lambda_2}
\right),
\label{eq:trans_ref_main}
\\
[\ell]^\refl_{\LL m;\lambda_1\lambda_2}
&=
\frac{1}{2}
\left(
T^\ell_{0,m;\lambda_1\lambda_2}
+\refl(-1)^m
T^\ell_{0,-m;\lambda_1\lambda_2}
\right).
\label{eq:long_ref_main}
\end{align}
The longitudinal amplitudes obey
\begin{equation}
[\ell]^\refl_{\LL m;\lambda_1\lambda_2}
=
\refl(-1)^m
[\ell]^\refl_{\LL,-m;\lambda_1\lambda_2}.
\label{eq:long_reflection_symmetry_main}
\end{equation}
Therefore, for $m=0$, only the positive-reflectivity longitudinal amplitude
is independent.

In the high-energy limit for the production of two spin-zero mesons, the
reflectivity eigenvalue identifies the naturality of the exchanged
$t$-channel trajectory.  
The corresponding derivation, inverse
relations, and complete amplitude counting are given in
Appendix~\ref{app:reflectivity}.

Parity conservation relates amplitudes in which both nucleon helicities are
reversed:
\begin{align}
[\ell]^\refl_{\TT m;-\lambda_1,-\lambda_2}
&=
\refl(-1)^{\lambda_1-\lambda_2}
[\ell]^\refl_{\TT m;\lambda_1\lambda_2},
\label{eq:paritylawT_main}
\\
[\ell]^\refl_{\LL m;-\lambda_1,-\lambda_2}
&=
\refl(-1)^{\lambda_1-\lambda_2}
[\ell]^\refl_{\LL m;\lambda_1\lambda_2}.
\label{eq:paritylawL_main}
\end{align}
Thus, only two of the four nucleon-helicity transitions are independent.  We
use the reduced index $k=\pm1$,
\begin{align}
[\ell]^\refl_{\TT m;1}
&\equiv
[\ell]^\refl_{\TT m;++},
&
[\ell]^\refl_{\TT m;-1}
&\equiv
[\ell]^\refl_{\TT m;+-},
\label{eq:kbasisT_main}
\\
[\ell]^\refl_{\LL m;1}
&\equiv
[\ell]^\refl_{\LL m;++},
&
[\ell]^\refl_{\LL m;-1}
&\equiv
[\ell]^\refl_{\LL m;+-},
\label{eq:kbasisL_main}
\end{align}
where $k=1$ and $k=-1$ denote nucleon helicity non-flip and flip,
respectively.  The remaining $--$ and $-+$ amplitudes follow from
Eqs.~\eqref{eq:paritylawT_main} and \eqref{eq:paritylawL_main}.

After summing over the unobserved target and recoil helicities, parity makes
the reflectivity sectors diagonal.  Consequently, there is no interference
between $\refl=+1$ and $\refl=-1$ for the observable quantities. The explicit proof and the reflectivity-basis SDME bilinears are also collected in
Appendix~\ref{app:reflectivity}.
 Furthermore, parity and Hermicity relations for the SDMEs and moments determine the real or imaginary character of the moments;
in particular, the $M=0$ moments vanish for
$\alpha=2,3,6,7$.

\subsection{Amplitude factorization and coherence}
\label{subsec:unpolarized_amplitudes}

So far, we have established the mathematical relationship mapping partial-wave amplitudes to SDMEs, moments, and ultimately physical intensities. We now consider the implications for extracting the underlying amplitudes from experimental observables with unpolarized nucleons. We begin here with the simplified case of a single dominant exchange and demonstrate how it can be analyzed using an unpolarized baryon dataset with the formulas derived above. In Sec.~\ref{sec:polarized_nucleons_main}, we expand the formalism to include full baryon polarization, which is a strict requirement for complete and unambiguous amplitude extraction.

For unpolarized nucleon data, the target and recoil helicities remain unobserved, meaning that the individual nucleon helicity-flip and non-flip transitions ($k=\pm 1$) cannot be resolved. The true physical intensity is therefore an inherently incoherent sum over these unobserved $k$ states.

Following a Regge factorization framework as in~\cite{PhysRevD.97.094003} for light vector  meson photoproduction, the physical partial-wave amplitudes for an individual exchange trajectory $x$ can be factorized into a virtual-photon transition ($\mathcal{U}$), and a target-nucleon exchange coupling ($\mathcal{B}$). The total amplitude for a given reflectivity $\refl$ is the coherent sum over all contributing trajectories within that naturality sector:
\begin{align}
    [\ell]^\refl_{m; k} & =\, \sum_x \mathcal{U}^\refl_{x,a} \, \, \mathcal{B}^\refl_{x,k} \ ,
\label{eq:factorization}
\end{align}
with $a \equiv (c, m)$, where $c \in \{\TT,\LL\}$. 

When the reaction kinematics within a specific reflectivity sector are dynamically dominated by a single exchange trajectory, the target coupling ($\, \mathcal{B}^\epsilon_k$) is completely decoupled from the virtual-photon state ($c$) and the resonance decay dynamics ($m$, $\ell$). As detailed in Appendix~\ref{app:factorization_coherence}, this specific condition allows the unobserved $k$-summation to factor out of all unpolarized observables as a purely real scalar. 
Because this scalar introduces no relative phase shift, the phase coherence between longitudinal and transverse responses, as well as between varying spin projections, is preserved. This mathematical simplification offers a substantial experimental payoff: the number of unknown parameters is halved, allowing the complete kinematic determination of partial waves without a beam-energy-varying Rosenbluth separation.

However, extracting a single effective amplitude becomes mathematically dangerous in kinematic regimes where this  factorization breaks down. It is important to distinguish between the Regge factorization of \textit{individual} exchanges (valid at asymptotic energies) and the factorization of the \textit{total} amplitude. Because opposite reflectivities do not interfere in this formalism, simultaneous exchanges of different naturalities (e.g., natural Pomeron and unnatural pion exchange) do not disrupt factorization. However, at intermediate energies, or in complex reactions, multiple exchange trajectories within the \textit{same} naturality sector (such as simultaneous Pomeron and $\rho$ exchange) can contribute simultaneously. If these competing exchanges couple to orthogonal nucleon helicity states (e.g., a Pomeron driving $k=1$ while a $\rho$-exchange drives $k=-1$), the target vertex cannot be factored out, even if Eq.~\eqref{eq:factorization} holds for each individual trajectory. 

As rigorously demonstrated in Appendix~\ref{app:factorization_coherence}, this breakdown results in a loss of quantum coherence in the unpolarized observables. Because the orthogonal nucleon helicity states do not interfere, the target trace incoherently mixes the independent upper-vertex transitions. This incoherence suppresses the interference terms in the moments (such as those governing $\LL/\TT$ or $m$-projection interference) relative to the pure diagonal intensities.

Therefore, extracting effective amplitudes from purely unpolarized data is only physically justified in kinematic regimes where a single exchange trajectory dynamically dominates a given naturality sector. In transitional regimes where this factorization fails, one may not simply parameterize the full incoherent $k$-summation using unpolarized data, as the system becomes severely underconstrained. Instead, a model-independent amplitude extraction must mathematically separate the complete $k$-basis, which experimentally requires the inclusion of target and recoil polarimetry, as detailed in Sec.~\ref{sec:polarized_nucleons_main}.

\subsection{Longitudinal to transverse cross section ratio}

In the traditional Schilling and Wolf framework~\cite{schilling_eprod}, the ratio of longitudinal to transverse cross sections, $R=\sigma_\LL/\sigma_\TT$, is treated as an external kinematic parameter that typically requires a beam-energy-varying Rosenbluth separation to evaluate, see~\cite{PhysRevLett.117.262001,Bebek:1977pe}. In contrast, the normalized spherical harmonic moments implicitly contain this information. By extracting the underlying partial-wave amplitudes, we can reconstruct $R$ directly from the amplitude solutions without relying on external parameterizations.

As discussed in Sec.~\ref{subsec:unpolarized_amplitudes}, analyzing unpolarized data requires assuming a single exchange trajectory per naturality.  The incoherent sum over $k$ is then effectively replaced by a single coherent amplitude per state. Because this factorization preserves the phase coherence between the longitudinal and transverse responses, the LT interference moments ($H^5$ and $H^6$) constrain the relative magnitudes of the L and T amplitudes.

Suppressing the implicitly factored $k$ index, the cross section ratio is reconstructed directly from the extracted effective amplitudes via:
\begin{equation}
R=\frac{\sigma_\LL}{\sigma_\TT}\equiv \frac{\displaystyle\sum_{\refl,\ell,m}\big|[\ell]^\refl_{\LL m}\big|^2}
{\displaystyle\sum_{\refl,\ell,m}\big|[\ell]^\refl_{\TT m}\big|^2} = \frac{H^4(00)}{H^0(00)}
.
\label{eq:Ramp_main}
\end{equation}
Effectively, the amplitude-level interference between the longitudinal and transverse photon states bypasses the need for a Rosenbluth separation, provided the underlying single-exchange assumption holds.

\section{Numerically inverting moments to amplitudes}
\label{sec:results}

In order to better understand which amplitudes we may extract directly from experimental data, we develop an inversion procedure similar to that used in Ref.~\cite{glazier_elliptical_2025}.

The inversion is formulated as a nonlinear least-squares optimization to solve for the amplitude parameters from the system of moment equations. Based on Eq.~\eqref{eq:momentformula_main}, the objective function $\chi^2$ minimizes the variance-weighted residuals between the measured input moments $\bar{H}^\alpha(LM)$, and the `theoretical' moments $H^{\alpha}(LM)$ calculated over the set of free complex amplitude parameters $\{[\ell]^\epsilon_{c,m}\}$, where $c \in \{\LL,\TT\}$. The residual is normalized by the corresponding experimental uncertainty, $\sigma_{\alpha,L,M}$:
\begin{equation}
    \chi^2 = \sum_{\alpha,L,M} \left(\frac{H^{\alpha}(LM)(\{[\ell]^\epsilon_{c,m}\}) - \bar{H}^{\alpha}(LM)}{\sigma_{\alpha,L,M}}\right)^2
\end{equation}
 This is minimized by the MINUIT2 package~\cite{Hatlo:2005zr} using the gradient-based MIGRAD routine. The uncertainties for the minimization are propagated through a bootstrap procedure. For this, the moment values are sampled from a Gaussian defined with the mean as the moment value and the standard deviation given by its uncertainty. The inversion is repeated for each of these moment samples, and the resulting distributions of the amplitudes are used to define their uncertainties.
\subsection{Fixed-amplitude closure test}
\label{sec:closure}

\begin{figure}
    \centering
    \includegraphics[width=0.65\linewidth]{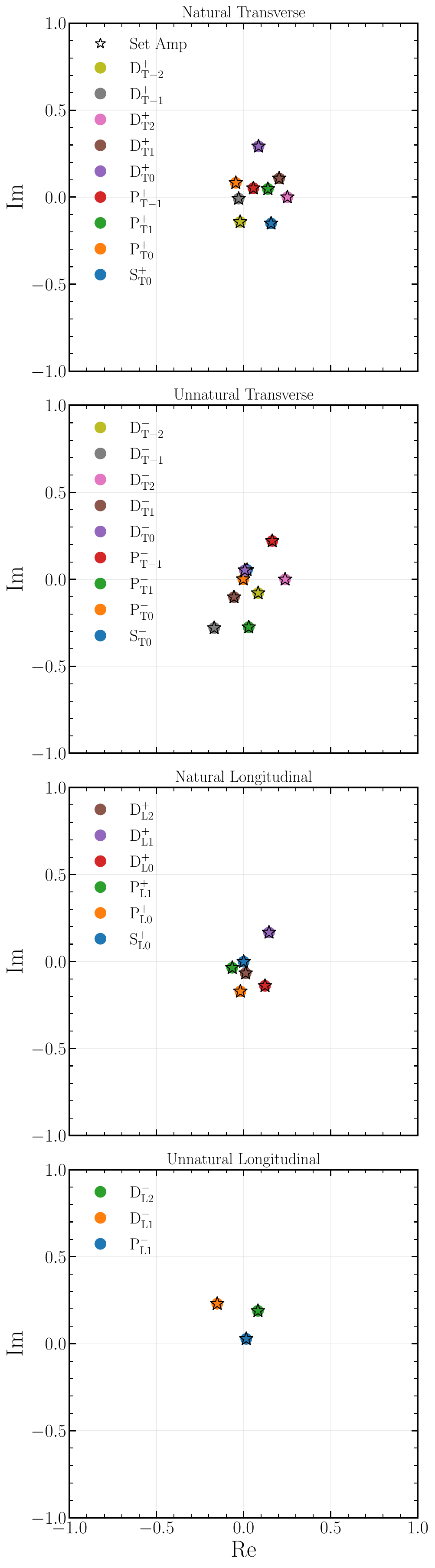}
    \caption{Fixed-amplitude closure test for a randomly generated amplitude configuration. The amplitudes extracted from the minimum-$\chi^2$ solution (colored dots) basically coincide with the fixed input values used to generate the moments (open stars).}
    \label{fig:fixedamps}
\end{figure}

\begin{figure*}
    \centering
    \includegraphics[width=0.98\linewidth]{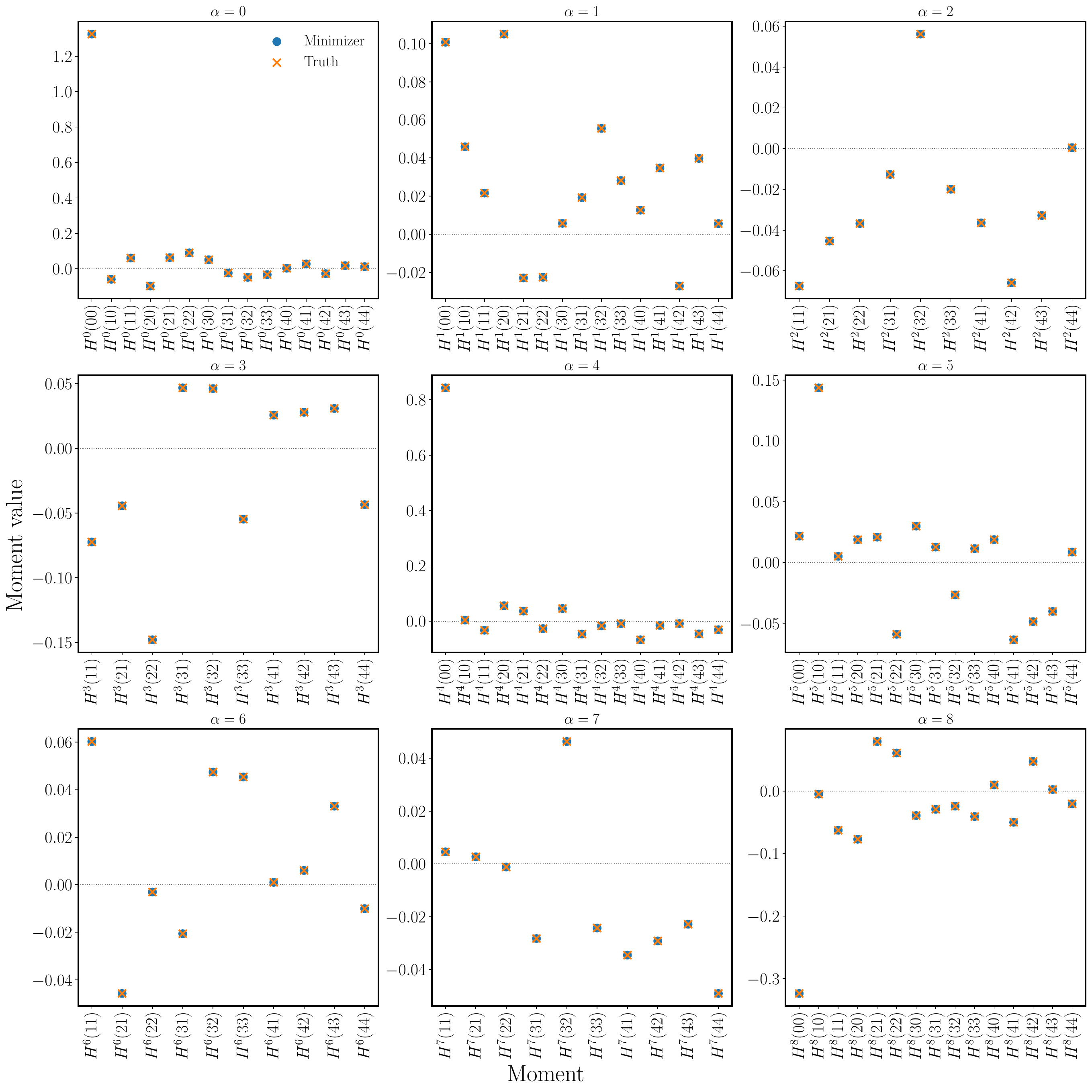}
    \caption{Moment-level closure test for the same fixed-amplitude configuration. The moments reconstructed from the minimum-$\chi^2$ amplitude solution (blue dots) are compared with the input moments (orange crosses).}
    \label{fig:fixed_moments_large_norm_rand.pdf}
\end{figure*}

To demonstrate that the nonlinear inversion successfully produces unique amplitude solutions, we perform a closure test starting from a known amplitude set. At this stage we are assuming Regge factorization and that it is therefore sufficient to consider single $k$ amplitudes and we suppress that index from here. Furthermore, for the assumed case here, there are two trivial ambiguities in the overall phase, for which the conventional choice is that the phases of the $[l_{\text{max}}]^\pm_{\TT;m_{\text{max}}}$ waves are set to 0. A set of complex amplitudes for $[\ell]\leq2$ is generated randomly in the four sectors: natural transverse, unnatural transverse, natural longitudinal, and unnatural longitudinal. These are then converted into moments, $\bar{H}^\alpha(LM)$, and inverted back to the amplitude level with the inversion procedure. For such a closure test, as there is no experimental uncertainty for the moments, instead an unweighted $\chi^2$, in which all error weights are set to unity, is applied.  The complete set of 27 random amplitudes used is,
\begin{equation}
\begin{aligned}
 &[D]^{\pm}_{\TT;\{2,1,0,-1,-2\}},[P]^{\pm}_{\TT;\{1,0,-1\}},[S]^{\pm}_{\TT;0},\\
&[D]^{\pm}_{\LL;\{2,1\}},[P]^{\pm}_{\LL;1},[D]^{+}_{\LL;0},[P]^{+}_{\LL;0},[S]^{+}_{\LL;0} .
\label{eq:practical_truncation_fixed}
\end{aligned}
\end{equation}
Hence, within this analysis the phases of $[D]^{+}_{\TT;2}$ and $[D]^{-}_{\TT;2}$ were used as our reference gauge and set to 0.
The results are shown in Fig.~\ref{fig:fixedamps}. The reconstructed moment values (the values the minimizer evaluates the constraints to be) are also compared with input truth values within Fig.~\ref{fig:fixed_moments_large_norm_rand.pdf}.
The reconstructed solutions coincide with the input values for the wave set considered here. The closure study therefore supports the practical uniqueness of the numerical solution in the general case.

In addition to randomly choosing all the amplitudes, further tests examining specific extreme scenarios were performed and are collected in Appendix~\ref{app:supplementary}. These scenarios yield similar results, demonstrating that it is fully possible to extract the amplitudes from a given set of moments. However, we note that the convergence behavior of each scenario can differ. Specifically, when the amplitude sets lack particular combinations and interference SDMEs vanish, the convergence efficiency decreases, and a larger number of minimizations is required to obtain the global minimum.

\subsection{Vector meson amplitudes from published results}
\label{subsec:amps_main}
We consider now the case of vector meson electroproduction, where P-waves dominate the SDMEs. At  high energies and at low to intermediate $Q^2$, the production of vector mesons is dominated by (spin--non-flip, natural)   Pomeron exchange and (spin-flip, unnatural) pion exchange. Hence we assume that only a single $k$ amplitude contributes to each naturality, allowing us to apply the Regge factorization assumption and suppress the $k$ index:

\begin{align}
\begin{split}
[P]^+_{\TT;\{1,0,-1\}},[P]^+_{\LL;\{1,0\}}, \\
[P]^-_{\TT;\{1,0,-1\}},[P]^-_{\LL;1}.
\label{eq:practical_truncation_main}
\end{split}
\end{align}
Here, the trivial phase convention defines the amplitudes $[P]^+_{\TT;1}$ and $[P]^-_{\TT;1}$ to be positive and real, i.e. the phases of these two amplitudes are chosen as the reference gauge and are set to 0.
This then leads to a system of equations which is overconstrained with 23 independent moment equations for 16 real parameters (accounting for the two overall phases of the noninterfering reflectivity sectors; see Appendix~\ref{app:counting} for more details on counting constraints and free parameters). Therefore, in principle the inversion should be able to obtain a unique value for each parameter, and this is confirmed by the inversion results which find distinct global minima.

\begin{figure*}
    \centering
    \subfloat[\label{fig:amp_erho}]{\includegraphics[width=0.24\textwidth]{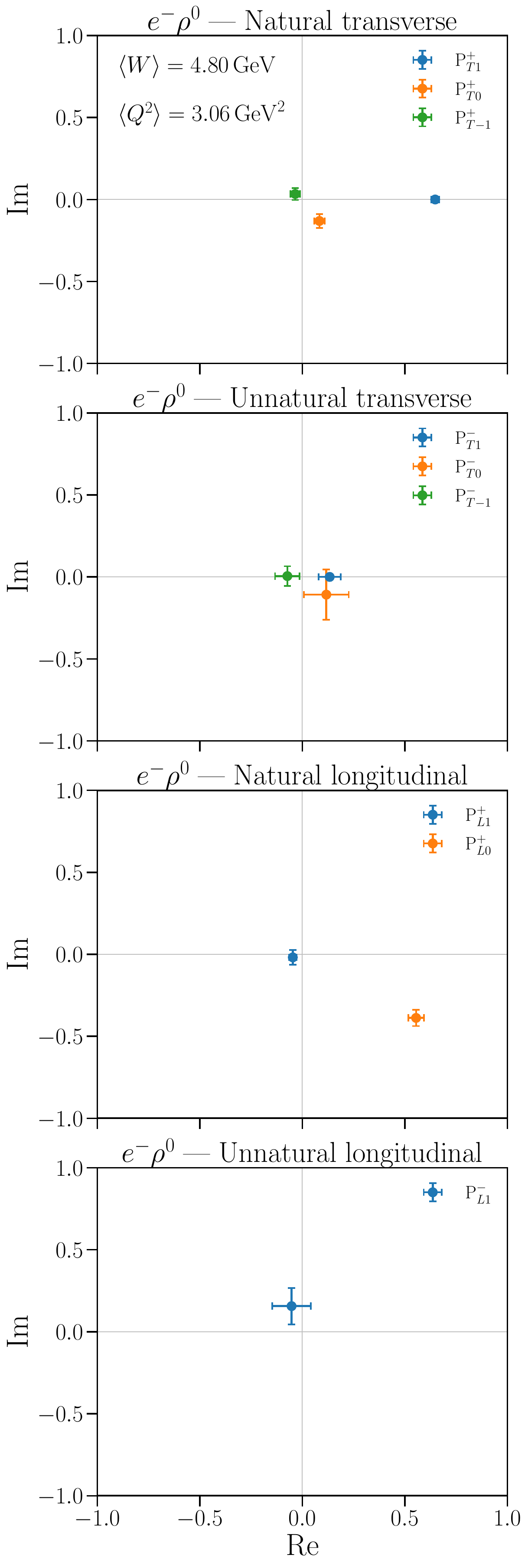}}\hfill
    \subfloat[\label{fig:amp_murho}]{\includegraphics[width=0.24\textwidth]{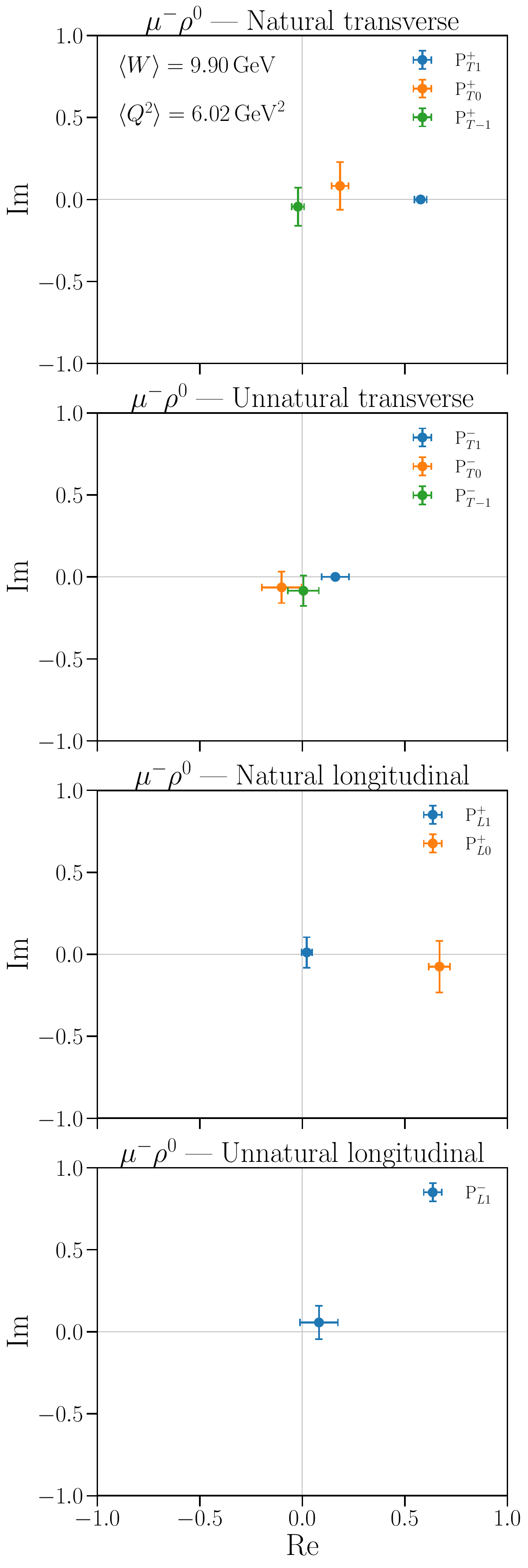}}\hfill
    \subfloat[\label{fig:amp_eomega}]{\includegraphics[width=0.24\textwidth]{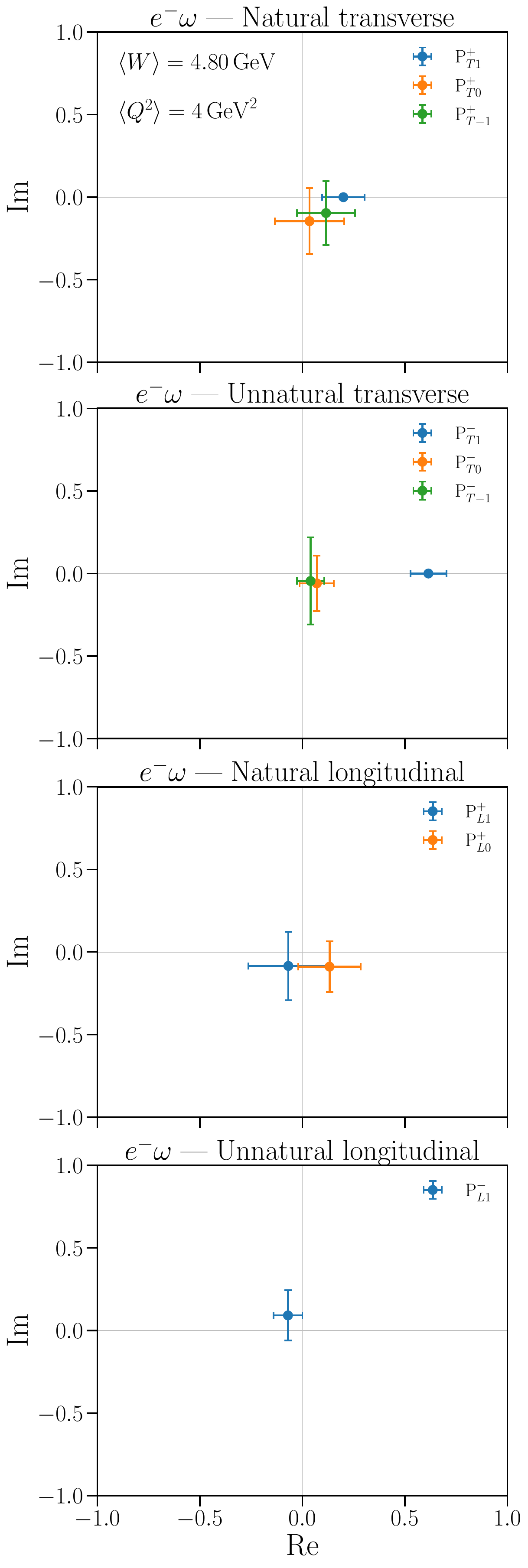}}\hfill
    \subfloat[\label{fig:amp_muomega}]{\includegraphics[width=0.24\textwidth]{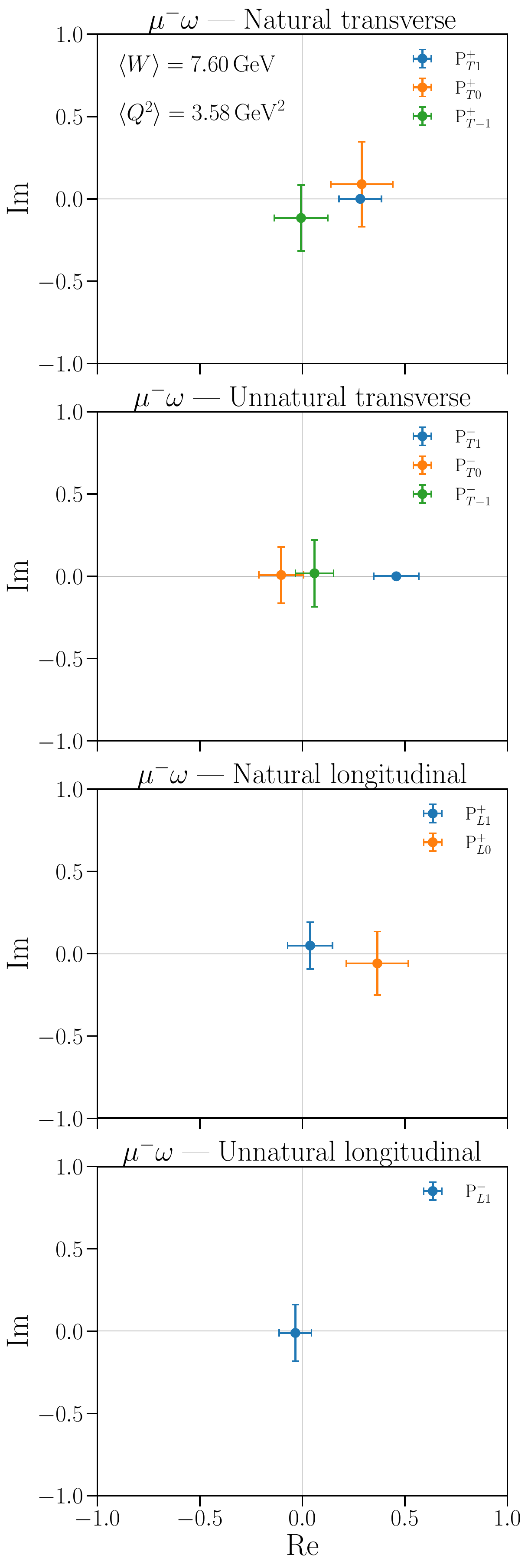}}
    
    \caption{Representative amplitude extractions for the two $\rho^0$ reaction classes~\cite{HERMESrho,COMPASSrho} and two $\omega$ reaction classes~\cite{HERMESomega,COMPASSomega} discussed in the text, note that the lack of vertical error bars on $P^\pm _{\TT 1}$ is due to the phase fixing convention, for which these two amplitudes are constrained to lie along the positive real axis. Within each subfigure we show the natural-transverse, unnatural-transverse, natural-longitudinal, and unnatural-longitudinal sectors.}
    \label{fig:amp_rho_main}
\end{figure*}

The inverted amplitude solutions for all four reaction classes are collected in Figs.~\ref{fig:amp_rho_main}: $e^-\rho^0$, $\mu^-\rho^0$, $e^-\omega$, and $\mu^-\omega$. The panels are chosen from the highest $Q^2$ bins of the available HERMES and COMPASS SDME sets~\cite{HERMESrho,HERMESomega,COMPASSrho,COMPASSomega}, whilst the bins $t$ and $W$ have been integrated.

A common pattern is immediately visible in the two $\rho^0$ datasets. In both the electron- and muon-induced cases, the dominant amplitude is the $[P]^+_{\TT; 1}$ contribution, while the longitudinal amplitudes are led by $[P]^+_{\LL; 0}$. These correspond to the two natural-exchange helicity--non-flip amplitudes, which are expected to dominate under $s$-channel helicity conservation (SCHC). The unnatural-transverse and unnatural-longitudinal amplitudes remain comparatively small.

The two $\omega$ panels look qualitatively different. The natural sector is still significant, but the unnatural contributions are no longer negligible. In particular, the unnatural-transverse amplitudes are visibly enhanced compared to the $\rho^0$ case. Therefore, the $\omega$ solutions depart more strongly from SCHC expectations. It is widely understood that $\omega$ meson production carries a significant pion-exchange contribution, which is reflected clearly in these enhanced unnatural partial waves.

We note that previous phenomenological studies, such as the handbag approach by Goloskokov and Kroll~\cite{Goloskokov2014FF}, required model-dependent assumptions regarding the natural background to estimate the unnatural pion pole contribution from SDME combinations. By contrast, our reflectivity formalism natively separates these exchanges, providing a direct confirmation of the large unnatural contribution to $\omega$ electroproduction.

The remaining electron and muon solutions are collected in Appendix~\ref{app:supplementary}, which also contains the corresponding minimizer-experimental moment comparisons. The qualitative distinction between predominantly natural $\rho^0$ solutions and more strongly mixed $\omega$ solutions persists in these additional figures, while the detailed hierarchy within a given naturality sector evolves with $Q^2$.

\subsection{Longitudinal-to-transverse ratio $R$}
\label{subsec:R_main}

\begin{figure*}[!tb]
    \centering
    \includegraphics[width=1.0\textwidth]{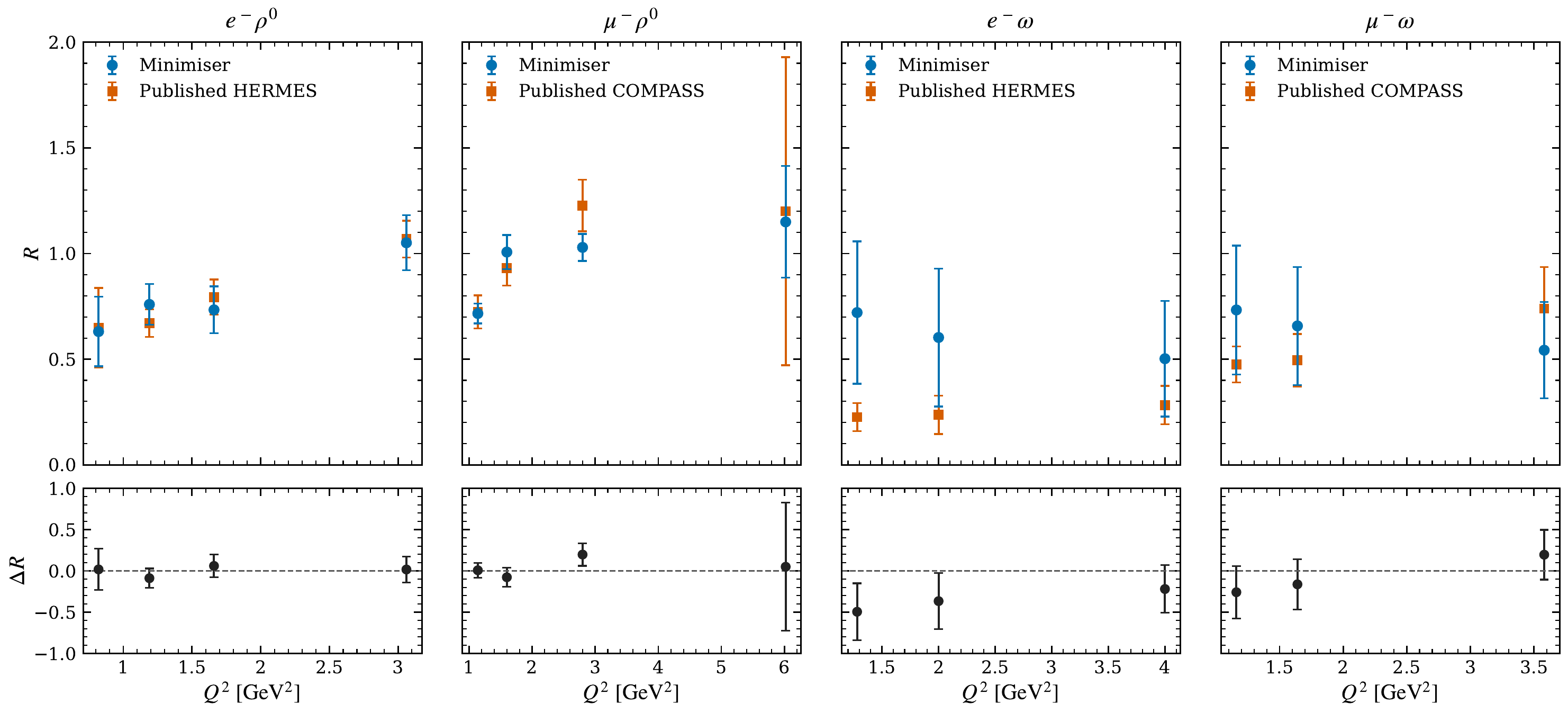}
    \caption{Comparison of the extracted longitudinal-to-transverse ratio $R=\sigma_\LL/\sigma_\TT$ with published values~\cite{HERMESrho,COMPASSrho,HERMESomega,COMPASSomega} in the $e^-\rho^0$ ($\langle W \rangle = 4.8~\mathrm{GeV}$), $\mu^-\rho^0$ ($\langle W \rangle = 9.9~\mathrm{GeV}$), $e^-\omega$ ($\langle W \rangle = 4.8~\mathrm{GeV}$), and $\mu^-\omega$ ($\langle W \rangle = 7.6~\mathrm{GeV}$) channels. The lower panels show $\Delta R$, defined as the minimizer value minus the published central value.}
    \label{fig:Rscaling}
\end{figure*}

We extract the ratio $R=\sigma_\LL/\sigma_\TT$ via Eq.~\eqref{eq:Ramp_main} and compare it with published values in Fig.~\ref{fig:Rscaling}. Published extractions rely on model-dependent approximations; more detail on this approximation can be found within the studies themselves~\cite{HERMESrho,HERMESomega,COMPASSomega,COMPASSrho}. In summary, these prior studies calculate an unpolarized proxy,  the ratio of the longitudinal and transverse meson polarization. Under strict $s$-channel helicity conservation (SCHC) this ratio is exactly $R$; however, when SCHC is broken, they differ by some correction parameter that depends on the underlying production amplitudes. These previous studies then estimated it by explicitly ignoring all unnatural helicity-flip amplitudes, thereby expressing the parameter solely in terms of experimentally accessible SDMEs. In this work, the amplitude inversion natively evaluates both natural and unnatural exchanges without such truncations. Our model instead assumes a single Reggeon exchange per reflectivity, each contributing to a different $k$ amplitude.

As seen in Fig.~\ref{fig:Rscaling}, the upper row shows the extracted $R$ value comparisons, while the lower row shows the difference $\Delta R$ between the minimizer and the published central values. Several features are apparent.

For $e^-\rho^0$, the agreement is excellent across the full range of $Q^2$. The inversion method tracks the published trend closely, and the corresponding $\Delta R$ values fluctuate around zero without a visible systematic offset. The $\mu^-\rho^0$ channel behaves similarly at low and high $Q^2$, though the minimizer lies somewhat below the published values in the central-$Q^2$ bin. However, even there, the discrepancy is modest compared with the quoted uncertainties.

The $\omega$ channels yield observationally different results, yet remain entirely consistent with the underlying physics of this channel. In $e^-\omega$, the extracted $R$ values are systematically larger than the published ones across the displayed $Q^2$ range, leading to a nearly constant negative offset in $\Delta R$. This systematic shift is driven by the large unnatural-transverse amplitude $[P]^-_{\TT; 1}$, which was inherently ignored in previous approximation schemes~\cite{HERMESrho,HERMESomega,COMPASSrho,COMPASSomega}. 

We also observe notably larger uncertainties on $R$ for the $\omega$ channel. This is a direct consequence of the lack of interference between reflectivities. In the $\rho^0$ case, the dominant natural-transverse and natural-longitudinal amplitudes interfere directly. In contrast, for $\omega$, the dominant amplitudes, unnatural-transverse and natural-longitudinal, reside in opposite reflectivity sectors and do not interfere. This fundamental lack of cross-reflectivity interference significantly reduces the experimental sensitivity to the LT interference terms. This isolation is a direct result of the longitudinal $m=0$ (spin non-flip) amplitude being purely natural, as derived in Sec.~\ref{subsec:moments_main}.

In $\mu^-\omega$, the discrepancy is smaller and evolves with $Q^2$: the minimizer starts above the published values at low $Q^2$, approaches them as $Q^2$ increases, and becomes compatible with them in the highest-$Q^2$ point. In particular, the scaling trend is negative; however, due to the limited statistics in this channel, this trend lacks high statistical significance.

\section{Generalization to polarized target and recoil nucleons}
\label{sec:polarized_nucleons_main}

The preceding sections detail the leptoproduction formalism with only the photon density matrix appearing explicitly, while the initial and recoil nucleon spins are summed over. A polarized initial nucleon target or recoil baryon polarization measurement can be included by inserting the corresponding $2\times2$ spin-density matrix for the spin-$\frac{1}{2}$ nucleon.

We evaluate these density matrices in the exact same frame that the unpolarized formalism was constructed, the rest frame of the produced meson pair. Specifically, the initial nucleon spin is quantized along its own momentum direction $\mathbf{p}^*$, while the recoil baryon spin is quantized along its momentum direction $\mathbf{p}'^*$. 
The initial and recoil polarization vectors in these respective frames are denoted by $\mathbf{S}_I$ and $\mathbf{S}_R$. Their density matrices are expanded in the Pauli basis $\sigma^\beta$ and $\sigma^\delta$, where $\beta,\delta\in\{0,1,2,3\}$ denote the identity matrix $\mathbb{I}$ and the Cartesian Pauli matrices $\bm{\sigma}=(\sigma_x,\sigma_y,\sigma_z)$:
\begin{align}
    \rho^I
    &=
    \frac{1}{2}\sum_{\beta=0}^{3}S_I^\beta\sigma^\beta
    =
    \frac{1}{2}
    \left(
        \mathbb{I}+\mathbf{S}_I\cdot\bm{\sigma}
    \right),
    \label{eq:initial_density_main}
    \\
    \rho^R
    &=
    \frac{1}{2}\sum_{\delta=0}^{3}S_R^\delta\sigma^\delta
    =
    \frac{1}{2}
    \left(
        \mathbb{I}+\mathbf{S}_R\cdot\bm{\sigma}
    \right),
    \label{eq:recoil_density_main}
\end{align}

\begin{figure}
    \centering
    \includegraphics[width=1.0\linewidth]{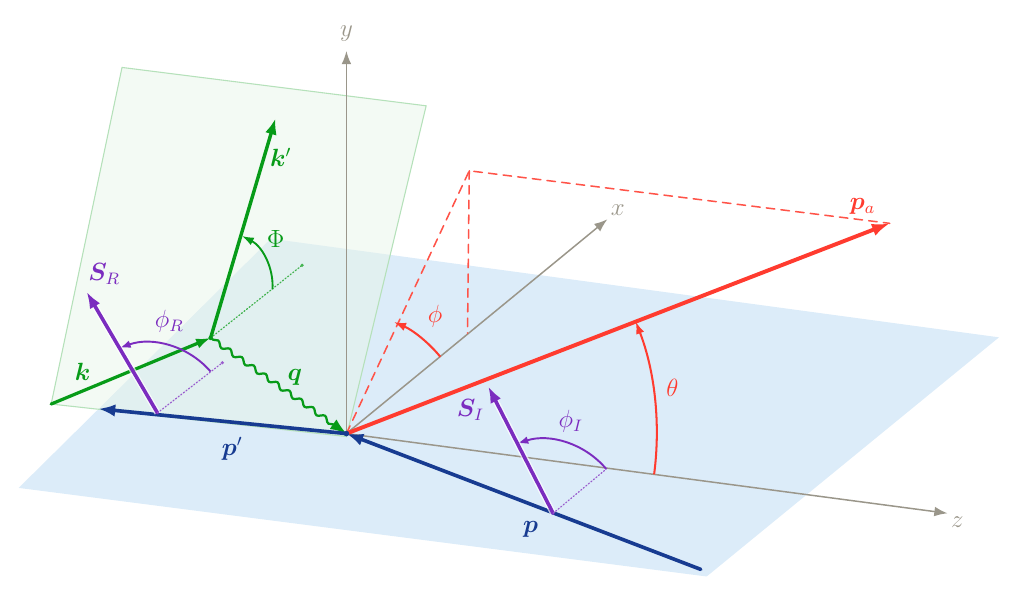}
        \caption{Helicity frame previously used for the decay analysis, now with the inclusion of the initial and recoil polarization vectors (given in purple) $\mathbf{S}_I$ and $\mathbf{S}_R$, as well as their transverse projection angles $\phi_I$ and $\phi_R$, respectively.}
    \label{fig:angle_fullpol_diagram}
\end{figure}

The components $S_I^\beta$ and $S_R^\delta$ must therefore be obtained by transforming the experimentally defined polarization vectors into the helicity frame axes. An updated version of Fig.~\ref{fig:angle_diagram} containing the updated kinematics, in which the polarization vectors for the initial and recoil nucleons are shown, is given within Fig.~\ref{fig:angle_fullpol_diagram}. The experimental prescription for determining $S_I^\beta$ and $S_R^\delta$ as per event quantities is detailed in Appendix~\ref{app:nuc_pol_trans}.

Note that our convention differs from the one of  Diehl~\cite{Diehl:2007hd} in that the target polarization was transformed via a 2D planar rotation strictly evaluated within the \emph{target rest frame}, and the $\gamma^* p$ production amplitudes are subsequently quantized in the overall $\gamma^* p$ center-of-mass frame. In contrast, our system evaluates both the production amplitudes and the meson decay partial waves directly in the \emph{two-meson rest frame}. Consequently, while the two frameworks are formally identical, the partial-wave amplitudes extracted with our conventions are numerically distinct from standard center-of-mass amplitudes, remaining related to them via kinematic Wigner rotations of the spin quantization axes.

\subsection{The full tensor intensity expansion}
With the initial and recoil spin tensors cleanly parameterized via their frame-transformed static components $S_I^\beta$ and $S_R^\delta$, we generalize the beam-response intensity distribution. The total differential intensity is expanded as a triple sum over the photon response class ($\alpha$), the initial polarization state ($\beta$), and the recoil polarization state ($\delta$):
\begin{equation}
I(\Omega,\Phi;\mathbf{S}_I,\mathbf{S}_R)
=\kappa\sum_{\alpha=0}^{8}\sum_{\beta=0}^{3}\sum_{\delta=0}^{3}
\Pi_\alpha(\Phi) S_I^\beta S_R^\delta I^{\alpha\beta\delta}(\Omega).
\label{eq:full_polarized_intensity}
\end{equation}
Here, the unbolded terms $S_I^\beta$ and $S_R^\delta$ denote the scalar components of the respective polarization vectors $(1, \mathbf{S}_I)$ and $(1, \mathbf{S}_R)$. The structural bilinear tensor SDMEs, fundamental moments, and partial intensities expand uniformly:

\begin{equation}
\rho^{\alpha\beta\delta,\ell\ell'}_{mm'}=\frac{1}{8}
\sum_{\substack{\lambda\lambda_1\lambda_2 \\ \lambda'\lambda'_1\lambda'_2}}
T^{\ell}_{\lambda m;\lambda_1\lambda_2}
\Sigma^\alpha_{\lambda\lambda'}
\sigma^\beta_{\lambda_1\lambda'_1}
\sigma^\delta_{\lambda'_2\lambda_2}
T^{\ell' *}_{\lambda' m';\lambda'_1\lambda'_2},
\label{eq:sdme_fullpol}
\end{equation}
\begin{equation}
H^{\alpha\beta\delta}(LM)=
\sum_{\ell\ell'mm'}
\left(\frac{2\ell'+1}{2\ell+1}\right)^{1/2}
C^{\ell'0}_{\ell0\,L0}
C^{\ell'm'}_{\ell m\,LM}
\rho^{\alpha\beta\delta,\ell\ell'}_{mm'},
\label{eq:H_fullpol}
\end{equation}
\begin{equation}
I^{\alpha\beta\delta}(\Omega)=
\sum_{LM}\left(\frac{2L+1}{4\pi}\right)\,H^{\alpha\beta\delta}(LM)Y_L^M(\Omega).
\label{eq:I_fullpol}
\end{equation}

The Pauli matrices and the photon response matrices are Hermitian, and so is every tensor SDME $\rho^{\alpha\beta\delta}$. Parity gives the same type of constraint as in the unpolarized-nucleon case: $\rho^{\alpha\beta\delta}_{-m,-m'}$ is related to $\rho^{\alpha\beta\delta}_{mm'}$ by the usual factor $(-1)^{m-m'}$ times a sign $\pi_{\alpha\beta\delta}=\pm1$. This sign is fixed by the photon index $\alpha$ and by the target and recoil Pauli indices $\beta$ and $\delta$. The explicit sign table is given in Appendix~\ref{app:fullpol}. Combining this parity relation with Hermiticity determines whether each moment is real or imaginary: $\pi_{\alpha\beta\delta}=+1$ gives real moments, while $\pi_{\alpha\beta\delta}=-1$ gives purely imaginary moments and removes the corresponding $M=0$ contribution.

This form also makes the common experimental reductions transparent. An unpolarized recoil with a longitudinally polarized target keeps only $\delta=0$ and $\beta=0,3$. A transversely polarized target keeps $\beta=1,2$ with weights $S^1_{I}$ and $S^2_{I}$. An unpolarized target with a longitudinally analysed recoil keeps only $\beta=0$ and $\delta=0,3$. A transversely analysed recoil keeps $\delta=1,2$ with weights $S^1_{R}$ and $S^2_{R}$. The compact reflectivity replacement rules for generating the full $9\times4\times4$ tensor are collected in Appendix~\ref{app:fullpol}.

\section{Symmetry constraints and amplitude ambiguities}
\label{sec:symmetry}

As formally derived in Appendix~\ref{app:symmetries}, the extraction of amplitudes is fundamentally constrained by continuous spatial ($SO(2)$ helicity-mixing) and complex ($U(1)$ global phase) redundancies. These continuous ambiguities must be physically broken or mathematically constrained to ensure fit convergence. In the unpolarized case, the $SO(2)$ mixing symmetry is explicitly broken by the physical assumption of a single dominant transition per reflectivity. This leaves independent $U(1)$ phase ambiguities in the non-interfering natural and unnatural sectors, which are resolved by fixing one reference amplitude phase to exactly zero in each. 

When extracting the full, untruncated $k$-basis using a singly polarized nucleon experiment, the observables retain one overall $U(1)$ phase and one unobservable $SO(2)$ mixing angle associated with the unmeasured nucleon. These two continuous flat directions are mathematically locked by applying two gauge constraints: fixing the phase of a primary reference amplitude to constrain the global rotation, and applying a secondary constraint to lock the helicity mixing. However, this effectively means that one cannot truly separate the nucleon helicity-flip and non-flip amplitudes model-independently with a polarized target alone. In contrast, in a fully polarized experiment, the simultaneous observation of both the target and recoil physically breaks all spatial $SO(2)$ redundancies through measurable interference terms. Consequently, only a single global $U(1)$ phase remains, requiring just one reference amplitude phase to be fixed to zero to unambiguously extract the complete partial-wave space. While singly polarized amplitude extractions require model-dependent assumptions to break the mixing symmetry, measuring bilinear observables, such as the polarized spherical harmonic moments, remains completely valid and model-independent. From these moments, it is still possible to derive physical quantities, such as mesonic form factors, without requiring amplitude-level symmetry breaking.

This specific topological hierarchy was recently deduced in the context of exclusive vector meson electroproduction~\cite{Singh2026}. In that framework, it was explicitly demonstrated that even with a fully polarized beam and target, the complete amplitude basis is determined only up to an exact recoil rotation unless explicit recoil polarimetry (or external threshold input) is introduced. 

The mathematical reality of this continuous mixing symmetry has a long history in the $s$-channel resonance partial-wave analysis literature. In the context of pseudoscalar meson photoproduction, Chiang and Tabakin formally proved that unobserved recoil spins generate generalized $SU(2)$ unitary transformations under which the experimental observables remain completely invariant~\cite{Chiang1997}. More recently, Wunderlich applied modern graph-theoretical criteria to the complete experiment problem in both photo- and electroproduction~\cite{PhysRevC.104.045203}. While these foundational works comprehensively examine the classification and resolution of \textit{discrete} phase ambiguities, they inherently establish that breaking the continuous $SU(2)$ or $SO(2)$ rotation—graphically represented as achieving a ``connected topology'' between disjoint amplitude subsets—is a mandatory prerequisite before any isolated amplitude can be uniquely evaluated.

Recent studies of mesonic resonance photoproduction using polarized photon beams have investigated the partial-wave amplitudes of intermediate states decaying into two spinless particles~\cite{PhysRevD.108.076001,kjcp-h8b9}. These works demonstrate that, assuming a dominant $k$-amplitude set, these partial waves can generally be extracted unambiguously. However, this extraction process is sensitive to experimental noise, which can introduce significant deviations in the resulting amplitudes. The effects of the symmetries discussed here were not considered in those works.

For experimental setups lacking either recoil polarimetry, or polarized inital nucleons, an alternative to imposing rigid bin-by-bin gauge constraints, may be to break the continuous symmetry globally across the kinematics. In the resonance region, Truncated Partial-Wave Analysis achieves this by enforcing a rigid $s$-channel angular structure (e.g., via Legendre polynomials), which analytically locks the angular dependence~\cite{PhysRevC.96.025210}. This global polynomial constraint destroys the continuous flat directions that plague isolated single-bin Complete Experiment Analyses, allowing amplitudes to be successfully extracted without requiring observables that mix target and recoil polarimetry.

A similar approach may be feasible in $t$-channel processes. For high-energy vector meson production dominated by $t$-channel exchanges, a technically equivalent global constraint can be established without relying on $s$-channel multipoles. Instead, the amplitudes can be parameterized as continuous analytic functions of the momentum transfer $t$ (e.g., diffractive exponentials), governed by strict kinematic threshold factors $\sqrt{-t}^{|\Delta \lambda|}$ for a net helicity flip $\Delta \lambda$. Because angular momentum conservation strictly forces the nucleon-flip amplitudes to zero at exactly forward scattering ($t \to 0$), the unobservable $SO(2)$ mixing angle is physically constrained at the origin. The analytic continuity of the $t$-dependent model would then propagate this structural constraint across the entire momentum transfer range. By fitting amplitudes to the extracted spherical harmonic moments globally across $t$, rather than in isolated local bins, the continuity of the physics itself may break the continuous gauge symmetry. This would theoretically permit the separation of flip and non-flip components without explicit recoil measurements, though we leave the investigation of this specific approach to future work.

\subsection{Model-independent longitudinal to transverse ratio}
\label{subsec:polarized_R}

Recall that in the unpolarized case (Sec.~\ref{subsec:R_main}), extracting the longitudinal-to-transverse cross section ratio $R$ without a Rosenbluth separation relied heavily on the assumption of Regge factorization. We were forced to suppress the unobserved nucleon helicity transitions to maintain phase coherence in the LT interference moments. 

When fully polarized data, incorporating beam, target, and recoil measurements, are available, this theoretical constraint is no longer required. The extended tensor of polarization observables provides the necessary constraints to independently extract both the helicity-non-flip ($k=1$) and helicity-flip ($k=-1$) amplitudes. 

Consequently, $R$ can be evaluated entirely model-independently by explicitly performing the incoherent sum over the measured $k$ states:
\begin{equation}
R = \frac{\sigma_\LL}{\sigma_\TT} = \frac{\displaystyle\sum_{\refl,\ell,m,k}\big|[\ell]^\refl_{\LL m;k}\big|^2}
{\displaystyle\sum_{\refl,\ell,m,k}\big|[\ell]^\refl_{\TT m;k}\big|^2} = \frac{H^{4}(00)}{H^{0}(00)},
\label{eq:Ramp_polarized}
\end{equation}
where $H^{4}(00)$ and $H^{0}(00)$ are the fully normalized, unpolarized scalar moments reconstructed from the complete amplitude solution. Because the polarized amplitude inversion natively extracts the incoherent $k$-basis, this extraction of $R$ remains rigorous even at lower energies or in kinematic regimes where multiple exchange trajectories may contribute to the total amplitudes.

\section{Relating measured moments to cross sections}
\label{sec:macro_cross_sections}

To connect the theoretical intensity distributions to the experimental observables, we map the moment expansion to the standard electroproduction cross sections~\cite{RASKIN198978}. Integrating the 7-fold differential cross section over the meson decay angles $\Omega = (\theta, \phi)$ yields the 5-fold cross section in terms of $W,Q^2,m,t$ and $\Phi$. Explicit derivations of these specific kinematic dependencies are detailed for unpolarized and polarized electron beams in Appendices \ref{app:phot_sdme} and \ref{app:pol_sdme}, respectively.

For an unpolarized initial target and unobserved recoil, several interference cross sections must strictly vanish. The cross section is parameterized by only the surviving terms, including those with an explicit dependence on the longitudinal lepton beam polarization $P_l$ (where the sign of $P_l$  denotes the beam helicity state and its magnitude represents the degree of polarization):
\begin{widetext}
\begin{equation}
\begin{aligned}
\frac{\mathrm{d}^5\sigma}{\mathrm{d}Q^2 \, \mathrm{d}W \, \mathrm{d}t \,\mathrm{d}m \, \mathrm{d}\Phi} = \Gamma(Q^2, W) \Bigg\{ 
& \frac{\mathrm{d}^2\sigma_\TT}{\mathrm{d}t \, \mathrm{d}m} + \eps \, \frac{\mathrm{d}^2\sigma_\LL}{\mathrm{d}t \, \mathrm{d}m} 
+ \eps \cos(2\Phi) \, \frac{\mathrm{d}^2\sigma_{\TT\TT}}{\mathrm{d}t \, \mathrm{d}m} \\
+ \, & \sqrt{2\eps(1+\eps)} \cos(\Phi) \, \frac{\mathrm{d}^2\sigma_{\LL\TT}}{\mathrm{d}t \, \mathrm{d}m} 
+ P_l \sqrt{2\eps(1-\eps)} \sin(\Phi) \, \frac{\mathrm{d}^2\sigma_{\LL\TT'}}{\mathrm{d}t \, \mathrm{d}m} \Bigg\}.
\end{aligned}
\label{eq:macro_cross_section_full}
\end{equation}
\end{widetext}

This restricted form mirrors the parity constraints of the partial-wave formalism. The cross sections are extracted by integrating the beam responses $I^\alpha(\Omega)$ over the solid angle $\mathrm{d}\Omega$. Because the spherical harmonics are orthogonal, the integration isolates the $L=0, M=0$ scalar moment:
\begin{equation}
\int Y_L^M(\Omega) \, \mathrm{d}\Omega = \sqrt{4\pi} \, \delta_{L0} \delta_{M0}.
\label{eq:orthonormal}
\end{equation}
Applying this property directly to Eq.~\eqref{eq:Ialpha_main} yields:
\begin{equation}
\int I^\alpha(\Omega) \, \mathrm{d}\Omega =  H^\alpha(00).
\label{eq:integrated_I_alpha}
\end{equation}

From Eq.~\eqref{eq:fullpol_parity_app}, the parity of a moment is defined by $\tau_\alpha \tau_\beta \tau_\delta$. For an unpolarized target and recoil ($\beta=\delta=0$), the responses $\alpha \in \{2, 3, 6, 7\}$ carry negative parity ($\tau_\alpha = -1$). Because purely imaginary states have vanishing $M=0$ moments, their corresponding scalar moments $H^2(00), H^3(00), H^6(00),$ and $H^7(00)$ are analytically zero. Consequently, their associated cross-section terms ($\sigma_{\TT\TT}^{\sin 2\Phi}, \sigma_{\TT\TT'}, \sigma_{\LL\TT}^{\sin \Phi},$ and $\sigma_{\LL\TT'}^{\cos \Phi}$) vanish entirely from Eq.~\eqref{eq:macro_cross_section_full}. These terms only become physically accessible when a polarized target or recoil introduces a compensating parity flip.

The surviving physical moments map directly onto the non-zero cross sections:
\begin{equation}
\begin{aligned}
\frac{\mathrm{d}^2\sigma_\TT}{\mathrm{d}t \, \mathrm{d}m} &=\kappa H^0(00), & \frac{\mathrm{d}^2\sigma_\LL}{\mathrm{d}t \, \mathrm{d}m} &= \kappa H^4(00), \\
\frac{\mathrm{d}^2\sigma_{\TT\TT}}{\mathrm{d}t \, \mathrm{d}m} &= \kappa H^1(00), & \frac{\mathrm{d}^2\sigma_{\LL\TT}}{\mathrm{d}t \, \mathrm{d}m} &= -\kappa H^5(00), \\
\frac{\mathrm{d}^2\sigma_{\LL\TT'}}{\mathrm{d}t \, \mathrm{d}m} &= -\kappa H^8(00).
\end{aligned}
\label{eq:cross_section_to_moments}
\end{equation}

Experimentally, the purely transverse ($\alpha=0$) and longitudinal ($\alpha=4$) fluxes cannot be cleanly separated without conducting a Rosenbluth separation at varying beam energies. Therefore, one may only directly measure the photon polarization-weighted moment sum:
\begin{equation}
    H^{04}(LM) = H^0(LM) + \eps H^4(LM).
\end{equation}

In an experimental analysis, the fundamental moments are typically extracted via an unbinned extended maximum likelihood fit~\cite{Barlow:1990vc} to the angular distributions. Such fits yield a set of relative, normalized moments, denoted here as $\widehat{H}^\alpha(LM)$, which we conventionally normalize such that the combined unpolarized scalar component evaluates to unity:
\begin{equation}
\widehat{H}^{04}(00) = \widehat{H}^0(00) + \eps \widehat{H}^4(00) = 1.
\label{eq:fit_normalization}
\end{equation}
To convert these fitted shapes into absolute physical cross sections, we must introduce an overall scale factor derived from the total experimental yield.

In a specific kinematic bin defined by widths $\Delta V = \Delta Q^2 \, \Delta W \, \Delta t \, \Delta m$, the total number of acceptance-corrected events, $N_{\text{events}}$, is strictly related to the integrated luminosity $\mathcal{L}$ and the bin-averaged photon flux $\bar{\Gamma}$:
\begin{equation}
N_{\text{events}} = \mathcal{L} \cdot \Delta V \cdot \bar{\Gamma} \int_0^{2\pi} \mathrm{d}\Phi \int \mathrm{d}\Omega \, I(\Omega, \Phi).
\label{eq:yield_integral}
\end{equation}
Integrating over the lepton azimuth $\Phi$ eliminates all interference terms, leaving only the combined $\alpha=04$ response scaled by a factor of $2\pi$. Subsequent integration over the solid angle $\mathrm{d}\Omega$ isolates the true physical scalar moments via Eq.~\eqref{eq:integrated_I_alpha}, providing the global yield constraint:
\begin{equation}
N_{\text{events}} = 2\pi \cdot \mathcal{L} \cdot \Delta V \cdot \bar{\Gamma} \cdot H^{04}(00).
\label{eq:norm_constraint}
\end{equation}

By recognizing that the true physical moments are proportional to the fitted moments ($H^\alpha = \mathcal{N} \widehat{H}^\alpha$), we substitute the fit normalization convention from Eq.~\eqref{eq:fit_normalization} into Eq.~\eqref{eq:norm_constraint}. This yields the experimental scale factor $\mathcal{N}$:
\begin{equation}
\mathcal{N} = \frac{N_{\text{events}}}{2\pi \cdot \mathcal{L} \cdot \Delta V \cdot \bar{\Gamma}}.
\label{eq:scale_factor}
\end{equation}
Notice the factor of $2\pi$ in the denominator, which naturally arises from the azimuthal integration over $\Phi$, combined with our $\widehat{H}^{04}(00) \equiv 1$ normalization convention. The isolated absolute differential cross sections are then obtained simply by multiplying the extracted scalar moments by this kinematic scale factor (accounting for the sign mappings of Eq.~\eqref{eq:cross_section_to_moments}):
\begin{equation}
\frac{\mathrm{d}^2\sigma^\alpha}{\mathrm{d}t \, \mathrm{d}m} = \pm \,\mathcal{N} \cdot \widehat{H}^\alpha(00).
\label{eq:final_scaled_cross_sections}
\end{equation}
Thus, by scaling the normalized likelihood fit to the physical event yield and integrated luminosity, the fitted angular moments directly determine the complete cross section basis of the exclusive reaction.

\subsection{Generalization to polarized initial and recoil cross sections}
\label{subsec:polarized_cross_sections}
The formalism mapping the unpolarized nucleon moments to the cross sections can be systematically extended to encompass measurements with polarized initial targets ($\beta \in \{1,2,3\}$) and observed polarized recoils ($\delta \in \{1,2,3\}$). Rather than expanding the 5-fold differential cross section into an unwieldy sum of hundreds of individual terms, it is elegantly expressed using the generalized polarization-weighted intensity.

Following the structure of Eq.~\eqref{eq:macro_cross_section_full}, the fully polarized cross section is given by the sum over the complete $\alpha \otimes \beta \otimes \delta$ response space:

\begin{equation}
\begin{split}
&\frac{\mathrm{d}^5\sigma}{\mathrm{d}Q^2 \, \mathrm{d}W \, \mathrm{d}t \,\mathrm{d}m \, \mathrm{d}\Phi} \\
&\quad = \Gamma(Q^2,W)\sum_{\alpha,\beta,\delta} \Pi^\alpha(\Phi, P_l) \, S_I^\beta \, S_R^\delta \, \frac{\mathrm{d}^2\sigma^{\alpha\beta\delta}}{\mathrm{d}t \, \mathrm{d}m},
\end{split}
\label{eq:full_polarized_cross_section}
\end{equation}

Here, $\Pi^\alpha(\Phi, P_l)$ defines the virtual photon polarization components (containing the kinematic $\Phi$ and lepton polarization $P_l$ dependencies), while $S_I^\beta = (1, \mathbf{S}_I)$ and $S_R^\delta = (1, \mathbf{S}_R)$ represent the four-component polarization vectors of the initial target and recoil baryon evaluated in their respective helicity frames.

As established in Sec.~\ref{sec:formalism}, the scalar moments $H^\alpha(00)$ for $\alpha \in \{2, 3, 6, 7\}$ vanish identically in the unpolarized limit due to parity conservation. However, in the generalized framework, the survival of a cross section $\sigma^{\alpha\beta\delta}$ is governed by the total parity product $\tau_\alpha \tau_\beta \tau_\delta = +1$.

When a polarized target ($\beta > 0$) or recoil ($\delta > 0$) is introduced, the $\tau$-sign flips provide the necessary parity compensation to mathematically activate these imaginary interference structure functions. For example, a target polarized normal to the scattering plane ($y$-axis, $\beta=2$) carries a negative naturality signature ($\tau_2 = -1$). When paired with the imaginary LT interference photon state ($\alpha=6$, $\tau_6 = -1$) and an unobserved recoil ($\delta=0, \tau_0 = +1$), the total parity is $(-1)(-1)(+1) = +1$. 

For example, the scalar moment $H^{620}(00)$ survives the parity constraint, and the recipe yields:
\begin{equation}
\frac{\mathrm{d}^2\sigma_{\LL\TT}^{\sin\Phi, y}}{\mathrm{d}t \, \mathrm{d}m} = \mathcal{N} \cdot \widehat{H}^{620}(00).
\end{equation}
This term directly isolates the target Single-Spin Asymmetry cross section. By mapping the observables back to the fundamental parity product $\tau_\alpha \tau_\beta \tau_\delta$, the partial-wave moment formalism naturally dictates the complete topography of standard polarized structure functions, unequivocally predicting which observables require single or double nucleon polarization to become experimentally accessible.
Finally, it is important to emphasize a critical operational distinction in the extraction of these cross sections. If experimental angular moments are extracted directly from the decay distributions without an underlying amplitude model, only the combined scalar moment $\widehat{H}^{0400}(00)$ is measurable at a single beam energy. Consequently, such an approach only yields the unseparated cross-section sum $\sigma_\TT + \eps\sigma_\LL$. 

Alternatively, if the data are fitted directly at the amplitude level, or if the experimental moments are subsequently inverted into partial-wave amplitudes, the underlying physical constraints of the amplitude model natively separate the photon responses. From the extracted amplitudes, the isolated scalar moments $H^{000}(00)$ and $H^{400}(00)$ can be independently recalculated. By then scaling the likelihood fit to the physical event yield and integrated luminosity via Eq.~\eqref{eq:scale_factor}, these recalculated amplitude-level moments determine the separated transverse and longitudinal cross sections, entirely circumventing the need for a beam-energy-varying Rosenbluth separation. Furthermore, this also works for the earlier unpolarized case, but it hinges upon some model or approximation, like the single $k$-amplitude contribution made here within the results of Sec.~\ref{sec:results}, to extract physically sensible results.

\section{Conclusions}
\label{sec:conclusion}
We have developed a comprehensive partial-wave formalism based on reflectivity amplitudes to extract production amplitudes for meson electroproduction. By fully generalizing this approach to include polarized initial targets and recoil baryon polarization measurements, we provide a complete, diagonalized kinematic description of these exclusive reactions. The formalism we derived may be applied to experiments in different ways. In some cases, it will make sense to extract the partial wave amplitudes directly. These can then be directly related to helicity amplitudes or Compton form factors depending on the application. On the other hand, one may prefer to extract only the moments $H^{\alpha\beta\delta}(LM)$ directly and then apply mass-dependent amplitude models via the given relationship between partial waves and moments. The latter method allows one to use correlations in the mass dependence to extract more reliable amplitudes with fewer artifacts from noise-related distortions in mass- or $t$-independent fits.

A major advantage of extracting amplitudes via these angular moments is the ability to natively separate longitudinal and transverse cross sections within a single experimental configuration, when Regge factorization occurs. This entirely circumvents the experimental overhead traditionally required for beam-energy-varying Rosenbluth separations. We demonstrated that at high energies, assuming a single Regge exchange per reflectivity, one may extract the longitudinal-to-transverse cross section ratio $R$ directly from the amplitudes. Applying this methodology to existing unpolarized Spin Density Matrix Elements from HERMES and COMPASS, we extracted amplitudes consistent with the expectations of Pomeron exchange in the natural-parity sector. Furthermore, for the specific case of $\omega$ meson production, the reflectivity basis successfully isolated the large unnatural-parity pion exchange contribution without requiring further model-dependent background approximations.

However, while this framework provides a powerful tool for separating these responses and evaluating the ratio $R = \sigma_\LL/\sigma_\TT$ via their contribution to interference moments, it is clear that fully polarized target and recoil measurements provide the most unbiased truth for the complete set of amplitudes. Measuring the fully polarized target and recoil observables is mathematically required to break the continuous $SO(2)$ parameter ambiguities that will otherwise hinder unpolarized or singly polarized amplitude extractions. With full polarization, the complete set of nucleon helicity-flip and non-flip transitions can be separated unambiguously, allowing $R$ and the underlying amplitudes to be extracted entirely model-independently. While we have established the complete theoretical foundation for these fully polarized observables here, performing a detailed numerical extraction of the full, incoherent $k$-basis amplitudes from polarized data lies beyond the scope of the present manuscript. The practical complexities of such an analysis, including the required multidimensional fits, continuous ambiguity resolution, and polarization systematics, warrant a dedicated future publication. By formally laying the mathematical groundwork now, we provide the necessary blueprint for these comprehensive amplitude extractions in upcoming experimental programs.

This framework offers significant utility across multiple domains of hadron physics. For meson spectroscopy, electroproduction provides an expanded set of measurable angular observables compared to real photoproduction, enabling tighter constraints on the partial waves associated with specific resonances. Mapping the invariant mass and $Q^2$ dependence of these isolated waves grants direct access to resonance poles, electrocouplings, and internal spatial structures. Simultaneously, extending these measurements to the high-$Q^2$ regime with polarized targets leverages this formalism to unravel the three-dimensional structure of the nucleon. Looking toward the high-$Q^2$ regime, extracting Compton form factors directly from these partial-wave amplitudes provides a rigorous pathway for three-dimensional nucleon tomography. Mapping the spatial distribution of partons via these Generalized Parton Distributions remains a central priority of the hadron physics community.

While the present formalism is explicitly derived for meson decays into two spin-zero mesons, a natural extension of this framework will incorporate Wigner $D$-matrices to handle decay products with intrinsic spin. This generalization is essential for analyzing $J/\psi$ electroproduction, as well as vector-pseudoscalar and vector-vector decay topologies. Because our methodology accommodates the production of mesons of arbitrary spin, it naturally opens the door for future measurements involving tensor mesons and hidden-charm exotic states. Expanding this formalism to these topologies will be a critical asset for driving experimental programs in exotic spectroscopy and the extraction of transition form factors, offering entirely new theoretical constraints for active and future facilities, including CLAS12, GlueX, and the Electron-Ion Collider.

Furthermore, the target and recoil polarization states are currently parameterized using Pauli matrices, restricting the strict exclusive derivations to spin-$1/2$ baryons. Expanding this spin-density matrix formalism to accommodate higher-spin recoil baryons, such as the $\Delta(1232)$ or other $N^*$ resonances, would yield a fully comprehensive set of polarization observables. Such an expansion would enable the simultaneous, unified study of baryonic excitations alongside the meson production mechanism. Finally we note that the effect of radiative effects on the distributions is beyond the scope of this work, but will need to be considered for a full analysis.

\acknowledgments
This material is based on work supported by the UK Science and Technology Facilities Council under grant ST/V00106X/1.
V.M. is a Serra Húnter Professor and acknowledges support from the Spanish national Grant CNS2022-136085. G.M. is a Beatriu de Pinós Fellow, supported by the grant \mbox{BP 2024 00189}, financed by AGAUR. The work of V.M. and G.M. is additionally supported by the Spanish MICIU grants \mbox{PID2023-147112NB-C21} and \mbox{CEX2024-001451-M} (Unidad de Excelencia ``María de Maeztu''). This material is based upon work supported by the U.S. Department of Energy, Office of Science, Office of Nuclear Physics under Contract No. 89243126CSC000213. This work contributes to the aims of the USDOE ExoHad Topical Collaboration, contract DE-SC0023598.  We acknowledge the use of AI Large Language Models for developing and reviewing the text.
\appendix

\section{ELECTROPRODUCTION FORMALISM}
\label{app:electroformalism}
\subsection{The virtual photon cross section}
\label{app:fieldtheory}
We consider the generic process
\begin{equation}
\ell(k,h)+N(p,\lambda_1)\to \ell'(k', h')+N'(p',\lambda_2)+M(q'),
\end{equation}
in which $\ell,\ell'$ are the leptons (electrons or muons), $N,N'$ are the nucleons and $M(q')$ denotes collectively all mesons produced whose total momentum is $q'$. If $n$ mesons are produced, the differential cross section reads 
\begin{align} \nonumber
    \diff \sigma & = \frac{\delta^4(P_i-P_f)}{4 m_N E (2\pi)^2} \frac{\diff^3 \bm k'}{2E'}\frac{\diff^3 \bm p'}{2E'_N} \frac{\diff^3 \bm q'}{2E'_M} 
    \\
    &\qquad \times 
    \frac{\diff m^2}{(2\pi)^{3n}} \prod_{i=1}^n \frac{\diff^3 \bm p_i}{2E_i}  \overline{|{\cal M}^\text{el}|^2} \delta^4(q' - P), 
    \label{eq:diff_xsec1}
\end{align}
with $\bm p_i$ the $n$ momenta of the mesons, their total momentum is $P = \sum_i p_i$ and $p_\text{lab}\equiv E$ is the electron beam momentum in the laboratory frame. There is a summation over the helicities not written explicitly and an average over the beam and target helicities in $\overline{|{\cal M}^\text{el}|^2}$. 

The first line represents the three-particle phase space that can be evaluated using standard techniques. We substitute the phase-space variables with the photon virtuality $Q^2 = -q^2 = -(k-k')^2$, the $\gamma^* N$ total invariant mass squared $s = W^2 = (q+p)^2$, the azimuthal angle between the leptonic and hadronic planes $\Phi$, and the angles $\Omega^*$ of the recoiling nucleon in the center-of-mass frame of the hadronic process:
\begin{align} 
    \diff \sigma & = \frac{\diff \Omega^*}{4\pi} \frac{\diff \Phi}{2\pi}\frac{2p^*_N\diff Q^2 \diff W }{(8m_N E)^2} 
    \times \diff \Phi_n, 
\end{align}
where we denote by $\diff \Phi_n$ the second line of Eq.~\eqref{eq:diff_xsec1}. We can further simplify the expression using $2p^*_\gamma p^*_N \diff \Omega^* = \diff \phi^* \diff t$, where the virtual photon momentum in the $\gamma^* N$ center-of-mass frame is given by $p^*_\gamma = \lambda^{1/2}(W^2,-Q^2,m_N^2)/2W$. We introduce the K\"all\'en function $\lambda(a,b,c) = a^2+b^2+c^2-2ab-2bc-2ca$ to define the virtual photon flux factor $\widetilde{F}^2_V = F_V\lambda^{1/2}(W^2,-Q^2,m_N^2)$, where $F_V = \lambda^{1/2}(W^2,0,m_N^2)$ is the Hand conventional flux. This allows us to define the electroproduction cross section conventionally as 
\begin{align} 
    \diff \sigma & = \frac{\diff \phi^*\diff t}{4\pi} \frac{\diff \Phi}{2\pi}\frac{2W\diff Q^2 \diff W }{\widetilde F^2_V(8m_N E)^2} 
    \times \diff \Phi_n.
\end{align}
There are 5 independent variables describing the inclusive electroproduction process: $(Q^2,\Phi,t, E,W)$. The differential cross section depends on the azimuthal angle of the recoiling proton $\phi^*$ only if the target is polarized. Typically, the invariant mass squared $s = W^2$ of the $\gamma^* N$ system is replaced by the Bjorken scaling variable $x = Q^2/(W^2-m_N^2+Q^2)$; here we present both whilst sticking only to $W$ within the main sections of the paper. The initial state energy is completely fixed by specifying the beam momentum $p_\text{lab}$ in the rest frame of the target nucleon, or equivalently by the virtual photon polarization parameter Eq.~\eqref{eq:epsilon_main}. If $n$ mesons are produced, there are $3(n-1)$ additional kinematic variables describing $\diff \Phi_n$.

In the single-photon approximation, the electroproduction amplitudes ${\cal M}^\text{el}$ are related to the virtual photon amplitudes ${\cal M}_{\lambda\lambda_1\lambda_2}$ by 
\begin{align}
    \mathcal M^\text{el}= ie \bar u(k',h')\gamma_\mu u(k,h) \varepsilon^{\mu*}_\lambda(q)\frac{\eta^{\lambda\lambda'}}{Q^2} {\cal M}_{\lambda'\lambda_1\lambda_2}. 
    \label{eq:Mfield_app}
\end{align}
The virtual photon amplitude is the contraction of the photon polarization vector with the hadronic current, ${\cal M}_{\lambda\lambda_1\lambda_2} = \varepsilon^{\mu}_\lambda(q) J^\mu_{\lambda_1,\lambda_2}$. The metric tensor $\eta_{\lambda\lambda'}$ is diagonal with its only non-zero elements being $\eta_{\pm1\pm1} = -\eta_{00} = 1$. We perform the sum over the final lepton helicities by introducing the virtual photon spin density matrix:
\begin{align}
    \frac{1}{4}\sum_\text{hel.} \left|{\cal M}^{\text{el}}\right|^2 & = \frac{e^2}{Q^4}  \sum_{\lambda,\lambda', \lambda_1,\lambda_2} {\cal M}_{\lambda\lambda_1\lambda_2} \rho^{\gamma^*}_{\lambda\lambda'}  {\cal M}^{*}_{\lambda'\lambda_1\lambda_2}. 
\end{align}
The virtual photon density matrix $\rho^{\gamma^*}_{\lambda\lambda'}$ depends on the lepton beam polarization and is a Hermitian $3\times3$ matrix that can be decomposed into a 9-dimensional basis:
\begin{equation}
\rho^{\gamma^*}_{\lambda\lambda'}(\Phi)= \frac{1}{2}
\sum_{\alpha=0}^{8}\Pi^\alpha(\Phi)\,\Sigma^\alpha_{\lambda\lambda'}. 
\label{eq:rhosigmabasis_app}
\end{equation}
To match this to the work of Schilling and Wolf the virtual photon density matrix has been normalized by the unit flux of transverse photons $Q^2/(1-\varepsilon)$. The coefficients $\Pi^\alpha(\Phi)$ depend on the experimental beam conditions and are evaluated in Appendix~\ref{app:phot_sdme}. The kinematic parameter $\varepsilon$, is defined in the standard manner, and the basis matrices $\Sigma^\alpha$ are explicitly listed in Appendix~\ref{app:basis}.

The virtual photon cross section is conventionally defined as 
\begin{align}
    \diff \sigma^* & = \frac{Q^2(1-\varepsilon)\diff\Phi\diff t}{32 \alpha \tilde F^2_V }\diff \Phi_n, 
\end{align}
with $\alpha = e^2/4\pi$. Collecting everything together, we arrive at the standard convention expressing the electroproduction cross section in terms of the virtual photon cross section:
\begin{align}
    \frac{\diff \sigma}{\diff x\diff\Phi \diff Q^2\diff t} & = \Gamma(Q^2,x)\frac{\diff \sigma^*}{\diff\Phi\diff t},\\
    \frac{\diff \sigma}{\diff W\diff\Phi \diff Q^2\diff t} & = \Gamma(Q^2,W) \frac{\diff \sigma^*}{\diff\Phi\diff t}, 
    \label{eq:app_factorised_cross_sectio}
\end{align}
with the virtual photon flux factor given by 
\begin{align}
    \Gamma(Q^2,x) & = \frac{\alpha}{2\pi} \frac{Q^2}{4m_N^2 E} \frac{1-x}{x^3} \frac{1}{1-\varepsilon},\\
    \Gamma(Q^2,W) & = \frac{\alpha}{2\pi} \frac{W\left(W^2 - m^2_N\right)}{2m_N^2 EQ^2} \frac{1}{1-\varepsilon},
    \label{eq:app_photon_flux}
\end{align}
and the corresponding virtual cross section producing $n$ mesons written as 
\begin{align} \nonumber
    \frac{\diff \sigma^*}{\diff\Phi\diff t} = \frac{1}{64\pi^2 \tilde F^2_V}  \frac{\diff m^2}{(2\pi)^{3n}} \prod_{i=1}^n \frac{\diff^3 \bm p_i}{2E_i} \delta^4(q' - P)
    \\
    \times 
    \sum_{\alpha=0}^{8} \Pi^\alpha(\Phi) \sum_{\lambda,\lambda', \lambda_1,\lambda_2} {\cal M}_{\lambda\lambda_1\lambda_2} \Sigma^\alpha_{\lambda\lambda'}  {\cal M}^{*}_{\lambda'\lambda_1\lambda_2}. 
\end{align}
For the specific case explored in this paper, the final state will always involve two pseudoscalar mesons, labeled $a$ and $b$ with an invariant mass of the total system given by $m$. Thus, we can fully expand the virtual cross section to account for the two body final state
\begin{align}
     \frac{\diff \sigma^*}{\diff\Phi\diff t} &= \frac{1}{64\pi^2 \tilde F^2_V}  \frac{\diff m^2}{(2\pi)^{6}}\diff\Omega\frac{\lambda^{1/2}(m^2, m^2_a,m^2_b)}{8m^2}\nonumber \\
    &\times 
    \sum_{\alpha=0}^{8} \Pi^\alpha(\Phi) \sum_{\lambda,\lambda', \lambda_1,\lambda_2} {\cal M}_{\lambda\lambda_1\lambda_2} \Sigma^\alpha_{\lambda\lambda'}  {\cal M}^{*}_{\lambda'\lambda_1\lambda_2}.
\end{align}
This can be seen to be just the standard 2 body phase space solution for particle decay $M(q')$ at rest. This means that the virtual photon cross section can then be expressed as
\begin{align}
    \frac{\diff \sigma^*}{\diff\Phi\diff t \diff m \diff\Omega} &= I(\Omega,\Phi) = \frac{\lambda^{1/2}(m^2, m^2_a,m^2_b)}{64(2\pi)^5\tilde F^2_V m}\nonumber \\
    &\times 
    \sum_{\alpha=0}^{8} \Pi^\alpha(\Phi) \sum_{\lambda,\lambda', \lambda_1,\lambda_2} {\cal M}_{\lambda\lambda_1\lambda_2} \Sigma^\alpha_{\lambda\lambda'}  {\cal M}^{*}_{\lambda'\lambda_1\lambda_2}\nonumber\\
    &=  \kappa\sum_{\alpha=0}^{8} \Pi^\alpha(\Phi)I^\alpha(\Omega),
    \label{eq:app_intensity_def}
\end{align}
where all the phase terms are collected into a single phase-space factor $\kappa$. For the specific case here where we have two pseudoscalar mesons in the final state,
\begin{equation}
    \kappa = \frac{\lambda^{1/2}(m^2, m^2_a,m^2_b)}{32(2\pi)^5\tilde F^2_V m}.
    \label{eq:app_phase_factor}
\end{equation}

\subsection{Unpolarized electron beam}
\label{app:phot_sdme}
In the case of an unpolarized electron beam, the spin density matrix of the virtual photon is given by 
\begin{align}
    \rho^{\gamma^*}_{\lambda\lambda'} & = 
    \left[ k_\mu k'_{\bar \mu} + k'_\mu k_{\bar \mu} -(Q^2/2) g_{\mu \bar \mu}\right] \varepsilon^\mu_{\sigma} \varepsilon^{\bar \mu*}_{\sigma'}  \eta^{\sigma \lambda}\eta^{\sigma' \lambda'}. 
\end{align}
The spin density is invariant under any boost along the $z$-axis. Let us evaluate it in the Breit frame, where the virtual photon momentum is purely space-like, $q^\mu = (0,0,0,Q)$. In this frame, the lepton momenta are 
\bsub\begin{align}
    k &= (E_B, k_x \cos \Phi, k_x \sin \Phi, +Q/2), 
    \\
    k' &= (E_B, k_x \cos \Phi, k_x \sin \Phi, -Q/2), 
\end{align}\esub
and the polarization vectors of the virtual photon are specified by 
\bsub\begin{align}
    \varepsilon_\pm &= (0,\mp1,-i,0)/\sqrt{2}, \\
    \varepsilon_0 & = (1,0,0,0).
\end{align}\esub
The Breit frame kinematic components can be expressed in terms of invariant parameters via 
\begin{align}
    E^2_B & = \frac{Q^2}{4} \frac{1+\eps+2\delta}{1-\eps}, 
    &
    k_x^2 & = \frac{Q^2}{2} \frac{\eps}{1-\eps}. 
\end{align}
Here we have introduced the parameter $\delta$ related directly to the non-zero lepton mass $m_\ell$:
\begin{align}
    \delta & = \frac{2 m_\ell^2}{Q^2} (1-\varepsilon).
\end{align}
For an unpolarized electron beam, the response vector coefficients are found to be
\begin{align}
    \Pi^{0-4}(\Phi)
    &=
    \left[
    1,
    -\eps\cos 2\Phi,
    -\eps\sin 2\Phi,
    0,
    \eps+\delta
    \right],
    \\
    \Pi^{5-8}(\Phi)
    &=
    \sqrt{2\eps(1+\eps+2\delta)}
    \left[
    \cos\Phi,
    \sin\Phi,
    0,
    0
    \right].
\end{align}
The full virtual photon cross section in Eq.~\eqref{eq:app_intensity_def} can therefore be fully specified from these kinematic terms as well as the angular response functions $I^\alpha$. These angular response functions can then be defined through their moment expansion,
\begin{align}
    I^{\alpha}(\Omega)
    &\equiv
    \sum_{\lambda,\lambda',\lambda_1,\lambda_2}
    {\cal M}_{\lambda\lambda_1\lambda_2}(\Omega)
    \Sigma^\alpha_{\lambda\lambda'}
    {\cal M}^{*}_{\lambda'\lambda_1\lambda_2}(\Omega)
    \nonumber\\
    &=
    \sum_{L,M}
    \left(\frac{2L+1}{4\pi}\right)\,
    H^\alpha(LM)Y_L^M(\Omega).
    \label{eq:app_response_moment_expansion}
\end{align}
The corresponding unpolarized contribution to the intensity is

\begin{widetext}
    \begin{align}
    I_{\mathrm{unpol}}(\Omega,\Phi)
    &\equiv
    \kappa\!\!\!\!\!\!\!\!\!\sum_{\alpha\in\{0,1,2,4,5,6\}}\!\!\!\!\!\!\!\!\!\!
    \Pi^\alpha(\Phi)I^\alpha(\Omega)\nonumber\\
    &=\sum_{L,M}
    \left(\frac{2L+1}{4\pi}\right)Y_L^M(\Omega)\Big[H^0(LM)-\eps H^1(LM)\cos2\Phi\nonumber -\eps H^2(LM)\sin2\Phi
    +(\eps+\delta)H^4(LM)\nonumber\\
    &\hspace*{6cm} + \sqrt{2\eps(1+\eps+2\delta)}
    \left(
        H^5(LM)\cos\Phi
        +H^6(LM)\sin\Phi
    \right)\Big].
    \label{eq:app_unpolarized_intensity_moments}
\end{align}
\end{widetext}

Since $Y_0^0=1/\sqrt{4\pi}$, integration over the complete decay solid angle selects the corresponding zeroth moment,
\begin{align}
    H^\alpha(00)
    &=
    \int\diff\Omega\,I^\alpha(\Omega),
\end{align}
which are directly related to the differential cross sections:
\bsub\begin{align}
    \frac{\diff\sigma_\TT}{\diff t\diff m}
    &=\kappa H^0(00),
    &
    \frac{\diff\sigma_{\TT\TT}}{\diff t\diff m}
    &=-\kappa H^1(00),
    \\
    \frac{\diff\sigma_\LL}{\diff t\diff m}
    &=\kappa H^4(00),
    &
    \frac{\diff\sigma_{\LL\TT}}{\diff t\diff m}
    &=\kappa H^5(00),
    \\
    \frac{\diff\widetilde\sigma_{\TT\TT}}{\diff t\diff m}
    &=-\kappa H^2(00),
    &
    \frac{\diff\widetilde\sigma_{\LL\TT}}{\diff t\diff m}
    &=\kappa H^6(00).
\end{align}\label{eq:app_integrated_cross_sections_moments}\esub
where $\kappa$ is the phase-space factor defined in
Eq.~\eqref{eq:app_phase_factor}. We now integrate directly over the complete decay solid angle and define the decay-angle-integrated hadronic tensor
\begin{align}
    W_{\lambda\lambda'}
    &=
    \sum_{\lambda_1,\lambda_2}
    \int\diff\Omega\,
    {\cal M}_{\lambda\lambda_1\lambda_2}(\Omega)
    {\cal M}^{*}_{\lambda'\lambda_1\lambda_2}(\Omega).
    \label{eq:app_integrated_hadronic_tensor}
\end{align}
The individual component cross sections are defined through the integrated helicity-amplitude products as
\begin{align}
    \frac{\diff\sigma_\TT}{\diff t\diff m}
    &=
    \frac{\kappa}{16\pi \widetilde F_V^2}\frac{1}{4}
    \left(
    W_{++}+W_{--}
    \right),
    \\
    \frac{\diff\sigma_{\TT\TT}}{\diff t\diff m}
    &=
    -\frac{\kappa}{16\pi \widetilde F_V^2}\frac{1}{2}
    \operatorname{Re}W_{+-},
    \\
    \frac{\diff\widetilde\sigma_{\TT\TT}}{\diff t\diff m}
    &=
    -\frac{\kappa}{16\pi \widetilde F_V^2}\frac{1}{2}
    \operatorname{Im}W_{+-}=0,
    \\
    \frac{\diff\sigma_\LL}{\diff t\diff m}
    &=
    \frac{\kappa}{16\pi \widetilde F_V^2}\frac{1}{2}
    W_{00},
    \\
    \frac{\diff\sigma_{\LL\TT}}{\diff t\diff m}
    &=
    \frac{\kappa}{16\pi \widetilde F_V^2}\frac{\sqrt{2}}{4}
    \operatorname{Re}
    \left(
    W_{0+}-W_{0-}
    \right),
    \\
    \frac{\diff\widetilde\sigma_{\LL\TT}}{\diff t\diff m}
    &=
    -\frac{\kappa}{16\pi \widetilde F_V^2}\frac{\sqrt{2}}{4}
    \operatorname{Im}
    \left(
    W_{0+}+W_{0-}
    \right)=0.
\end{align}
Parity relates the amplitude at $\Omega$ to that at the corresponding parity-reflected decay angle $\Omega\rightarrow\Omega' = (\pi-\theta,\pm\pi+\phi)$. Since the complete angular measure is invariant under this transformation, the angular dependence has been integrated out, parity conservation and hermiticity imply
\begin{align}
    W_{-\lambda,-\lambda'}
    &=
    (-1)^{\lambda-\lambda'}W_{\lambda\lambda'},
    &
    W_{\lambda'\lambda}
    &=
    W^*_{\lambda\lambda'}.
    \label{eq:app_hadronic_tensor_symmetries}
\end{align}
For the transverse--transverse interference, these relations give
\begin{align}
    W_{-+}
    &=
    W_{+-}
    =
    W^*_{+-},
    \label{eq:app_W_1}
\end{align}
implying $\operatorname{Im}W_{+-}=0$.
Hence, the $\alpha=2$ response, which is proportional to
$\operatorname{Im}W_{+-}$ and multiplies $\sin2\Phi$, vanishes. Similarly, parity gives
\begin{align}
    W_{0-}
    &=
    -W_{0+},
    \label{eq:app_W_2}
\end{align}
so that
the $\alpha=6$ response, which is proportional to
$\operatorname{Im}(W_{0+}+W_{0-})$ and multiplies $\sin\Phi$, therefore also vanishes. These contributions need not vanish at a fixed $\Omega$, since parity relates different decay-angle configurations, but cancel when integrated over the complete decay solid angle.

The decay-angle-integrated virtual photon cross section can consequently be written in the standard structural form (where the null $\alpha=2$ and $\alpha=6$ terms no longer contribute)
\begin{align}
    \frac{\diff\sigma^*_{\mathrm{unpol}}}
    {\diff\Phi\diff t\diff m}
    &=
    \kappa\int\diff\Omega\,
    I_{\mathrm{unpol}}(\Omega,\Phi)
    \nonumber\\
    &=
    \frac{\diff\sigma_\TT}{\diff t\diff m}
    +(\eps+\delta)
    \frac{\diff\sigma_\LL}{\diff t\diff m}
    +\eps
    \frac{\diff\sigma_{\TT\TT}}{\diff t\diff m}
    \cos2\Phi
    \nonumber\\
    &\quad
    +\sqrt{2\eps(1+\eps2\delta)}
    \frac{\diff\sigma_{\LL\TT}}{\diff t\diff m}
    \cos\Phi.
    \label{eq:app_unpolarized_structural_cross_section}
\end{align}

For the energies explored here, the lepton-mass parameter can be neglected since $E^2\gg m_\ell^2$, and it will therefore be omitted hereafter.

\subsection{Polarized electron beam}
\label{app:pol_sdme}
A fermion with momentum $\bm k$ has two helicity states. The standard helicity $\pm \frac{1}{2}$ states are quantized along the direction of its 3-momentum. For these states, the spin 4-vector is $\pm s^\mu$ with $s^\mu = (1/m_\ell) (|\bm k |, E \hat{\bm k} )$. The lepton beam can be polarized such that its state is a normalized linear combination of the pure helicity states:
\begin{align}
 u(k, P_l) & = A_+ u_+(k) + A_- u_-(k), 
 & A_+^2 +A_-^2 = 1.
\end{align}
The relation between the polarization vector $P$ and the helicity vector $s$ is given by $A_\pm^2  =(1/2) (1 \mp P\cdot s)$. Note that since $s^2 = -1$, a state with $P = \pm s$ leads to $A_\pm = 1$ and $A_\mp = 0$. If the lepton at rest is quantized along the direction $\bm n$ ($\bm n^2 = 1$), after the boost to bring its momentum to $k=(E,\bm k)$, the polarization vector becomes 
\begin{align} \label{eq:PfromN}
 P^\mu & = \left(\frac{\bm k \cdot \bm n}{m_\ell}, \bm n + \frac{\bm k \cdot \bm n}{m_\ell(E+m_\ell)} \bm k \right) \simeq (P_l/m_\ell) k^\mu, 
\end{align}
where $P_l = \bm k\cdot \bm n/|\bm k|$ denotes the degree of longitudinal polarization. The polarization vector $s^\mu$ matches $P^\mu$ with $\bm n = \bm k/|\bm k|$, meaning $P_l=1$ for a fully longitudinally polarized lepton. For a given polarization vector $P^\mu$, the leptonic cross section calculation is proportional to the projection operator:
\begin{align}
\sum_ku(k, P) \bar u(k, P) & = \frac{1}{2} \left( 1+ \gamma_5 \slashed{P} \right)  \left( \slashed{k} + m_\ell \right). 
\end{align}
The spin density matrix for the virtual photon gains an extra term depending directly on this lepton polarization:
\begin{align}
    \rho^{\gamma^*}_{\lambda\lambda'} \to \rho^{\gamma^*}_{\lambda\lambda'} + \tilde \rho^{\gamma^*}_{\lambda\lambda'}. 
\end{align}
Its explicit expression reads 
\begin{align}
 \tilde \rho^{\gamma^*}_{\lambda\lambda'} & = m_\ell i\epsilon_{\alpha\bar\alpha \mu\nu}  q^\mu P^\nu \varepsilon^\alpha_{\sigma}\varepsilon^{\bar \alpha*}_{\sigma'}  \eta^{\sigma \lambda}\eta^{\sigma' \lambda'}. 
\end{align}
In the Breit frame, if we implement the ultrarelativistic approximation $P^\mu \simeq (P_l/m_\ell) k^\mu$ and neglect the lepton mass terms, the polarization-dependent part of the density matrix simplifies to:

\begin{align} \label{eq:rho_pol}
    \tilde \rho^{\gamma^*}_{\lambda\lambda'} & =  \frac{P_l}{2}\sqrt{1-\eps} \begin{pmatrix}
     \sqrt{1+\eps} & \sqrt{\eps} e^{-i \Phi} & 0 \\
      \sqrt{\eps} e^{i \Phi} & 0 &  \sqrt{\eps} e^{-i \Phi} \\
    0 & \sqrt{\eps} e^{i \Phi} &- \sqrt{1+\eps}
    \end{pmatrix}. 
\end{align}
The response vector components $\Pi^\alpha(\Phi)$ induced by the beam polarization are
\bsub\begin{align}
    \Pi^{3}(\Phi) & = P_l \sqrt{1-\eps^2},\\
    \Pi^{7}(\Phi) & = P_l \sqrt{2\eps(1-\eps)} \cos\Phi,\\
    \Pi^{8}(\Phi) & = P_l \sqrt{2\eps(1-\eps)} \sin\Phi,
\end{align}\esub
The corresponding polarized contribution to the intensity is
\begin{widetext}
    \begin{align}
    &I_{\mathrm{pol}}(\Omega,\Phi)
    \equiv
    \Pi^3(\Phi)I^3(\Omega)
    +\Pi^7(\Phi)I^7(\Omega)
    +\Pi^8(\Phi)I^8(\Omega)
    \nonumber\\
    &=
    \sum_{L,M}
    \left(\frac{2L+1}{4\pi}\right)\,
    Y_L^M(\Omega)P_l\Big[
        \sqrt{1-\eps^2}\,H^3(LM)
        +\sqrt{2\eps(1-\eps)}
        \left(
            H^7(LM)\cos\Phi
            +H^8(LM)\sin\Phi
        \right)
    \Big].
    \label{eq:app_full_polarized_intensity_moments}
\end{align}
\end{widetext}

The total intensity and its corresponding moments are therefore given by
\begin{align}
    I(\Omega,\Phi)
    &\equiv
    I_{\mathrm{unpol}}(\Omega,\Phi)
    +I_{\mathrm{pol}}(\Omega,\Phi).
    \label{eq:app_total_intensity_moments}
\end{align}
The decay-angle integral of the polarized contribution is determined by the three zeroth moments,
\begin{align}
    \frac{\diff\sigma^*_{\mathrm{pol}}}
    {\diff\Phi\diff t\diff m}
    &=
    \kappa
    \int\diff\Omega\,
    I_{\mathrm{pol}}(\Omega,\Phi)
    \nonumber\\
    &=
    \kappa P_l
    \Big[
        \sqrt{1-\eps^2}\,H^3(00)
        \nonumber\\
    &+\sqrt{2\eps(1-\eps)}
        \left(
            H^7(00)\cos\Phi
            +H^8(00)\sin\Phi
        \right)
    \Big].
    \label{eq:app_integrated_polarized_cross_section_moments}
\end{align}

Following integration over the complete decay solid angle, the polarized responses can be expressed in terms of the hadronic tensor defined in Eq.~\eqref{eq:app_integrated_hadronic_tensor}:
\begin{align}
    \frac{\diff\sigma_{\TT\TT'}}{\diff t\diff m}
    &=
    \kappa H^3(00)
    \nonumber\\
    &=
    \frac{\kappa}{16\pi \widetilde F_V^2}\frac{1}{4}
    \left(
        W_{++}-W_{--}
    \right)
    =
    0,
    \\
    \frac{\diff\widetilde\sigma_{\LL\TT'}}
    {\diff t\diff m}
    &=
    \kappa H^7(00)
    \nonumber\\
    &=
    \frac{\kappa}{16\pi \widetilde F_V^2}\frac{\sqrt{2}}{4}
    \operatorname{Re}
    \left(
        W_{0+}+W_{0-}
    \right)
    =
    0,
    \\
    \frac{\diff\sigma_{\LL\TT'}}
    {\diff t\diff m}
    &=
    \kappa H^8(00)
    \nonumber\\
    &=
    -\frac{\kappa}{16\pi \widetilde F_V^2}
    \frac{1}{4\sqrt{2}}\,
    \operatorname{Im}
    \left(
        W_{0+}-W_{0-}
    \right).
    \label{eq:app_polarized_lt_cross_section}
\end{align}
These can easily be inferred from the relations within Eqs.~\eqref{eq:app_W_1}~and~\eqref{eq:app_W_2}.
Including the single non-zero additional contribution, the decay-angle-integrated virtual photon cross section becomes
\begin{align}
    \frac{\diff\sigma^*}
    {\diff\Phi\diff t\diff m}
    &=
    \kappa\int\diff\Omega\,
    I(\Omega,\Phi)
    \nonumber\\
    &=
    \frac{\diff\sigma_\TT}{\diff t\diff m}
    +\eps
    \frac{\diff\sigma_\LL}{\diff t\diff m}
    +\eps
    \frac{\diff\sigma_{\TT\TT}}{\diff t\diff m}
    \cos2\Phi
    \nonumber\\
    &\quad
    +\sqrt{2\eps(1+\eps)}
    \frac{\diff\sigma_{\LL\TT}}{\diff t\diff m}
    \cos\Phi
    \nonumber\\
    &\quad
    +P_l\sqrt{2\eps(1-\eps)}
    \frac{\diff\sigma_{\LL\TT'}}
    {\diff t\diff m}
    \sin\Phi.
    \label{eq:app_polarized_structural_cross_section}
\end{align}
\subsection{Response-basis decomposition}
\label{app:basis}

The nine Hermitian basis matrices entering Eq.~\eqref{eq:rhosigmabasis_app} are chosen according to the decomposition introduced by Schilling and Wolf~\cite{schilling_eprod}:
\begin{widetext}
\begin{equation}
\begin{aligned}
\Sigma^0&=\begin{pmatrix}1&0&0\\0&0&0\\0&0&1\end{pmatrix},
&\Sigma^1&=\begin{pmatrix}0&0&1\\0&0&0\\1&0&0\end{pmatrix},
&\Sigma^2&=\begin{pmatrix}0&0&-\ii\\0&0&0\\\ii&0&0\end{pmatrix},\\[1.2ex]
\Sigma^3&=\begin{pmatrix}1&0&0\\0&0&0\\0&0&-1\end{pmatrix},
&\Sigma^4&=2\begin{pmatrix}0&0&0\\0&1&0\\0&0&0\end{pmatrix},
&\Sigma^5&=\frac{1}{\sqrt2}\begin{pmatrix}0&1&0\\1&0&-1\\0&-1&0\end{pmatrix},\\[1.2ex]
\Sigma^6&=\frac{1}{\sqrt2}\begin{pmatrix}0&-\ii&0\\\ii&0&\ii\\0&-\ii&0\end{pmatrix},
&\Sigma^7&=\frac{1}{\sqrt2}\begin{pmatrix}0&1&0\\1&0&1\\0&1&0\end{pmatrix},
&\Sigma^8&=\frac{1}{\sqrt2}\begin{pmatrix}0&-\ii&0\\\ii&0&-\ii\\0&\ii&0\end{pmatrix}. 
\end{aligned}
\label{eq:SigmaBasis_app}
\end{equation}
\end{widetext}
Here the column/row indices go in the order $1,0,-1$. The first four matrices span the purely transverse sector, $\Sigma^4$ isolates the purely longitudinal response, and $\Sigma^{5,6,7,8}$ generate the four LT interference components.

\section{ANGULAR DISTRIBUTIONS AS INTENSITIES AND MOMENTS}
\label{app:angular_ints_moments}
The previous system can be expanded in terms of partial-wave amplitudes $T^\ell_{m\lambda\lambda_1\lambda_2}$ multiplying a Wigner $D$-matrix, which in the two-pseudoscalar case investigated here simplifies to an expansion in shperical harmonics,
\begin{equation}
   \mathcal{M}_{\lambda\lambda_1\lambda_2}(\Omega)=\sum_{\ell m} T^\ell_{m\lambda\lambda_1\lambda_2}Y^\ell_m(\Omega)
\end{equation}
where $\Omega = (\theta,\phi)$ are the decay angles for one of the two final-state mesons in the two-meson rest frame.  The spin density matrix for the virtual photon can then be decomposed in $\alpha$, as was done in the previous appendices, separating out each polarization contribution. With the basis vectors of Eq.~\eqref{eq:SigmaBasis_app} and the expansion coefficents $\Pi^\alpha$ can be written in terms of angular response functions, which in turn depend on the moments of each spherical harmonic contribution $H^\alpha(LM)$. As for these moments, they can then be expressed as a linear sum over the SDMEs
\begin{equation}
    H^\alpha(LM) = \sum_{\ell\ell'mm'}\left(\frac{2\ell'+1}{2\ell+1}\right)^{1/2} C^{l0}_{\ell'0L0}C^{\ell m}_{\ell'm'LM}\rho^{\alpha,\ell\ell'}_{mm'},
\end{equation}
where $C^{\ell 0}_{\ell'0L0}$ and $C^{\ell m}_{\ell'm'LM}$ are Clebsch-Gordan coefficients that impose the selection rules of our process --  they impose that $L+\ell+\ell'$ be an even integer and restrict the summation to $M+m'=m$. The spin density matrices are then given as bilinears of the partial-wave amplitudes,
\begin{equation}
    \rho^{\alpha,\ell\ell'}_{mm'} = \frac{1}{2}\sum_{\substack{\lambda\lambda' \\ \lambda_1\lambda_2}}\sum_{\substack{\ell\ell' \\ mm'}}T^{\ell}_{\lambda m; \lambda_1 \lambda_2} \Sigma^\alpha_{\lambda\lambda'} T^{\ell'*}_{\lambda' m'; \lambda_1 \lambda_2},
    \label{eq:app_Moments_SDME}
\end{equation}
where we can specifically expand out over the basis matrices of Eq.~\eqref{eq:SigmaBasis_app} to get the explicit SDMEs. The first four components are already known~\cite{Mathieu2019}
\bsub\begin{align}
\label{eq:sdme_T}
    \rho^{0,\ell\ell'}_{mm'}&=\frac{1}{2}\sum_{\lambda=\pm1}\sum_{\lambda_1\lambda_2}T^{\ell}_{\lambda,m;\lambda_1\lambda_2}T^{\ell' *}_{\lambda,m';\lambda_1\lambda_2},
    \\
    \rho^{1,\ell\ell'}_{mm'}&=\frac{1}{2}\sum_{\lambda=\pm1}\sum_{\lambda_1\lambda_2}T^{\ell}_{-\lambda,m;\lambda_1\lambda_2}T^{\ell' *}_{\lambda,m';\lambda_1\lambda_2},
    \\
    \rho^{2,\ell\ell'}_{mm'}&=\frac{\ii}{2}\sum_{\lambda=\pm1}\sum_{\lambda_1\lambda_2}\lambda\,T^{\ell}_{-\lambda,m;\lambda_1\lambda_2}T^{\ell' *}_{\lambda,m';\lambda_1\lambda_2},
    \\
    \rho^{3,\ell\ell'}_{mm'}&=\frac{1}{2}\sum_{\lambda=\pm1}\sum_{\lambda_1\lambda_2}\lambda\,T^{\ell}_{\lambda,m;\lambda_1\lambda_2}T^{\ell' *}_{\lambda,m';\lambda_1\lambda_2}.
    \label{eq:sdme_T_end}
\end{align}\esub
The other five components depend on the longitudinal components of the virtual photon and only contribute to electroproduction processes:
\begin{widetext}
\bsub\begin{align}
\rho^{4,\ell\ell'}_{mm'}&=\sum_{\lambda_1\lambda_2} T^{\ell}_{0,m;\lambda_1\lambda_2}T^{\ell' *}_{0,m';\lambda_1\lambda_2},
    \\
    \rho^{5,\ell\ell'}_{mm'}&=\frac{1}{2\sqrt2}\sum_{\lambda=\pm1}\sum_{\lambda_1\lambda_2}\lambda\Big(T^{\ell}_{0,m;\lambda_1\lambda_2}T^{\ell' *}_{\lambda,m';\lambda_1\lambda_2}+T^{\ell}_{\lambda,m;\lambda_1\lambda_2}T^{\ell' *}_{0,m';\lambda_1\lambda_2}\Big),
    \\
    \rho^{6,\ell\ell'}_{mm'}&=\frac{\ii}{2\sqrt2}\sum_{\lambda=\pm1}\sum_{\lambda_1\lambda_2}\Big(T^{\ell}_{0,m;\lambda_1\lambda_2}T^{\ell' *}_{\lambda,m';\lambda_1\lambda_2}-T^{\ell}_{\lambda,m;\lambda_1\lambda_2}T^{\ell' *}_{0,m';\lambda_1\lambda_2}\Big),
    \\
    \rho^{7,\ell\ell'}_{mm'}&=\frac{1}{2\sqrt2}\sum_{\lambda=\pm1}\sum_{\lambda_1\lambda_2}\Big(T^{\ell}_{0,m;\lambda_1\lambda_2}T^{\ell' *}_{\lambda,m';\lambda_1\lambda_2}+T^{\ell}_{\lambda,m;\lambda_1\lambda_2}T^{\ell' *}_{0,m';\lambda_1\lambda_2}\Big),
    \\
    \rho^{8,\ell\ell'}_{mm'}&=\frac{\ii}{2\sqrt2}\sum_{\lambda=\pm1}\sum_{\lambda_1\lambda_2}\lambda\Big(T^{\ell}_{0,m;\lambda_1\lambda_2}T^{\ell' *}_{\lambda,m';\lambda_1\lambda_2} - T^{\ell}_{\lambda,m;\lambda_1\lambda_2}T^{\ell' *}_{0,m';\lambda_1\lambda_2}\Big).
    \label{eq:sdme_L}
\end{align}\esub
\end{widetext}
The equations here are defined within the helicity frame, as the amplitudes and SDMEs are dependent on which frame is used. However, the formalism is equally applicable to any frame in which the produced resonance is at rest, for example the Gottfried-Jackson frame. These frames are simply related by a rotation around the $y$-axis and can thus be related by
\begin{align}
\left.\rho^{\alpha,\ell\ell'}_{mm'}\right|_{\mathrm{GJ}}
&=
\sum_{\lambda\lambda'}
d^{\ell}_{m\lambda}(\theta_q)\,
\left.\rho^{\alpha,\ell\ell'}_{\lambda\lambda'}\right|_{\mathrm{hel}}\,
d^{\ell'}_{m'\lambda'}(\theta_q),
\\[1ex]
\left.H^\alpha(LM)\right|_{\mathrm{GJ}}
&=
\sum_{M'}
\left.H^\alpha(LM')\right|_{\mathrm{hel}}\,
d^{L}_{MM'}(\theta_q).
\end{align}
The cosine of the angle between the frames is given by $\cos\theta_q=(\beta - z_s)/(\beta z_s - 1)$ with the boost parameter given by  $\beta=\lambda^{1/2}\!\left(s,m_N^2,m^2\right)/
\left(s - m_N^2 + m^2\right)$,
and $z_s = \cos\theta_s$ being the cosine of the scattering angle between the target and recoiling nucleon in the centre-of-mass frame.

By definition, the spin density matrix is hermitian $\left(\rho^{\alpha,\ell'\ell}_{m'm}\right)^* = \rho^{\alpha,\ell\ell'}_{mm'}$, this then means that for the moments $\left[H^\alpha(LM)\right]^* = (-1)^M H^\alpha(L-M)$. The decay angles transform under parity as $(\theta,\phi)\rightarrow(\pi-\theta,\pm\pi+\phi)$ which results in the spherical harmonics transforming as $Y^m_\ell(\Omega)\rightarrow(-1)^\ell Y^m_\ell(\Omega)$. By taking into account the intrinsic parity of our pseudoscalar particles, our partial-waves transform under parity as
\begin{equation}
\label{eq:parity_amp}
    T^\ell_{-\lambda-m;-\lambda_1-\lambda_2} = (-1)^{\lambda+m+\lambda_1-\lambda_2}T^\ell_{\lambda m;\lambda_1\lambda_2}.
\end{equation}
By exploiting parity and the properties of the Clebsch-Gordan coefficients, a set of relations for the SDMEs can be derived:
\begin{align}
    \rho^{\alpha,\ell\ell'}_{mm'} = \begin{cases}
      \phantom{+}(-1)^{m-m'}\rho^{\alpha,\ell\ell'}_{-m-m'}   &\alpha\in\{0,1,4,5,8\}\\
      -(-1)^{m-m'}\rho^{\alpha,\ell\ell'}_{-m-m'}  & \alpha\in\{2,3,6,7\},
    \end{cases}
\end{align}
for which similar relations for the moments follow:
\begin{align}
    H^\alpha(LM) = \begin{cases}
      \phantom{+}(-1)^MH^\alpha(L\,{-M})  &\alpha\in\{0,1,4,5,8\}\\
      -(-1)^{M}H^\alpha(L\,{-M})   & \alpha\in\{2,3,6,7\}.
    \end{cases}
\end{align}
From these relations, the moments are then purely real for $\alpha=0,1,4,5,8$ and imaginary for $\alpha=2,3,6,7$. The intensities can then be written as
\begin{align} 
    I^\alpha(\Omega)  = \sum_{L,M\geq0}& \left( \frac{2L+1}{4\pi}\right)\tau(M)H^\alpha(LM)
    d^L_{M0}(\theta)\cos M\phi
\end{align}
for $\alpha = 0,1,4,5,8$ and 
\begin{align} 
    I^\alpha(\Omega)  = -2 \sum_{L,M>0}& \left( \frac{2L+1}{4\pi}\right)\text{Im } H^\alpha(LM)
    d^L_{M0}(\theta)\sin M\phi
\end{align}
for $\alpha = 2,3,6,7$. The symbol $\tau(M) = 2-\delta_{M,0}$. These can then be inverted to calculate the moments by performing an inverse integral transform
\bsub\begin{align}
    H^0(LM) &= N^0\int_{\circ} I(\Omega,\Phi) d^L_{M0}(\theta)\cos M\phi \\
    H^1(LM) &= N^1\int_{\circ} I(\Omega,\Phi) d^L_{M0}(\theta)\cos M\phi \cos 2\Phi \\
    H^2(LM) &= N^2\int_{\circ} I(\Omega,\Phi) d^L_{M0}(\theta)\sin M\phi \sin 2\Phi \\
    H^3(LM) &= N^3\int_{\circ} I(\Omega,\Phi) d^L_{M0}(\theta)\sin M\phi\\
    H^4(LM) &= N^4\int_{\circ} I(\Omega,\Phi) d^L_{M0}(\theta)\cos M\phi \\
    H^5(LM) &= N^5\int_{\circ} I(\Omega,\Phi) d^L_{M0}(\theta)\cos M\phi \cos\Phi \\
    H^6(LM) &= N^6\int_{\circ} I(\Omega,\Phi) d^L_{M0}(\theta)\sin M\phi \sin\Phi \\
    H^7(LM) &= N^7 \int_{\circ} I(\Omega,\Phi) d^L_{M0}(\theta)\sin M\phi\cos\Phi\\
    H^8(LM) &= N^8 \int_{\circ} I(\Omega,\Phi) d^L_{M0}(\theta)\cos M\phi\sin\Phi
\end{align}\label{eq:app_moments_fourier}\esub
where $\int_\circ = \int^1_{-1} \diff(\cos\theta)\int^{2\pi}_0 \diff\phi \int^{2\pi}_0\diff\Phi$. The normalization factors $N^\alpha$ can be read off the response function 
\bsub\begin{align}
    N^0 & = \frac{1}{2\pi}, &
    N^1 & = \frac{-1}{\pi\varepsilon}, \\
    N^2 & = \frac{1}{\pi\varepsilon}, &
    N^3 & = \frac{-1}{\pi\sqrt{1-\varepsilon^2} P_l}, \\
    N^4 & = \frac{1}{2\pi\varepsilon}, &
    N^5 & = \frac{1}{\pi\sqrt{2\varepsilon(1+\varepsilon)}},\\
    N^6 & = \frac{-1}{\pi\sqrt{2\varepsilon(1+\varepsilon)}},&
    N^7 & = \frac{-1}{\pi\sqrt{2\varepsilon(1-\varepsilon)}P_l},\\
    N^8 & = \frac{1}{\pi\sqrt{2\varepsilon(1-\varepsilon)}P_l}
\end{align}\label{eq:app_moments_fourier_norms}\esub
Only the combination $H^0 + \varepsilon H^4$ is measurable unless the $\varepsilon$ dependence is studied via Rosenbluth separation. The other polarization components have distinguishable angular dependence.

\section{THE REFLECTIVITY BASIS}
\label{app:reflectivity}
The reflectivity basis was introduced in Ref.~\cite{Mathieu2019} for purely transverse photon. We extend this definition to longitudinal photon as
\bsub\begin{align}
[\ell]^{\refl}_{\TT m;\lambda_1\lambda_2}
&= \frac{1}{2}\left(T^{\ell}_{1m; \lambda_1\lambda_2} - 
\refl(-1)^{m}T^{\ell}_{-1-m; \lambda_1\lambda_2}\right), 
\\
[\ell]^\refl _{\LL m;\lambda_1\lambda_2}
&= \frac{1}{2}\left(T^{\ell}_{0m;\lambda_1\lambda_2} + 
\refl(-1)^{m}T^{\ell}_{0-m; \lambda_1\lambda_2}\right),
\end{align}\esub
with inverse relations given by
\begin{align} \nonumber
    T^\ell_{+1 m;\lambda_1\lambda_2} &= [\ell]^{(+)}_{\TT m;\lambda_1\lambda_2} + [\ell]^{(-)}_{\TT m;\lambda_1\lambda_2}
    \\  \nonumber
    T^\ell_{-1m;\lambda_1\lambda_2} &= (-1)^m\left(
     [\ell]^{(-)}_{\TT -m;\lambda_1\lambda_2} - [\ell]^{(+)}_{\TT -m;\lambda_1\lambda_2}\right)
    \\
    T^\ell_{0m;\lambda_1\lambda_2} &= [\ell]^{(+)}_{\LL m;\lambda_1\lambda_2} + [\ell]^{(-)}_{\LL m;\lambda_1\lambda_2}
\end{align}
With this definition, the longitudinal state possesses an inherent symmetry such that $[\ell]^{\refl} _{\LL -m;\lambda_1\lambda_2} = \epsilon (-1)^m \, [\ell]^{\refl} _{\LL m;\lambda_1\lambda_2}$. Consequently, for longitudinal photon states with $m=0$, the reflected partner is not mathematically independent. The $12(2\ell+1)$ original amplitudes $T^\ell_{\lambda m;\lambda_1 \lambda_2}$ are now divided into $8(2\ell+1)$ amplitudes $[\ell]^{\refl} _{\TT m;\lambda_1\lambda_2}$ with all projections $-\ell \ge m \ge \ell$, $4 (\ell+1)$ amplitudes $[\ell]^{(+)}_{\LL m;\lambda_1\lambda_2}$ with positive projections $\ell \ge m\ge 0$ and $4 \ell$ amplitudes $[\ell]^{(-)}_{\LL m;\lambda_1\lambda_2}$ with strictly positive projection $\ell \ge m> 0$. Of course, although we are using the same notation $m$, the meaning of the projections $m$ in the helicity basis $T^\ell_{\lambda m;\lambda_1\lambda_2}$ and in the reflectivity basis $[\ell]^{\refl} _{\TT/\LL m;\lambda_1\lambda_2}$ are not the same. 

In the high-energy limit, for the specific case of a meson decaying to two pseudoscalars, the reflectivity eigenvalue tracks the naturality of the exchanged $t$-channel trajectory, as demonstrated for photoproduction in  Ref.~\cite{Mathieu2019}. For electroproduction this just requires accounting for the $\lambda=0$ helicity state, which just results in a sign change to the overall relation. Hence, this is why there is a sign flip in the definition of the longitudinal reflectivity basis Eq.~\eqref{eq:long_ref_main}, to account for this sign change and preserve the naturality-reflectivity equivalence. Thus, within electroproduction we have that transverse states and longitudinal states with $m \neq 0$ are given by $\eta_{\mathrm{ex}} = \refl \eta_M$, where $\eta$ denotes the naturalities. However, because the longitudinal $m=0$ state is mathematically restricted to positive reflectivity ($\refl = +1$), it isolates exchanges where the naturality strictly matches that of the produced meson ($\eta_{\mathrm{ex}} = \eta_M$).
This is the main motivation for performing this transformation, the reflectivity basis tracks the naturality of the exchange at high energies.

Parity conservation for the virtual-photon subprocess imposes a strict constraint on the target and recoil nucleon helicities. The conservation of parity \eqref{eq:parity_amp} yields:
\begin{align}
[\ell]^\refl _{\TT/\LL m;-\lambda_1-\lambda_2} & = \epsilon (-1)^{\lambda_1-\lambda_2} [\ell]^\refl _{\TT/\LL m;\lambda_1\lambda_2}
\end{align}
Because flipping both nucleon helicities simply reproduces the original amplitude (up to a phase), only two of the four possible initial-to-final nucleon helicity combinations are mathematically independent.

To streamline the formalism, we formally define a reduced nucleon basis indexed by $k \in \{-1, 1\}$, which explicitly isolates these two independent transitions. The production amplitudes with transverse (T)/longitudinal(L) photon will be denoted as $[\ell]^{(\epsilon)}_{\TT/\LL m;k}$. The index $k$ directly identifies the net helicity flip of the nucleon vertex. We define the helicity-non-flip transition as $k=1$ and the helicity-flip transition as $k=-1$:
\begin{align} \label{eq:kbasisT_main}
[\ell]^\refl _{\TT/\LL m;\pm 1}&\equiv[\ell]^\refl _{\TT/\LL m;+\pm}
\end{align}
The two remaining unlisted helicity states ($--$ and $-+$) are fully determined by the parity relations in Eq.~\eqref{eq:paritylawL_main}. Note that this convention differs slightly from previous photoproduction frameworks~\cite{Mathieu2019} by explicitly assigning $k=-1$ to the spin-flip state, a choice that greatly simplifies the generalized tensor projections for polarized nucleons developed in Sec.~\ref{sec:polarized_nucleons_main}.

By transforming the amplitudes into this reduced $k$-basis, another strength of this basis can be exploited in that it establishes a fully diagonalized framework where the reflectivity $\refl$ acts as a strict eigenvalue. To see this diagonalization for both $\epsilon$ and $k$, first expand out the sum of any bilinear combination of reflectivity amplitudes, then apply the parity relation to recast the reflectivity interference terms to obtain the condition
\begin{align}
    \sum_{\lambda_1\lambda_2} ^\refl [\ell]^\refl_{\substack{\lambda m\\ \lambda_1 \lambda_2}} [\ell']^{\refl'*}_{\substack{\lambda' m'\\ \lambda_1 \lambda_2}}
    & = 2\delta_{\refl,\refl'}\sum_{k}[\ell]^\refl _{\lambda m;k}[\ell']^{\refl'*}_{\lambda' m';k}.
\end{align}
The indices $\lambda,\lambda'$ runs over the four combinations TT, LL, TL and LT. 
More details on this and a proof of the diagonalization (as well as more details on the more general case where nucleon polarization is accounted for) is provided in Appendix~\ref{app:fullpol}. As a consequence of this diagonality, the interference terms vanish and all experimental moments, intensities, and SDMEs can be constructed by summing incoherently over the reflectivity states:
\begin{equation}
\begin{aligned}
\rho^{\alpha,\ell\ell'}_{mm'}&=\ ^{(+)}\rho^{\alpha,\ell\ell'}_{mm'}+\ {}^{(-)}\rho^{\alpha,\ell\ell'}_{mm'},\\
I^{\alpha}(\Omega)&=\ {}^{(+)}I^{\alpha}(\Omega)+\ {}^{(-)}I^{\alpha}(\Omega),\\
H^{\alpha}(LM)&=\ {}^{(+)}H^{\alpha}(LM)+\ {}^{(-)}H^{\alpha}(LM),
\end{aligned}
\label{eq:reflsum_main}
\end{equation}
with no quantum interference permitted between the two reflectivity components due to parity conservation of the strong interactions. 

In this basis, the SDMEs can be read as
\begin{widetext}
\bsub\label{eq:refl_billinears_full}\begin{align}
\rho^{0,\ell\ell'}_{mm'}
&=
\sum_{k,\refl}
\Big(
[\ell]^{\refl}_{\TT m;k}
[\ell']^{\refl*}_{\TT m';k}
+
(-1)^{m'-m}
[\ell]^{\refl}_{\TT,-m;k}
[\ell']^{\refl*}_{\TT,-m';k}
\Big),
\\
\rho^{1,\ell\ell'}_{mm'}
&=
-\sum_{k,\refl}\refl
\Big(
(-1)^m
[\ell]^{\refl}_{\TT,-m;k}
[\ell']^{\refl*}_{\TT m';k}
+
(-1)^{m'}
[\ell]^{\refl}_{\TT m;k}
[\ell']^{\refl*}_{\TT,-m';k}
\Big),
\\
\rho^{2,\ell\ell'}_{mm'}
&=
-\ii\sum_{k,\refl}\refl
\Big(
(-1)^m
[\ell]^{\refl}_{\TT,-m;k}
[\ell']^{\refl*}_{\TT m';k}
-
(-1)^{m'}
[\ell]^{\refl}_{\TT m;k}
[\ell']^{\refl*}_{\TT,-m';k}
\Big),
\\
\rho^{3,\ell\ell'}_{mm'}
&=
\sum_{k,\refl}
\Big(
[\ell]^{\refl}_{\TT m;k}
[\ell']^{\refl*}_{\TT m';k}
-
(-1)^{m'-m}
[\ell]^{\refl}_{\TT,-m;k}
[\ell']^{\refl*}_{\TT,-m';k}
\Big),
\\
\rho^{4,\ell\ell'}_{mm'}
&=
2\sum_{k,\refl}
[\ell]^{\refl}_{\LL m;k}
[\ell']^{\refl*}_{\LL m';k},
\\
\rho^{5,\ell\ell'}_{mm'}
&=
\frac{1}{\sqrt{2}}\sum_{k,\refl}
\Bigg[
[\ell]^{\refl}_{\LL m;k}
[\ell']^{\refl*}_{\TT m';k}
+
[\ell]^{\refl}_{\TT m;k}
[\ell']^{\refl*}_{\LL m';k}
\notag\\
&\hspace{31mm}
+\refl\Big(
(-1)^{m'}
[\ell]^{\refl}_{\LL m;k}
[\ell']^{\refl*}_{\TT,-m';k}
+
(-1)^m
[\ell]^{\refl}_{\TT,-m;k}
[\ell']^{\refl*}_{\LL m';k}
\Big)
\Bigg],
\\
\rho^{6,\ell\ell'}_{mm'}
&=
\frac{\ii}{\sqrt{2}}\sum_{k,\refl}
\Bigg[
[\ell]^{\refl}_{\LL m;k}
[\ell']^{\refl*}_{\TT m';k}
-
[\ell]^{\refl}_{\TT m;k}
[\ell']^{\refl*}_{\LL m';k}
\notag\\
&\hspace{31mm}
-\refl\Big(
(-1)^{m'}
[\ell]^{\refl}_{\LL m;k}
[\ell']^{\refl*}_{\TT,-m';k}
-
(-1)^m
[\ell]^{\refl}_{\TT,-m;k}
[\ell']^{\refl*}_{\LL m';k}
\Big)
\Bigg],
\\
\rho^{7,\ell\ell'}_{mm'}
&=
\frac{1}{\sqrt{2}}\sum_{k,\refl}
\Bigg[
[\ell]^{\refl}_{\LL m;k}
[\ell']^{\refl*}_{\TT m';k}
+
[\ell]^{\refl}_{\TT m;k}
[\ell']^{\refl*}_{\LL m';k}
\notag\\
&\hspace{31mm}
-\refl\Big(
(-1)^{m'}
[\ell]^{\refl}_{\LL m;k}
[\ell']^{\refl*}_{\TT,-m';k}
+
(-1)^m
[\ell]^{\refl}_{\TT,-m;k}
[\ell']^{\refl*}_{\LL m';k}
\Big)
\Bigg],
\\
\rho^{8,\ell\ell'}_{mm'}
&=
\frac{\ii}{\sqrt{2}}\sum_{k,\refl}
\Bigg[
[\ell]^{\refl}_{\LL m;k}
[\ell']^{\refl*}_{\TT m';k}
-
[\ell]^{\refl}_{\TT m;k}
[\ell']^{\refl*}_{\LL m';k}
\notag\\
&\hspace{31mm}
+\refl\Big(
(-1)^{m'}
[\ell]^{\refl}_{\LL m;k}
[\ell']^{\refl*}_{\TT,-m';k}
-
(-1)^m
[\ell]^{\refl}_{\TT,-m;k}
[\ell']^{\refl*}_{\LL m';k}
\Big)
\Bigg].
\label{eq:reflbilinears_app}
\end{align}\esub
\end{widetext}
As can be seen above, those with $\alpha=0-3$ are given purely by transverse partial wave amplitudes; $\alpha=4$ by purely longitudinal; $\alpha=5,6$ by unpolarized beam LT interference and, $\alpha=7,8$ by polarized beam interference of L/T partial waves.

\section{AMPLITUDE FACTORIZATION FOR UNPOLARIZED NUCLEONS}
\label{app:factorization_coherence}

In the analysis of unpolarized electroproduction and photoproduction data, the target and recoil nucleon helicities remain unobserved. Consequently, the physical intensities and SDMEs must be constructed by summing incoherently over the nucleon helicity transitions, $k \in \{-1, 1\}$. 

To circumvent the ambiguities of an underconstrained system, amplitude analyses frequently parameterize the problem using a single effective amplitude. This formulation projects the unpolarized target sum onto a fully coherent state, allowing the complete kinematic determination of partial waves without a Rosenbluth separation. This appendix outlines the strict mathematical conditions under which this approximation is justified via Regge factorization, and explicitly details the mechanism by which the assumption fails when factorization breaks down.

\subsection{Regge factorization and the preservation of coherence}

Let the generalized state of the upper vertex and meson decay be denoted by a composite index $a \equiv (c, m)$, where $c \in \{\TT, \LL\}$ denotes the virtual-photon polarization state and $m$ denotes the spin projection of the decaying resonance. 

If high-energy Regge factorization holds, and the kinematics are dominated by a single exchange trajectory, the partial-wave amplitude for a given reflectivity $\epsilon$ factorizes completely into an upper-vertex transition term ($\mathcal{U}^\epsilon_a$) and a target-nucleon exchange coupling ($\mathcal{B}^\epsilon_k$):
\begin{equation}\label{eq:app_UB}
[\ell]^\epsilon_{a; k} = \mathcal{U}^\epsilon_a \mathcal{B}^\epsilon_k \ .
\end{equation}

The lower vertex $\mathcal{B}^\epsilon_k$ depends only on the $t$-channel exchange trajectory and the nucleon helicities. It is entirely independent of the virtual-photon state $c$ and the spin projection $m$.

To examine the coherence of the target summation, we define a fundamental incoherent bilinear $\mathcal{I}^\epsilon_{ab}$ for any two generalized states $a$ and $b$:

\begin{equation}
\mathcal{I}^\epsilon_{ab} \equiv \sum_{k \in \{-1,1\}} [\ell]^{\epsilon}_{a; k} [\ell]^{\epsilon *}_{b; k} \ .
\end{equation}

As detailed in Eq.~\eqref{eq:refl_billinears_full}, the physical SDMEs are constructed from linear combinations of these fundamental bilinears, incorporating parity symmetries and kinematic prefactors. 

Substituting the factorized amplitudes \eqref{eq:app_UB} into this bilinear, the upper-vertex terms factor entirely out of the incoherent target summation:
\begin{equation}  
\mathcal{I}^\epsilon_{ab} = \mathcal{U}^\epsilon_a \mathcal{U}_b^{\epsilon *} \sum_{k \in \{-1,1\}} \big|\mathcal{B}^\epsilon_k\big|^2.
\end{equation}

Because the $k$-summation consists strictly of absolute squares, it defines a purely real, positive scalar quantity, $S^{\epsilon} \equiv \sum_k |\mathcal{B}^\epsilon_k|^2$. The bilinear simplifies to:
$$\mathcal{I}^\epsilon_{ab} = S^{\epsilon} \left( \mathcal{U}^\epsilon_a \mathcal{U}_b^{\epsilon *} \right) .$$

To determine whether the underlying system is factorizable, we evaluate the determinant of the $2 \times 2$ sub-matrix for states $a$ and $b$. The condition for absolute coherence is the exact saturation of the inequality, $\Delta \ge0$, which requires $\Delta = \mathcal{I}^\epsilon_{aa} \mathcal{I}^\epsilon_{bb} - |\mathcal{I}^\epsilon_{ab}|^2 = 0$. 
\begin{align}
\mathcal{I}^\epsilon_{aa} \mathcal{I}^\epsilon_{bb} &= \left( S^{\epsilon} \big|\mathcal{U}^\epsilon_a\big|^2 \right) \left( S^{\epsilon} \big|\mathcal{U}^\epsilon_b\big|^2 \right) = (S^{\epsilon})^2 \big|\mathcal{U}^\epsilon_a\big|^2 \big|\mathcal{U}^\epsilon_b\big|^2 , \nonumber \\
|\mathcal{I}^\epsilon_{ab}|^2 &= \Big| S^{\epsilon} \mathcal{U}^\epsilon_a \mathcal{U}_b^{\epsilon *} \Big|^2 = (S^{\epsilon})^2 \big|\mathcal{U}^\epsilon_a\big|^2 \big|\mathcal{U}^\epsilon_b\big|^2 .
\end{align}
Therefore, $\mathcal{I}^\epsilon_{aa} \mathcal{I}^\epsilon_{bb} = |\mathcal{I}^\epsilon_{ab}|^2$, and the determinant is exactly zero. 

In practical amplitude analysis frameworks, this state of absolute coherence is routinely enforced by setting half of the nucleon helicity transitions to zero (e.g., requiring $[\ell]^\epsilon_{a; k=-1} = 0$). The mathematical saturation of the inequality demonstrates why this artificial truncation is a valid mapping of an underconstrained, unpolarized system, provided the factorization assumption holds across the bilinears defining the observables in Eq.~\eqref{eq:refl_billinears_full}.

\subsection{Breakdown of factorization and onset of decoherence}

The mathematical justification for the coherent amplitude failing is when the reaction dynamics violate strict vertex factorization across the unobserved target states. It should be noted that if multiple exchanges populate the \textit{same} nucleon helicity transition, they sum coherently at the amplitude level prior to the target trace, maintaining $\Delta = 0$. Loss of factorization strictly arises when multiple exchange trajectories populate \textit{different} nucleon helicity states.

Consider a naturality sector $\epsilon$ driven by two competing exchanges: Exchange $X$ (e.g., Pomeron), which couples exclusively to the non-flip transition ($k=1$), and Exchange $Y$ (e.g., $\rho$), which couples exclusively to the spin-flip transition ($k=-1$). The total amplitude for state $a$ is the coherent sum of these exchanges prior to the $k$-summation:
\begin{equation}
[\ell]^\epsilon_{a; k} = \mathcal{U}^\epsilon_{X,a} \mathcal{B}^\epsilon_{X,k} + \mathcal{U}^\epsilon_{Y,a} \mathcal{B}^\epsilon_{Y,k}.
\end{equation}

By defining orthogonal target couplings ($\mathcal{B}_{X,-1} = \mathcal{B}_{Y,1} = 0$), the fundamental bilinear becomes:
\begin{align}
\mathcal{I}^\epsilon_{ab} &=\!\!\!\!\!\!\sum_{k \in \{-1,1\}}\!\!\!\!\!\! \left( \mathcal{U}^\epsilon_{X,a} \mathcal{B}^\epsilon_{X,k} + \mathcal{U}^\epsilon_{Y,a} \mathcal{B}^\epsilon_{Y,k} \right) \left( \mathcal{U}^\epsilon_{X,b} \mathcal{B}^\epsilon_{X,k} + \mathcal{U}^\epsilon_{Y,b} \mathcal{B}^\epsilon_{Y,k} \right)^* \nonumber \\
&= \mathcal{U}^\epsilon_{X,a} \mathcal{U}_{X,b}^{\epsilon *} \big|\mathcal{B}^\epsilon_{X,1}\big|^2 + \mathcal{U}^\epsilon_{Y,a} \mathcal{U}_{Y,b}^{\epsilon *} \big|\mathcal{B}^\epsilon_{Y,-1}\big|^2.
\end{align}
Unlike the fully factorized or single-helicity case, a single universal scalar $S^{\epsilon}$ can no longer be factored out. The $k=1$ and $k=-1$ sectors are driven by fundamentally different upper-vertex kinematics ($\mathcal{U}_X$ vs. $\mathcal{U}_Y$).

Evaluating the determinant $\Delta = \mathcal{I}^\epsilon_{aa} \mathcal{I}^\epsilon_{bb} - |\mathcal{I}^\epsilon_{ab}|^2$ for this system yields:
\begin{align}
\Delta &= \big|\mathcal{B}^\epsilon_{X,1}\big|^2 \big|\mathcal{B}^\epsilon_{Y,-1}\big|^2 \nonumber \\
&\quad \times \Big( \big|\mathcal{U}^\epsilon_{X,a}\big|^2 \big|\mathcal{U}^\epsilon_{Y,b}\big|^2 + \big|\mathcal{U}^\epsilon_{Y,a}\big|^2 \big|\mathcal{U}^\epsilon_{X,b}\big|^2  \\
& \hspace*{3.15cm}-2 \text{Re}\left[ \mathcal{U}^\epsilon_{X,a} \mathcal{U}_{X,b}^{\epsilon *} \mathcal{U}_{Y,a}^{\epsilon *} \mathcal{U}^\epsilon_{Y,b} \right] \Big) \nonumber \\
&= \big|\mathcal{B}^\epsilon_{X,1}\big|^2 \big|\mathcal{B}^\epsilon_{Y,-1}\big|^2 \Big| \mathcal{U}^\epsilon_{X,a} \mathcal{U}^\epsilon_{Y,b} - \mathcal{U}^\epsilon_{Y,a} \mathcal{U}^\epsilon_{X,b} \Big|^2.
\end{align}

Because the term $\big| \mathcal{U}^\epsilon_{X,a} \mathcal{U}^\epsilon_{Y,b} - \mathcal{U}^\epsilon_{Y,a} \mathcal{U}^\epsilon_{X,b} \big|^2$ is strictly positive for any two distinct dynamic processes, $\Delta > 0$. The inequality is not saturated. The system is proven to be a non-factorizable state, where both nucleon helicity amplitudes must be considered to accurately construct the physical SDMEs defined in Eq.~\eqref{eq:refl_billinears_full}. 

This non-factorization applies equivalently to both physical domains:
\begin{enumerate}
    \item \textbf{Failure in $\LL/\TT$ interference:} If $a = (\LL, m)$ and $b = (\TT, m)$, the difference in longitudinal and transverse couplings between Exchange $X$ and Exchange $Y$ destroys the phase relationship. Unpolarized observables will exhibit suppressed LT interference moments.
    \item \textbf{Failure in $m$ interference:} If $a = (\TT, m)$ and $b = (\TT, m')$, differing spatial distributions (e.g., forward vs. non-forward peaking) of the two exchanges suppress the off-diagonal $m \neq m'$ SDMEs. 
\end{enumerate}

When the amplitudes cannot be factorized, forcing an amplitude fitter to extract a single effective amplitude can result in mathematical distortions. Because the model assumes $\Delta = 0$, the fitter will artificially compress the absolute magnitudes of the primary partial waves to geometrically satisfy the suppressed interference moments, or it will invent unphysical background phases to try and better fit the data. Consequently, the use of unpolarized effective amplitudes is only valid under kinematics where either a single exchange mechanism dominates, or the contributing exchanges within a given naturality sector predominantly populate the same nucleon helicity transition.

\section{GENERALIZING TO NUCLEON POLARIZATION}
\label{app:fullpol}

The fully polarized formalism is largely invariant to what has already been discussed, only expanding the formalism presented earlier in Appendix~\ref{app:angular_ints_moments} by now explicitly summing the virtual-photon, target nucleon, and recoil nucleon spin density matrices. Again, this formalism is constructed in the helicity frame and as a result the polarization vectors for the nucleons are frame specific, however, it is only the coefficients of the SDMEs that mix. The underlying Pauli matrix structure and the expansion presented here is invariant. The relationship between the helicity frame and the nucleon rest frame was covered in Sec.~\ref{sec:polarized_nucleons_main}. Here, the focus will be on the effect of including polarized nucleon observables at the amplitude level.
The tensor SDMEs, accounting for the initial and recoil nucleon spins, are now
\begin{equation}
\rho^{\alpha\beta\delta,\ell\ell'}_{mm'}=
\frac{1}{8}\sum_{\substack{\lambda\lambda_1\lambda_2 \\ \lambda'\lambda_1'\lambda_2'}}
T^{\ell}_{\substack{\lambda m\\ \lambda_1\lambda_2}}
\Sigma^\alpha_{\lambda\lambda'}\sigma^\beta_{\lambda_1\lambda_1'}
\sigma^\delta_{\lambda_2\lambda_2'}
T^{\ell' *}_{\substack{\lambda' m'\\ \lambda_1'\lambda_2'}},
\label{eq:rhotensor_app}
\end{equation}
with the corresponding moments being largely the same, up to accounting for the $\beta$ and $\delta$ degrees of freedom,
\begin{align} \nonumber
H^{\alpha\beta\delta}(LM) =
& \sum_{\ell\ell'mm'}
\left(\frac{2\ell'+1}{2\ell+1}\right)^{1/2} 
\\
& \times 
C^{\ell'0}_{\ell0\,L0}
C^{\ell'm'}_{\ell m\,LM}
\rho^{\alpha\beta\delta,\ell\ell'}_{mm'}.
\label{eq:fullmoments_app}
\end{align}
Similarly the intensity is also expanded to take this into account,
\begin{equation}
   I^{\alpha\beta\delta}(\Omega)=
\sum_{LM}\left(\frac{2L+1}{4\pi}\right)H^{\alpha\beta\delta}(LM)Y_L^M(\Omega) ,
\end{equation}
with the total angular distribution being
\begin{equation}
    I(\Omega,\Phi) = \kappa\sum_{\alpha=0}^8\sum_{\beta,\delta=0}^3 I^{\alpha\beta\delta}(\Omega)\Pi^{\alpha\beta\delta}(\Phi).
\end{equation}
Here, $\Pi^{\alpha\beta\delta}(\Phi) = \Pi^\alpha(\Phi) S^\beta_I S^\delta_R$ is just the product of all the polarization parameters.

As before there are still constraints that have to be taken into account to further simplify this structure. All the basis matrices in Eq.~\eqref{eq:rhotensor_app} are Hermitian, so every $\rho^{\alpha\beta\delta}$ is Hermitian and the strong force is parity conserving. These constraints, as they applied in Appendix~\ref{app:angular_ints_moments}, result in a global sign relating parity partners. Here we define, 
\bsub\begin{align}
\tau_\alpha&=\begin{cases}
+1, & \alpha=0,1,4,5,8\\
-1, & \alpha=2,3,6,7
\end{cases}
\\
\tau_{\beta,\delta} &=\begin{cases}
+1, & \beta,\delta=0,1\\
-1, & \beta,\delta=2,3
\end{cases}
\label{eq:parity_sign_defs_app}
\end{align}\esub
such that for a fully polarized beam-target-recoil moment the total parity sign is
\begin{equation}
\pi_{\alpha\beta\delta}=\tau_\alpha\tau_\beta\tau_\delta.
\label{eq:app_parity_full}
\end{equation}
By parity and hermicity we then get for the underlying spin density matrix that
\begin{equation}
\rho^{\alpha\beta\delta,\ell\ell'}_{-m,-m'}=
\pi_{\alpha\beta\delta}(-1)^{m-m'}
\rho^{\alpha\beta\delta,\ell\ell'}_{mm'}.
\label{eq:fullpol_parity_app}
\end{equation}
This then implies for the moments
\begin{equation}
H^{\alpha\beta\delta}(L\,{-M})=
\pi_{\alpha\beta\delta}(-1)^M H^{\alpha\beta\delta}(LM).
\label{eq:fullpol_reality_app}
\end{equation}
Therefore, $\pi_{\alpha\beta\delta}=+1$ gives a real moment, while $\pi_{\alpha\beta\delta}=-1$ gives a purely imaginary moment; the latter class has vanishing $M=0$ moments. This is the exact same relation as seen previously in Appendix~\ref{app:angular_ints_moments}, just extended to the higher dimensionality offered by the polarized nucleons. Table~\ref{tab:single_nucleon_parity} gives the one-nucleon sign table used to build Eq.~\eqref{eq:fullpol_parity_app}.  It applies identically to the target index $\beta$ and the recoil index $\delta$; to be concise, the values for the virtual-photon polarization and a singular nucleon polarization are shown.

\begin{table}[tbp]
\caption{One-nucleon parity factors. Here $\pi_{\alpha \beta,\delta}$ is the resulting one-nucleon parity sign. To include both nucleon parities, one would then multiply $\tau_{\alpha \{\beta,\delta\}}$ by the parity of the remaining polarized nucleon state.}
\label{tab:single_nucleon_parity}
\scriptsize
\begin{ruledtabular}
\begin{tabular}{cccc}
$(\{\beta,\delta\},\alpha)$
& $\tau_{\beta,\delta}$
& $\tau_\alpha$
& $\pi_{\alpha \{\beta,\delta\}}$ \\
\hline
$(0,0)$ & $+$ & $+$ & $+$ \\
$(0,1)$ & $+$ & $+$ & $+$ \\
$(0,2)$ & $+$ & $-$ & $-$ \\
$(0,3)$ & $+$ & $-$ & $-$ \\
$(0,4)$ & $+$ & $+$ & $+$ \\
$(0,5)$ & $+$ & $-$ & $-$ \\
$(0,6)$ & $+$ & $+$ & $+$ \\
$(0,7)$ & $+$ & $+$ & $+$ \\
$(0,8)$ & $+$ & $-$ & $-$ \\
\hline
$(1,0)$ & $+$ & $+$ & $+$ \\
$(1,1)$ & $+$ & $+$ & $+$ \\
$(1,2)$ & $+$ & $-$ & $-$ \\
$(1,3)$ & $+$ & $-$ & $-$ \\
$(1,4)$ & $+$ & $+$ & $+$ \\
$(1,5)$ & $+$ & $-$ & $-$ \\
$(1,6)$ & $+$ & $+$ & $+$ \\
$(1,7)$ & $+$ & $+$ & $+$ \\
$(1,8)$ & $+$ & $-$ & $-$ \\
\hline
$(2,0)$ & $-$ & $+$ & $-$ \\
$(2,1)$ & $-$ & $+$ & $-$ \\
$(2,2)$ & $-$ & $-$ & $+$ \\
$(2,3)$ & $-$ & $-$ & $+$ \\
$(2,4)$ & $-$ & $+$ & $-$ \\
$(2,5)$ & $-$ & $-$ & $+$ \\
$(2,6)$ & $-$ & $+$ & $-$ \\
$(2,7)$ & $-$ & $+$ & $-$ \\
$(2,8)$ & $-$ & $-$ & $+$ \\
\hline
$(3,0)$ & $-$ & $+$ & $-$ \\
$(3,1)$ & $-$ & $+$ & $-$ \\
$(3,2)$ & $-$ & $-$ & $+$ \\
$(3,3)$ & $-$ & $-$ & $+$ \\
$(3,4)$ & $-$ & $+$ & $-$ \\
$(3,5)$ & $-$ & $-$ & $+$ \\
$(3,6)$ & $-$ & $+$ & $-$ \\
$(3,7)$ & $-$ & $+$ & $-$ \\
$(3,8)$ & $-$ & $-$ & $+$ \\
\end{tabular}
\end{ruledtabular}
\end{table}
The reflectivity basis can also be readily applied to this generalised formalism and is discussed in the next section. To keep things simple and succinct, the explicit form of the polarized intensities $I^{\alpha\beta\delta}$, the transformation integral to obtain the moments $H^{\alpha\beta\delta}$, as well as the explicit forms of the spin density matrices in terms of the partial and reflectivity amplitudes $\rho^{\alpha\beta\delta, ll'}_{mm'}$ are not stated. This is due to there being 144 total equations for each $\alpha$, $\beta$ and it being far easier to condense into the formalism presented here. 

\label{subsec:app_beta_gamma_rules}
The Pauli matrices can be seen to act as follows: $\sigma^0$ leaves the amplitudes unchanged, $\sigma^1$ flips the sign of the first index and has a negative factor, $\sigma^2$ also flips the sign of the first index with a phase $\ii$ and a helicity weighting, and $\sigma^3$ leaves it unchanged but weights it by its
helicity sign. This is fully equivalent to the transverse sector of the virtual-photon expansion within Eqs.~\eqref{eq:sdme_T}~to~\eqref{eq:sdme_T_end}, with the corresponding nucleon helicity being the index and weighting of interest. For a general nucleon helicity $\lambda_i$ this can be explicitly seen as
\bsub\begin{align}   \sum_{\lambda_i\lambda'_i}T^{\ell}_{\lambda m; \lambda_i} \sigma^0_{\lambda_i\lambda'_i} T^{\ell'}_{\lambda' m';\lambda'_i},&=\sum_{\lambda_i}T^{\ell}_{\lambda m;\lambda_i}T^{\ell' *}_{\lambda',m';\lambda_i},\\
\sum_{\lambda_i\lambda'_i}T^{\ell}_{\lambda m; \lambda_i} \sigma^1_{\lambda_i\lambda'_i} T^{\ell'}_{\lambda' m';\lambda'_i}&=-\sum_{\lambda_i}T^{\ell}_{\lambda m;-\lambda_i}T^{\ell' *}_{\lambda',m';\lambda_i},\\
\sum_{\lambda_i\lambda'_i}T^{\ell}_{\lambda m; \lambda_i} \sigma^2_{\lambda_i\lambda'_i} T^{\ell'}_{\lambda' m';\lambda'_i}&=\ii\sum_{\lambda_i}\lambda_i\,T^{\ell}_{\lambda,m;-\lambda_i}T^{\ell' *}_{\lambda' m';\lambda_i},\\
\sum_{\lambda_i\lambda'_i}T^{\ell}_{\lambda m; \lambda_i} \sigma^3_{\lambda_i\lambda'_i} T^{\ell'}_{\lambda' m';\lambda'_i}&=\sum_{\lambda_i}\lambda_i\,T^{\ell}_{\lambda m;\lambda_i}T^{\ell' *}_{\lambda' m';\lambda_i},
\end{align}\esub
where $i\in\{1,2\}$ for the target and recoil states considered here. 
For a generic bilinear
$A^{\refl}_{X m;k}B^{\refl' *}_{Y m';k'}$, with either photon sector $X=\TT,\LL$, the general target--recoil helicity sum can be written as
\begin{align} \nonumber \sum_{\lambda_1^{(')},\lambda_2^{(')}}
\sum_{\refl\refl'} &
[\ell]^{\refl}_{\substack{X m\\\lambda_1\lambda_2}}
\sigma^\beta_{\lambda_1\lambda_1'}
\sigma^\delta_{\lambda_2\lambda_2'}
[\ell']^{\refl'*}_{\substack{Y m'\\\lambda_1'\lambda_2'}}
\\
&=2\sum_{\refl,k}
C_{\beta\delta}(\refl,k)
[\ell]^{\refl}_{X m;k}
[\ell']^{\refl'*}_{Y m';k'}
\label{eq:app_master_beta_gamma_rule}
\end{align}
where the values for $\refl'$ and $k'$ and $C_{\beta\delta}$ are listed in Table~\ref{tab:app_beta_gamma_rules}.  This
single table is the compact replacement rule for all target and recoil
polarization combinations.  It applies independently of the photon sector
selected by $\alpha$, Eqs.~\eqref{eq:sdme_T}~to~\eqref{eq:sdme_L}.
\begin{table}
\caption{Target--recoil replacement rules in the reduced $k$ basis, shows how each $(\beta,\delta)$ state affects the $\refl'$ and $k'$ indices as well as giving the prefactor $C_{\beta\delta}$.}
\centering
\label{tab:app_beta_gamma_rules}
\scriptsize
\begin{ruledtabular}
\begin{tabular}{crrr}
$(\beta,\delta)$
& $\refl'$
& $k'$
& $C_{\beta\delta}$ \\
\hline
$(0,0)$ & $\refl$  & $k$  & $1$ \\
$(0,1)$ & $-\refl$ & $-k$ & $-1$ \\
$(0,2)$ & $\refl$  & $-k$ & $-\ii k$ \\
$(0,3)$ & $-\refl$ & $k$  & $k$ \\
\hline
$(1,0)$ & $-\refl$ & $-k$ & $-\refl k$ \\
$(1,1)$ & $\refl$  & $k$  & $\refl k$ \\
$(1,2)$ & $-\refl$ & $k$  & $\ii\refl$ \\
$(1,3)$ & $\refl$  & $-k$ & $-\refl$ \\
\hline
$(2,0)$ & $\refl$  & $-k$ & $\ii\refl k$ \\
$(2,1)$ & $-\refl$ & $k$  & $-\ii\refl k$ \\
$(2,2)$ & $\refl$  & $k$  & $\refl$ \\
$(2,3)$ & $-\refl$ & $-k$ & $\ii\refl$ \\
\hline
$(3,0)$ & $-\refl$ & $k$  & $1$ \\
$(3,1)$ & $\refl$  & $-k$ & $-1$ \\
$(3,2)$ & $-\refl$ & $-k$ & $-\ii k$ \\
$(3,3)$ & $\refl$  & $k$  & $k$ \\
\end{tabular}
\end{ruledtabular}
\end{table}

The first row of Table~\ref{tab:app_beta_gamma_rules} is the unpolarized target
and recoil case, and reproduces the same results as
Eqs.~\eqref{eq:sdme_T}~to~\eqref{eq:sdme_L}.  Rows with $\beta=1,2$ or
$\delta=1,2$ contain a helicity flip from $\sigma_x$ or $\sigma_y$ with $\sigma_y$ also having a helicity weight; rows with
$\beta=3$ or $\delta=3$ contain the helicity weight from $\sigma_z$.  The
selection of $\refl$ and $k$ in each row is therefore fixed entirely by the
Pauli matrix and the parity of
Eq.~\eqref{eq:app_parity_full}.  Consequently, no new amplitudes
are introduced by target or recoil polarization: all
$9\times4\times4$ observables are bilinears of the same
$[\ell]^{\refl}_{\TT m;k}$ and $[\ell]^{\refl}_{\LL m;k}$ amplitudes. Any state can then be constructed from both the Table~\ref{tab:app_beta_gamma_rules} and the ``basis" in Eqs.~\eqref{eq:sdme_T}~to~\eqref{eq:sdme_L}. By choosing the target to be longitudinal or transverse one can also immediately reduce the 144 equations, as the non-corresponding $\alpha$, $\beta$
and $\delta$ states reduce to be identically 0.

One thing that does not get explored above are any redundancies in the parameter space. Previously, it has been shown that amplitudes can be extracted for the unpolarized case when each reflectivity is assumed to be dominated by a single nucleon transition, and two reference phases are chosen. These arise due to the inherent symmetries underlying the given system. Such symmetries are also present in the case of single nucleon polarization, and the symmetries are only fully broken when all nucleon spins are analyzed.

\section{NUCLEON POLARIZATION TRANSFORMATIONS}\label{app:nuc_pol_trans}

In this section, we outline how one would experimentally measure the nucleon spin vectors $S_I^\beta$ and $S_R^\delta$ event-by-event to allow measurement of the moments $H^{\alpha\beta\delta}(L,M)$.

\subsection{Initial nucleon polarization transformation}
\label{subsec:initial_pol}
Let $\mathbf{S}_{\mathrm{rest}}$ denote the constant initial nucleon spin 3-vector evaluated in its rest frame. 
$\mathbf{S}_{\mathrm{rest}} = P_I \widehat{\mathbf{n}}_{\mathrm{rest}}$ with $P_I$ the degree of target polarization and $\widehat{\mathbf{n}}_{\mathrm{rest}}$ its orientation.
For fixed-target experiments, the laboratory frame coincides with this rest frame. For collider experiments, the measured event 4-momenta must first be transformed into the initial nucleon rest frame via a Lorentz boost. Because this boost does not rotate the spatial axes, the Cartesian coordinate system in the rest frame remains perfectly aligned with the fixed laboratory detector axes. The components of $\mathbf{S}_{\mathrm{rest}}$ must be expressed in this shared Cartesian coordinate system, explicitly accounting for the true orientation of the incident proton momentum (e.g., accommodating a beam crossing angle).

While the nucleon spin polarization is defined in its rest frame, the amplitudes and decay angles are defined in the rest frame of the produced two-meson system. The event 4-momenta must therefore be boosted into this frame before the relevant spin-quantization axes are constructed. Let $q' = p + q - p'$ denote the four-momentum of the produced meson system. We apply a Lorentz boost to transform the initial nucleon ($p$), recoiling nucleon ($p'$), and virtual photon ($q$) 4-momenta from the initial nucleon rest frame into the $q'$ rest frame. Note, below a star denotes a quantity evaluated in this two-meson rest frame.

The helicity axes used to define the decay angles and the spin projection $m$ of the produced two-hadron system are
\begin{equation}
    \widehat{\mathbf z}_{H} = -\frac{\mathbf p'^*}{|\mathbf p'^*|}, \qquad
    \widehat{\mathbf y}_{H} = \frac{\mathbf p'^*\times\mathbf q^*}{|\mathbf p'^*\times\mathbf q^*|}, \qquad
    \widehat{\mathbf x}_{H} = \widehat{\mathbf y}_{H} \times \widehat{\mathbf z}_{H}.
    \label{eq:helicity_axes}
\end{equation}

The spin-density matrix is defined using the canonical spin 3-vector evaluated in the particle's own rest frame. Under a Lorentz transformation, a spin state undergoes a Wigner rotation; however, for a particle initially at rest, there is no Wigner rotation induced by a boost~\cite{Leader2001}. 

Consequently, the spin vector is completely unaffected by the boost. Its components in the initial-nucleon helicity basis are simply the dot products of the original unrotated spin vector $\mathbf{S}_{\mathrm{rest}}$ with the helicity axes:
\begin{align}
    S_I^1 &= P_I (\widehat{\mathbf{n}}_{\mathrm{rest}} \cdot \widehat{\mathbf{x}}_H), \nonumber\\
    S_I^2 &= P_I (\widehat{\mathbf{n}}_{\mathrm{rest}} \cdot \widehat{\mathbf{y}}_H), \nonumber\\
    S_I^3 &= P_I (\widehat{\mathbf{n}}_{\mathrm{rest}} \cdot \widehat{\mathbf{z}}_H).
    \label{eq:explicit_SI_components}
\end{align}

In summary, an experimental analysis avoids complex chains of rotation matrices entirely and instead executes the following streamlined event-by-event procedure: 
\begin{itemize}
    \item Define the initial nucleon polarization vector $\mathbf{S}_{\mathrm{rest}}$ as a constant 3-vector in its rest frame, ensuring its components are expressed using the identical Cartesian laboratory axes as the reconstructed 4-momenta.
    \item Boost the measured event 4-momenta into the produced two-meson rest frame to dynamically construct the helicity axes ($\widehat{\mathbf{x}}_H, \widehat{\mathbf{y}}_H, \widehat{\mathbf{z}}_H$) via Eq.~\eqref{eq:helicity_axes}. 
    \item Project the unrotated spin vector $\mathbf{S}_{\mathrm{rest}}$ directly onto these new axes using the dot products in Eq.~\eqref{eq:explicit_SI_components} to obtain the helicity-frame spin components $S_I^1$, $S_I^2$, and $S_I^3$. 
\end{itemize}
These three components, combined with the required density-matrix identity term, form the exact four-component object $S_I^\beta = (1, S_I^1, S_I^2, S_I^3)$ that serves as the initial-nucleon polarization weight in the  intensity expansion in Eq.~\eqref{eq:full_polarized_intensity}.

\subsection{Recoil polarization transformation}
\label{subsec:recoil_pol}

The polarization of the recoiling baryon can present an experimental challenge. In the special case of a $\Lambda$ hyperon however it can be determined experimentally through its parity-violating weak decay, $\Lambda\rightarrow p\pi^-$. In the $\Lambda$ rest frame, the angular distribution of the decay proton depends linearly on the $\Lambda$ polarization vector, providing a direct measurement of the recoil spin-density matrix. As before the helicity axes are defined through Eq.~\eqref{eq:helicity_axes}.
 
 To extract the polarization, the measured momentum of the decay proton is boosted into the $\Lambda$ rest frame. Because this final boost is purely collinear with the $\Lambda$ momentum $\mathbf{p'}^{*} = -\mathbf{z}_H$, it induces no spatial rotation of the helicity axes. Let $\mathbf p_p^{\,(\Lambda)}$ denote this decay proton momentum in the $\Lambda$ rest frame. Its direction relative to the helicity axes is given by the unit vector:
\begin{equation}
    \widehat{\mathbf n}_p = \frac{\mathbf p_p^{\,(\Lambda)}}{|\mathbf p_p^{\,(\Lambda)}|} =
    \begin{pmatrix}
        \sin\theta_p\cos\phi_p\\
        \sin\theta_p\sin\phi_p\\
        \cos\theta_p
    \end{pmatrix}_{\!\!H},
    \label{eq:lambda_decay_proton_direction}
\end{equation}
where the decay angles $(\theta_p, \phi_p)$ are defined strictly relative to the $(\widehat{\mathbf x}_H, \widehat{\mathbf y}_H, \widehat{\mathbf z}_H)$ basis:
\begin{equation}
    \cos\theta_p = \widehat{\mathbf n}_p \cdot \widehat{\mathbf z}_H, \qquad
    \phi_p = \operatorname{atan}_2 \left( \widehat{\mathbf n}_p \cdot \widehat{\mathbf y}_H, \widehat{\mathbf n}_p \cdot \widehat{\mathbf x}_H \right).
    \label{eq:lambda_decay_angles}
\end{equation}

The proton angular distribution is given by
\begin{equation}
    I_{\Lambda}(\theta_p,\phi_p) \propto 1 + \mathbf P_\Lambda \cdot (\alpha_{\Lambda} \widehat{\mathbf n}_p),
    \label{eq:lambda_decay_distribution}
\end{equation}

where $\mathbf P_\Lambda$ is the $\Lambda$ polarization vector in the helicity coordinate system and $\alpha_{\Lambda}$ is the $\Lambda$ weak decay parameter~\cite{PhysRevLett.129.131801}. 

In the intensity expansion of our formalism in Eq.~\eqref{eq:full_polarized_intensity}, the recoil spin state enters through the 4-component tensor $S_R^\delta$. Because the joint production-and-decay distribution effectively projects the helicity frame polarization onto the weak-decay analyzing vector, the event-by-event evaluation of the formalism is achieved by substituting the helicity frame polarization components with the measured decay directional cosines.

Thus, the effective components $S_R^1, S_R^2, S_R^3$ used to evaluate the intensity weight $S_R^\delta$ for a given event are directly given by:
\begin{align}\label{eq:S_R}
    S_R^1 &= \alpha_\Lambda \sin\theta_p \cos\phi_p, \nonumber\\
    S_R^2 &= \alpha_\Lambda \sin\theta_p \sin\phi_p, \nonumber\\
    S_R^3 &= \alpha_\Lambda \cos\theta_p.
\end{align}

In summary, an experimental analysis executes the following streamlined event-by-event procedure for the recoil sector:
\begin{itemize}
    \item Boost the measured event 4-momenta into the two-meson rest frame to dynamically construct the helicity axes ($\widehat{\mathbf x}_H, \widehat{\mathbf y}_H, \widehat{\mathbf z}_H$) via Eq.~\eqref{eq:helicity_axes}.
    \item Boost the measured decay proton momentum into the $\Lambda$ rest frame and calculate its angles $(\theta_p, \phi_p)$ relative to these axes via Eq.~\eqref{eq:lambda_decay_angles}.
    \item Compute the effective recoil components directly via Eq.~\eqref{eq:S_R}.
\end{itemize}

These three components, combined with the required density-matrix identity term, form the exact four-component object $S_R^\delta = (1, \alpha_\Lambda \widehat{\mathbf{n}}_p)$ that must be inserted into the intensity expansion in Eq.~\eqref{eq:full_polarized_intensity} to properly weight the event.

\section{POLARIZATION SYMMETRIES AND PARAMETER REDUNDANCIES}
\label{app:symmetries}
Within the scope of the paper so far, no continuous symmetries of the underlying amplitude parametrization have been considered. Only the discrete parity symmetries used in the construction of the formalism in Appendices~\ref{app:angular_ints_moments} and \ref{app:fullpol} have been imposed. Here, we consider unitary transformations in the target- and recoil-nucleon spin spaces (but not in the lepton-beam space, since the beam is assumed throughout to be polarized). These transformations lead to different parameter redundancies in the three cases of interest: unpolarized nucleons, a singly polarized nucleon, and both nucleons polarized.

In the reflectivity basis, a partial-wave amplitude can be organized as a $2\times2$ matrix acting from the target-nucleon helicity space to the recoil-nucleon helicity space:
\begin{align} \nonumber
    \, ^{\epsilon}T^{\ell}_{\lambda m} &=
    \begin{pmatrix}
    \, ^{\epsilon}T^{\ell}_{\lambda m;++} & \, ^{\epsilon}T^{\ell}_{\lambda m;-+}\\
    \, ^{\epsilon}T^{\ell}_{\lambda m;+-} & \, ^{\epsilon}T^{\ell}_{\lambda m;--}
    \end{pmatrix}
    \\ 
    & = \begin{pmatrix}
[\ell]^\epsilon_{\lambda m;1} & -\epsilon[\ell]^\epsilon_{\lambda m;-1}\\
[\ell]^\epsilon_{\lambda m;-1} & \epsilon[\ell]^\epsilon_{\lambda m;1}
\end{pmatrix}.
\label{eq:nucleon_spin_amp_matrix}
\end{align}

The entries of this matrix are complex, so $\, ^{\epsilon}T^{\ell}_{\lambda m}\in M_2(\mathbb C)$. The most general change of basis in either two-dimensional nucleon-spin space is therefore described by an element of $U(2)$. However, the transformed amplitudes must retain the parity-constrained form of Eq.~\eqref{eq:nucleon_spin_amp_matrix}, which restricts the allowed transformations to a subgroup of $U(2)$.

To see this, write
\begin{equation}
U=
\begin{pmatrix}
a & b\\
c & d
\end{pmatrix}
\in U(2),
\end{equation}
and require that $U\, ^{\epsilon}T^{\ell}_{\lambda m}$ have the same form as Eq.~\eqref{eq:nucleon_spin_amp_matrix}. Matching the transformed matrix elements gives $d=a$ and $c=-b$. An allowed transformation can therefore be written as
\begin{equation}
U=
\begin{pmatrix}
a & -b\\
b & a
\end{pmatrix}.
\end{equation}
The unitarity condition then gives
\begin{equation}
|a|^2+|b|^2=1,
\qquad
\Ima(ab^*)=0,
\end{equation}
so $a$ and $b$ have a common phase, up to a relative sign. The parity-preserving subgroup, denoted by $\widetilde U(2) = SO(2)\times U(1)$, consequently contains two continuous parameters and can be written as
\begin{equation}
\widetilde U(2)
=
e^{\i\phi}
\begin{pmatrix}
\cos\theta & -\sin\theta\\
\sin\theta & \cos\theta
\end{pmatrix}
\simeq U(1)\times SO(2).
\label{eq:parity_preserving_u2}
\end{equation}
An equivalent restriction is obtained for transformations acting from the right.

The spin-density matrix of Eq.~\eqref{eq:sdme_fullpol} can be expressed as a trace over the nucleon-spin spaces,
\begin{equation}
\rho^{\alpha\beta\delta,\ell\ell'}_{mm'}
=
\frac{1}{8}
\sum_{\lambda\lambda'}
\Sigma^\alpha_{\lambda\lambda'}
\sum_{\epsilon\epsilon'=\pm1}
\Tr\!\left[
\sigma^\delta
\, ^{\epsilon}T^{\ell}_{\lambda m}
\sigma^\beta
\left(\, ^{\epsilon '}T^{\ell'}_{\lambda' m'}\right)^\dagger
\right].
\label{eq:sdme_nucleon_trace}
\end{equation}
Here, $\beta$ and $\delta$ label the target- and recoil-polarization components, respectively. A change of basis acts on an amplitude matrix as
\begin{equation}
\, ^{\epsilon}T^{\ell}_{\lambda m}
\longrightarrow
U_R \, ^{\epsilon}T^{\ell}_{\lambda m}U_T^\dagger,
\end{equation}
where $U_T$ and $U_R$ act in the target- and recoil-spin spaces. The redundancies associated with these transformations depend on which nucleon polarizations are measured. The three cases are considered below. In summary, the fully unpolarized problem has eight continuous redundancies when both nucleon-helicity sectors are retained, reducing to two independent reference phases when a single dominant nucleon-helicity transition $k$ is assumed, as in Sec.~\ref{sec:results}. If either the target or the recoil is polarized, two redundancies remain: one overall phase and one arbitrary spin-basis angle in the unobserved nucleon-spin space. If both nucleons are polarized, only the overall reference phase remains.

\subsection{Unpolarized recoil and target}
For an unpolarized target and recoil, $\beta=\delta=0$, so $\sigma^\beta=\sigma^\delta=\mathbb I$. Parity conservation also removes interference between the two reflectivity sectors, and the SDME becomes
\begin{equation}
\rho^{\alpha,\ell\ell'}_{mm'}
=
\frac{1}{2}
\sum_{\lambda\lambda'}
\Sigma^\alpha_{\lambda\lambda'}
\sum_{\epsilon=\pm1}
\Tr\!\left[
\, ^{\epsilon}T^{\ell}_{\lambda m}
\left(T^{\ell',\epsilon}_{\lambda'm'}\right)^\dagger
\right].
\label{eq:unpol_trace}
\end{equation}
The full invariance is most easily seen by collecting the two independent nucleon-helicity-transition amplitudes into the vector
\begin{equation}
\label{eq:app_gram}
\boldsymbol{t}^{\,\epsilon}_{\lambda m}
=
\begin{pmatrix}
[\ell]^\epsilon_{\lambda m;1}\\
[\ell]^\epsilon_{\lambda m;-1}
\end{pmatrix}.
\end{equation}
For two amplitude labels $a=(\ell,\lambda,m)$ and $b=(\ell',\lambda',m')$, the trace appearing in Eq.~\eqref{eq:unpol_trace} reduces to
\begin{equation}
\Tr\!\left[T_a^\epsilon(T_b^\epsilon)^\dagger\right]
=2\left(\boldsymbol{t}^{\,\epsilon}_b\right)^\dagger
\boldsymbol{t}^{\,\epsilon}_a.
\end{equation}
It is therefore invariant under a common transformation
\begin{equation}
\boldsymbol{t}^{\,\epsilon}_a
\longrightarrow
V^\epsilon\boldsymbol{t}^{\,\epsilon}_a,
\qquad
V^\epsilon\in U(2),
\end{equation}
applied to every amplitude label $a$ in that reflectivity sector. Since the two reflectivities enter incoherently, $V^+$ and $V^-$ may be chosen independently. The full unpolarized symmetry is consequently
\begin{equation}
U_+(2)\times U_-(2),
\end{equation}
which has dimension $4+4=8$. The left- and right-sided parity-preserving spin-basis transformations discussed above form a subset of this more general invariance of the unpolarized bilinears. The reason for the symmetry becoming more general, expanding the parity-preserving transformations of the physical nucleon-spin spaces $\widetilde U(2)\times \widetilde U(2)$ subgroup, is that the unpolarized observables can simply be reduced to the inner product of two orthogonal vectors (Eq~\ref{eq:app_gram}). In our polarized cases, we will have reflectivities mixing and further billinears mixing terms that break this construction. Thus, for unpolarized nucleons the complete observational ambiguity is enlarged to $U(2) \times U(2)$.

If a single dominant $k$ sector is imposed, the nucleon-spin amplitude matrix is restricted to be either diagonal or anti-diagonal. A generic $SO(2)$ mixing would take the amplitude outside this restricted subspace and is therefore no longer an allowed transformation of the reduced model, as this imposition breaks the symmetry. The magnitude of every amplitude remains fixed, while an independent common phase can still be applied within each reflectivity sector. The remaining symmetry is thus
\begin{equation}
U_+(1)\times U_-(1),
\end{equation}
corresponding to the two reference phases that must be fixed in an unpolarized, single-$k$ fit.

\subsection{Polarized recoil or target}
For a singly polarized nucleon, we consider either a polarized target, with $\beta=1,2,3$ and $\delta=0$, or a polarized recoil, with $\beta=0$ and $\delta=1,2,3$. Only the target case is shown explicitly; the recoil case follows analogously. For a polarized target,
\begin{equation}
\rho^{\alpha\beta,\ell\ell'}_{mm'}
=
\frac{1}{4}
\sum_{\lambda\lambda'}
\Sigma^\alpha_{\lambda\lambda'}
\sum_{\epsilon\epsilon'=\pm1}
\Tr\!\left[
\, ^{\epsilon}T^{\ell}_{\lambda m}
\sigma^\beta
\left(\, {^{\epsilon'}}T^{\ell'}_{\lambda'm'}\right)^\dagger
\right].
\label{eq:target_pol_trace}
\end{equation}
The Pauli matrix prevents an arbitrary transformation in the measured target-spin space. By contrast, a common left transformation in the unobserved recoil-spin space remains invisible:
\begin{align}
&\Tr\!\left[
U_R \,
T^{\ell,\epsilon}_{\lambda m}
\sigma^\beta
\left(
T^{\ell',\epsilon'}_{\lambda' m'}
\right)^\dagger
U_R^\dagger
\right]
\nonumber\\
&\hspace{3cm}=
\Tr\!\left[
T^{\ell,\epsilon}_{\lambda m}
\sigma^\beta
\left(
T^{\ell',\epsilon'}_{\lambda' m'}
\right)^\dagger
\right].
\end{align}
Because the polarized observables contain interference between the two reflectivity sectors, the same $U_R$ must act on both values of $\epsilon$. The symmetry is therefore a single $\widetilde U_R(2)$, with two continuous parameters: an overall phase and a rotation angle specifying the unobserved recoil-spin basis. Conversely, for a polarized recoil, the remaining symmetry is a single $\widetilde U_T(2)$ acting from the right, giving an overall phase and an arbitrary target-spin basis angle. Thus, either singly polarized case contains two continuous parameter redundancies.

\subsection{Polarized recoil and target}
When both nucleons are polarized, $\beta=1,2,3$ and $\delta=1,2,3$, and the transformed SDME contains
\begin{equation}
\Tr\!\left[
\sigma^\delta
U_R T^\epsilon U_T^\dagger
\sigma^\beta
U_T\left(T'^{\epsilon'}\right)^\dagger U_R^\dagger
\right].
\end{equation}
For the complete set of target and recoil polarization components to remain invariant, $U_T$ and $U_R$ must commute with all three Pauli matrices. The only $2\times2$ matrices with this property are proportional to the identity, so
\begin{equation}
U_T=e^{\i\phi_T}\mathbb I,
\qquad
U_R=e^{\i\phi_R}\mathbb I.
\end{equation}
Their action on the amplitude depends only on the phase difference,
\begin{equation}
T\longrightarrow e^{\i(\phi_R-\phi_T)}T.
\end{equation}
Consequently, the only non-trivial continuous redundancy is the single overall reference phase, described by $U(1)$.

These redundancies are important because every independent continuous symmetry generates a null direction of the Jacobian. If the corresponding freedom is not fixed, different parameter vectors give identical observables and the amplitude solution is not identifiable. The appropriate number of reference conditions is therefore two for the unpolarized single-$k$ model, two when one nucleon is polarized, and one when both nucleons are polarized.

\section{LEPTOPRODUCTION CONSTRAINT COUNTING}
\label{app:counting}
The outcome of an inversion depends on both the number of experimentally available constraints and the number of identifiable real parameters. It is desirable for the system to be overconstrained, in the sense that there are more independent measured moments than free parameters. This counting is a useful necessary check, but it is not by itself sufficient to guarantee a unique amplitude solution: nonlinear amplitude equations may retain discrete ambiguities or become locally insensitive to particular parameter combinations. The orthogonality of the angular basis ensures that distinct measured moments are independent coefficients of the angular distribution; the continuous parameter redundancies identified above must nevertheless be removed separately.

We first count the amplitudes. There are three virtual-photon helicities, two independent nucleon-helicity transitions (four target--recoil helicity combinations reduced to two by parity), and
\begin{equation}
\sum_{\ell=0}^{\ell_{\max}}(2\ell+1)
=(\ell_{\max}+1)^2
\end{equation}
possible $(\ell,m)$ states. Retaining both nucleon-helicity sectors therefore gives
\begin{equation}
N_{\rm amps}^{\rm full}
=6(\ell_{\max}+1)^2
\end{equation}
complex amplitudes, or $12(\ell_{\max}+1)^2$ raw real parameters.

The number of identifiable parameters depends on the polarization information. With unpolarized nucleons and both $k$ sectors retained, the eight-dimensional symmetry found above would give
\begin{equation}
N_{\rm real}^{\rm unpol,full}
=12(\ell_{\max}+1)^2-8.
\end{equation}
In the practical unpolarized analysis used in Sec.~\ref{sec:results}, however, the two $k$ sectors cannot be separated and a single dominant $k$ contribution is assumed. The number of complex amplitudes is then halved,
\begin{equation}
N_{\rm amps}^{\rm unpol}
=3(\ell_{\max}+1)^2,
\end{equation}
and the two independent reflectivity phases must be fixed. Hence,
\begin{equation}
N_{\rm real}^{\rm unpol}
=6(\ell_{\max}+1)^2-2.
\label{eq:nreal_unpol}
\end{equation}

The number of moment constraints can be counted in a similar way. If $\ell_{\max}$ is the largest angular momentum in the two-meson amplitude, then the angular distribution contains moments up to $L_{\max}=2\ell_{\max}$. For $\alpha=0,1,4,5,8$, the allowed values are $M=0,\ldots,L$, giving
\begin{equation}
\sum_{L=0}^{2\ell_{\max}}(L+1)
=(\ell_{\max}+1)(2\ell_{\max}+1)
\end{equation}
moments for each such photon-polarization component. For $\alpha=2,3,6,7$, the $M=0$ moment is absent, and therefore
\begin{equation}
\sum_{L=1}^{2\ell_{\max}}L
=\ell_{\max}(2\ell_{\max}+1)
\end{equation}
moments occur for each component.

Unless a Rosenbluth separation is performed, only the combination $H^0+\epsilon H^4$ is observable. After this combination is made, there are four effective photon-polarization sectors with $M=0$ moments and four without them. Before fixing the overall normalization, the total is therefore
\begin{align}
N_{H,\rm raw}^{\rm unpol}
&=4(\ell_{\max}+1)(2\ell_{\max}+1)
+4\ell_{\max}(2\ell_{\max}+1)\nonumber\\
&=4(2\ell_{\max}+1)^2.
\end{align}
The normalization moment is fixed rather than fitted, leaving
\begin{equation}
N_H^{\rm unpol}
=4(2\ell_{\max}+1)^2-1.
\label{eq:nH_unpol}
\end{equation}
The resulting counts are shown in Table~\ref{tab:counting}. For the P-wave-only row, only $L=0$ and $L=2$ contribute, so the odd-$L$ moments associated with S--P interference are absent.

\begin{table}[h]
\caption{Illustrative counting for the unpolarized, single-$k$ electroproduction truncation used in the inversion. Here $N_{\rm amps}=3(\ell_{\max}+1)^2$ counts the complex amplitudes, $N_{\rm real}=2N_{\rm amps}-2$ counts the identifiable real fit parameters after fixing one reference phase in each reflectivity sector, and $N_H=4(2\ell_{\max}+1)^2-1$ counts the experimentally accessible, non-fixed moments. The $\ell_{\max}=2$ row is the case used in the fixed-amplitude study. Also shown is the P-wave-only case used when analysing vector-meson decay to two spin-0 particles.}
\begin{ruledtabular}
\begin{tabular}{cccc}
$\ell_{\max}$ & $N_{\rm amps}$ & $N_{\rm real}$ & $N_H$\\
0 & 3   & 4   & 3\\
1 & 12  & 22  & 35\\
1 (no S-wave) & 9 & 16 & 23\\
2 & 27  & 52  & 99\\
3 & 48  & 94  & 195\\
4 & 75  & 148 & 323\\
5 & 108 & 214 & 483
\end{tabular}
\end{ruledtabular}
\label{tab:counting}
\end{table}

The counting can be generalized to the case in which either one or both nucleons are polarized. Polarizing one nucleon separates the two $k$ sectors, so the full set of
\begin{equation}
N_{\rm amps}^{\rm single\ pol.}
=6(\ell_{\max}+1)^2
\end{equation}
complex amplitudes becomes accessible. As shown in the previous section, a singly polarized measurement still leaves two continuous redundancies: the overall phase and the arbitrary spin-basis angle of the unobserved nucleon. Two real reference conditions must therefore be imposed, giving
\begin{equation}
N_{\rm real}^{\rm single\ pol.}
=12(\ell_{\max}+1)^2-2.
\label{eq:nreal_single_pol}
\end{equation}

The moment counting is governed by the parity of the combined $\alpha$, $\beta$, and $\delta$ indices, as given in Eq.~\eqref{eq:app_parity_full}. This parity determines whether the corresponding moment is real or imaginary; the imaginary class has no $M=0$ moment. Adding one nucleon-polarization index introduces four Pauli components. Half preserve and half reverse the parity class, so the total raw moment count is four times the unpolarized value before its fixed normalization is removed. With the normalized-moment convention used here, two $L=0$ moments are fixed and are not included as experimental constraints. Thus,
\begin{equation}
N_H^{\rm single\ pol.}
=16(2\ell_{\max}+1)^2-2.
\label{eq:nH_single_pol}
\end{equation}
The resulting counts are shown in Table~\ref{tab:counting_single_pol}.

\begin{table}[h]
\caption{Illustrative counting for the case in which either the target or recoil nucleon is polarized. The polarization separates the two $k$ sectors, giving access to $N_{\rm amps}=6(\ell_{\max}+1)^2$ complex amplitudes. Fixing the two continuous redundancies gives $N_{\rm real}=2N_{\rm amps}-2=12(\ell_{\max}+1)^2-2$. The number of experimentally accessible, non-fixed moments is $N_H=16(2\ell_{\max}+1)^2-2$.}
\begin{ruledtabular}
\begin{tabular}{cccc}
$\ell_{\max}$ & $N_{\rm amps}$ & $N_{\rm real}$ & $N_H$\\
0 & 6   & 10  & 14\\
1 & 24  & 46  & 142\\
1 (no S-wave) & 18 & 34 & 94\\
2 & 54  & 106 & 398\\
3 & 96  & 190 & 782\\
4 & 150 & 298 & 1294\\
5 & 216 & 430 & 1934
\end{tabular}
\end{ruledtabular}
\label{tab:counting_single_pol}
\end{table}

When both nucleons are polarized, the number of amplitudes is unchanged because the two $k$ sectors have already been separated. However, the additional recoil-polarization information removes the arbitrary spin-basis angle, leaving only the single overall reference phase. Therefore,
\begin{align}
N_{\rm amps}^{\rm double\ pol.}
&=6(\ell_{\max}+1)^2,\\
N_{\rm real}^{\rm double\ pol.}
&=12(\ell_{\max}+1)^2-1.
\label{eq:nreal_double_pol}
\end{align}
The second nucleon-polarization index introduces a further factor of four in the raw number of moments. With the eight fixed $L=0$ moments removed according to the same normalization convention,
\begin{equation}
N_H^{\rm double\ pol.}
=64(2\ell_{\max}+1)^2-8.
\label{eq:nH_double_pol}
\end{equation}
The corresponding values are shown in Table~\ref{tab:counting_double_pol}.

\begin{table}[h]
\caption{Illustrative counting for the case in which both the target and recoil nucleons are polarized. The number of amplitudes is the same as in the singly polarized case, $N_{\rm amps}=6(\ell_{\max}+1)^2$. Only the overall reference phase remains redundant, giving $N_{\rm real}=2N_{\rm amps}-1=12(\ell_{\max}+1)^2-1$. The number of experimentally accessible, non-fixed moments is $N_H=64(2\ell_{\max}+1)^2-8$.}
\begin{ruledtabular}
\begin{tabular}{cccc}
$\ell_{\max}$ & $N_{\rm amps}$ & $N_{\rm real}$ & $N_H$\\
0 & 6   & 11  & 56\\
1 & 24  & 47  & 568\\
1 (no S-wave) & 18 & 35 & 376\\
2 & 54  & 107 & 1592\\
3 & 96  & 191 & 3128\\
4 & 150 & 299 & 5176\\
5 & 216 & 431 & 7736
\end{tabular}
\end{ruledtabular}
\label{tab:counting_double_pol}
\end{table}

\section{SUPPLEMENTARY PLOTS}
\label{app:supplementary}

\subsection{Fixed-amplitude closure plots}

\begin{figure*}[p]
\centering
\subfloat[]{\includegraphics[width=0.32\textwidth]{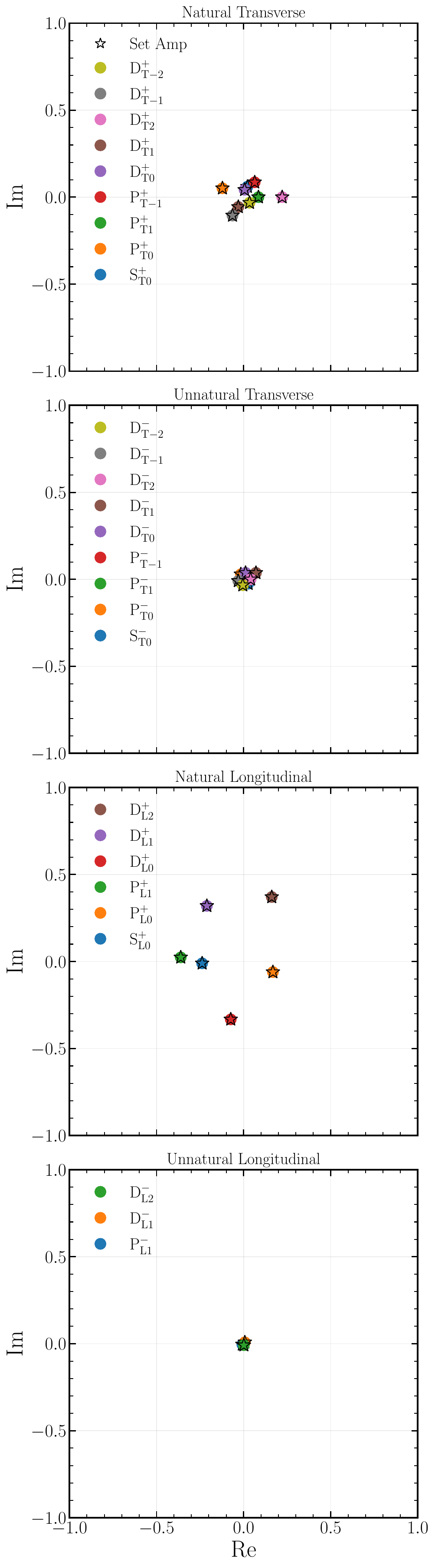}%
\label{fig:fixed-a}}
\hfill
\subfloat[]{\includegraphics[width=0.32\textwidth]{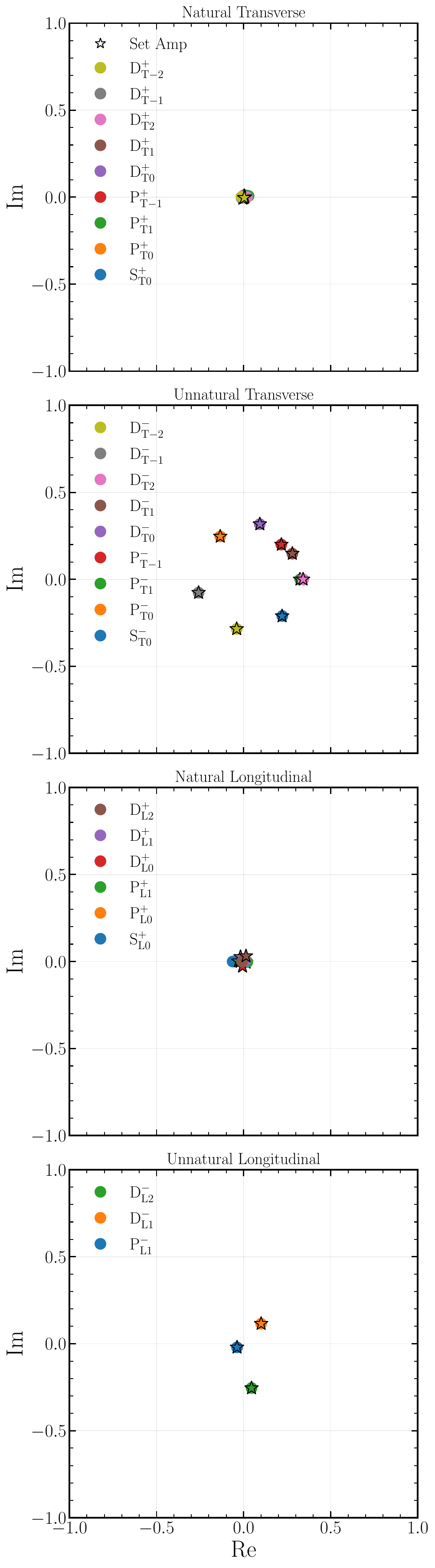}%
\label{fig:fixed-b}}
\hfill
\subfloat[]{\includegraphics[width=0.32\textwidth]{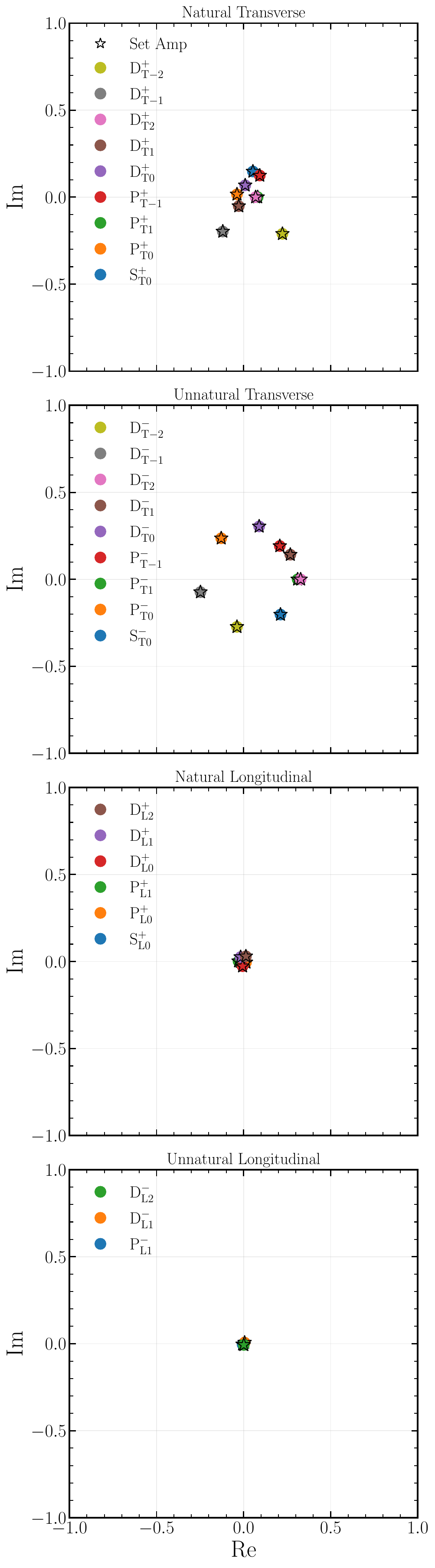}%
\label{fig:fixed-T}}

\caption{Fixed-amplitude closure tests for (a) large
natural and small unnatural amplitudes, (b) large unnatural and small
natural amplitudes, and (c) large transverse and small
longitudinal amplitudes.}
\label{fig:fixed-amplitude-summary}

\end{figure*}

\begin{figure*}[p]
\centering
\setcounter{subfigure}{3}
\subfloat[]{\includegraphics[width=0.31\textwidth]{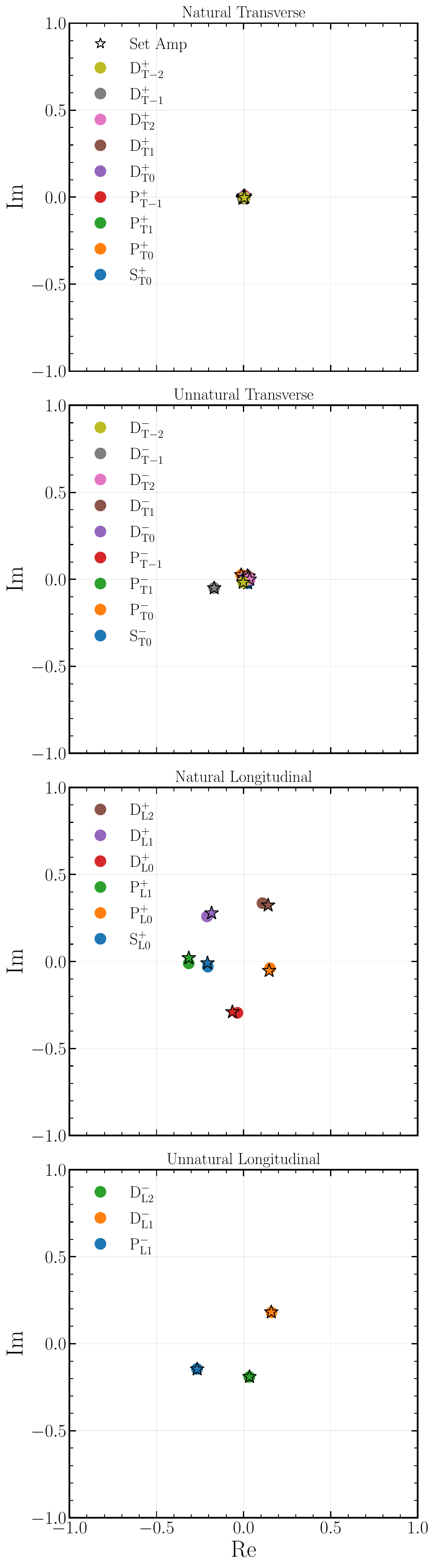}%
\label{fig:fixed-L}}
\hfill
\subfloat[]{\includegraphics[width=0.31\textwidth]{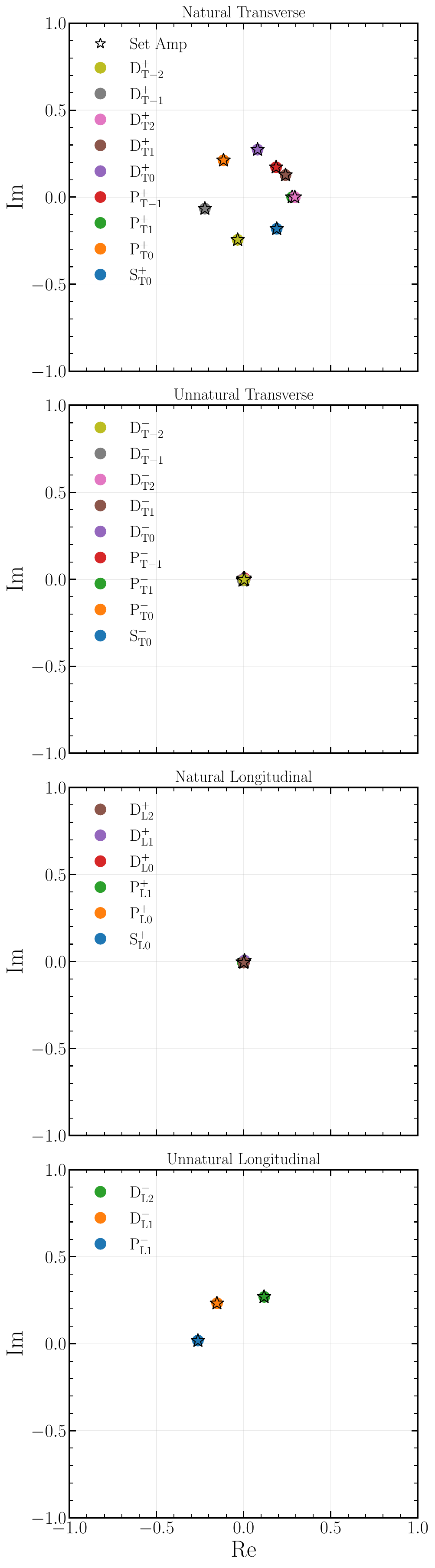}%
\label{fig:fixed-aTbL}}
\hfill
\subfloat[]{\includegraphics[width=0.31\textwidth]{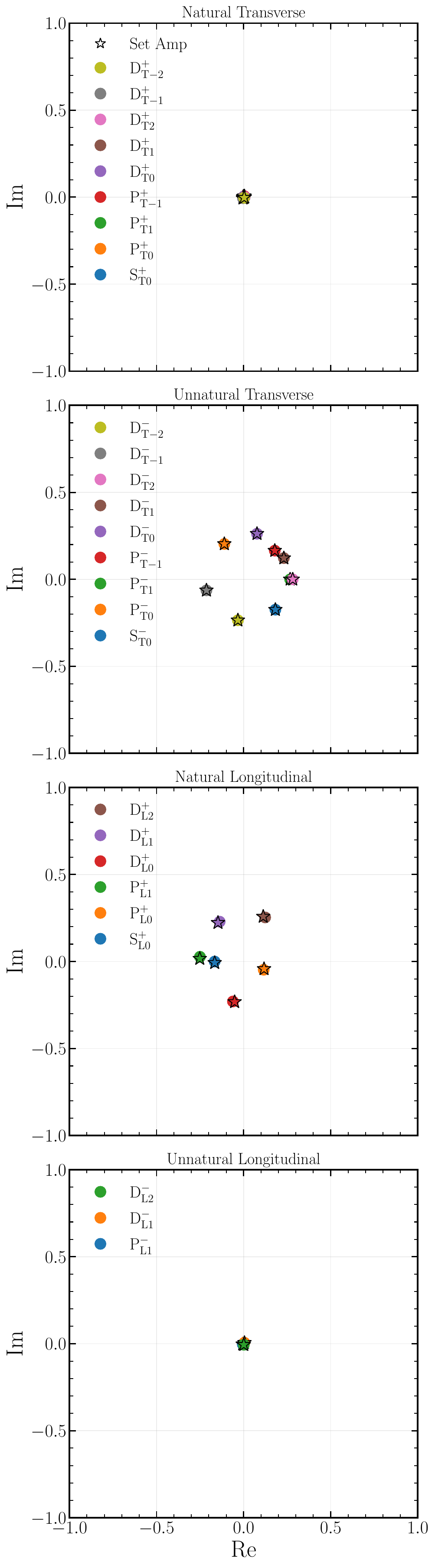}%
\label{fig:fixed-bTaL}}

\caption{Fixed-amplitude closure tests for (d) large longitudinal and small transverse
amplitudes, (e) large natural-transverse and unnatural-longitudinal
amplitudes with small natural-longitudinal and unnatural-transverse
amplitudes, and (f) large natural-longitudinal and unnatural-transverse
amplitudes with small natural-transverse and unnatural-longitudinal
amplitudes.}
\label{fig:fixed-amplitude-summary-second}

\end{figure*}

The main text shows one representative amplitude solution when the data is generated using known amplitudes. That specific case tested the minimizer on completely randomly generated values, but here we collect extra plots showing specific cases that were discussed within the main paper. These cases are: large natural and small unnatural amplitudes, shown in Fig.~\ref{fig:fixed-a}; large unnatural and small natural amplitudes, shown in Fig.~\ref{fig:fixed-b}; large transverse and small longitudinal amplitudes, shown in Fig.~\ref{fig:fixed-T}; large longitudinal and small transverse amplitudes, shown in Fig.~\ref{fig:fixed-L}; large natural-transverse and unnatural-longitudinal amplitudes with small natural-longitudinal and unnatural-transverse amplitudes, shown in Fig.~\ref{fig:fixed-aTbL}; and the converse configuration, shown in Fig.~\ref{fig:fixed-bTaL}.

\begin{figure*}[p]
\centering
\includegraphics[width=\textwidth]{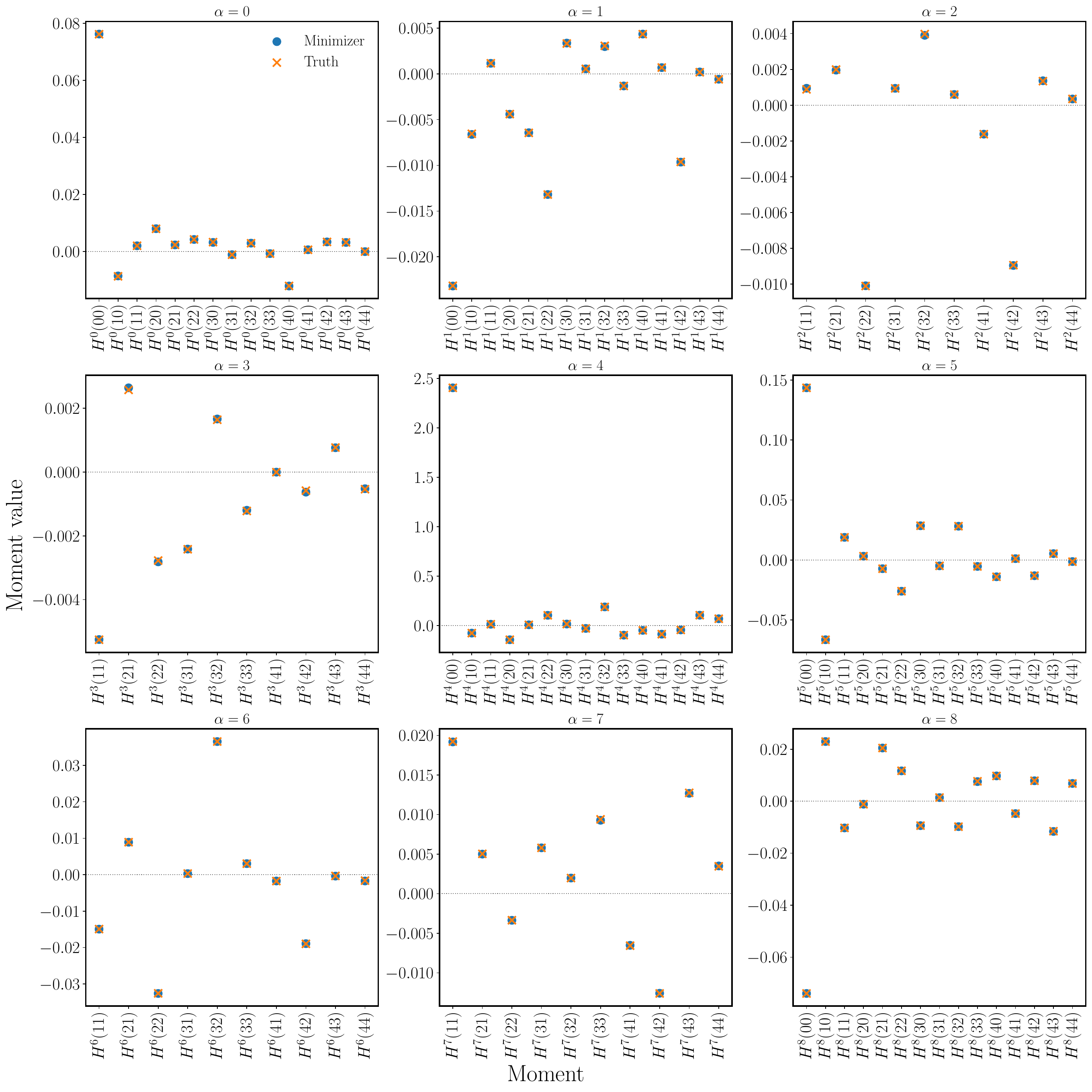}
\caption{Moment closure test for a fixed-amplitude configuration dominated by longitudinal amplitudes.}
\label{fig:fixedmoments_large_L}
\end{figure*}

\begin{figure*}[p]
\centering
\includegraphics[width=\textwidth]{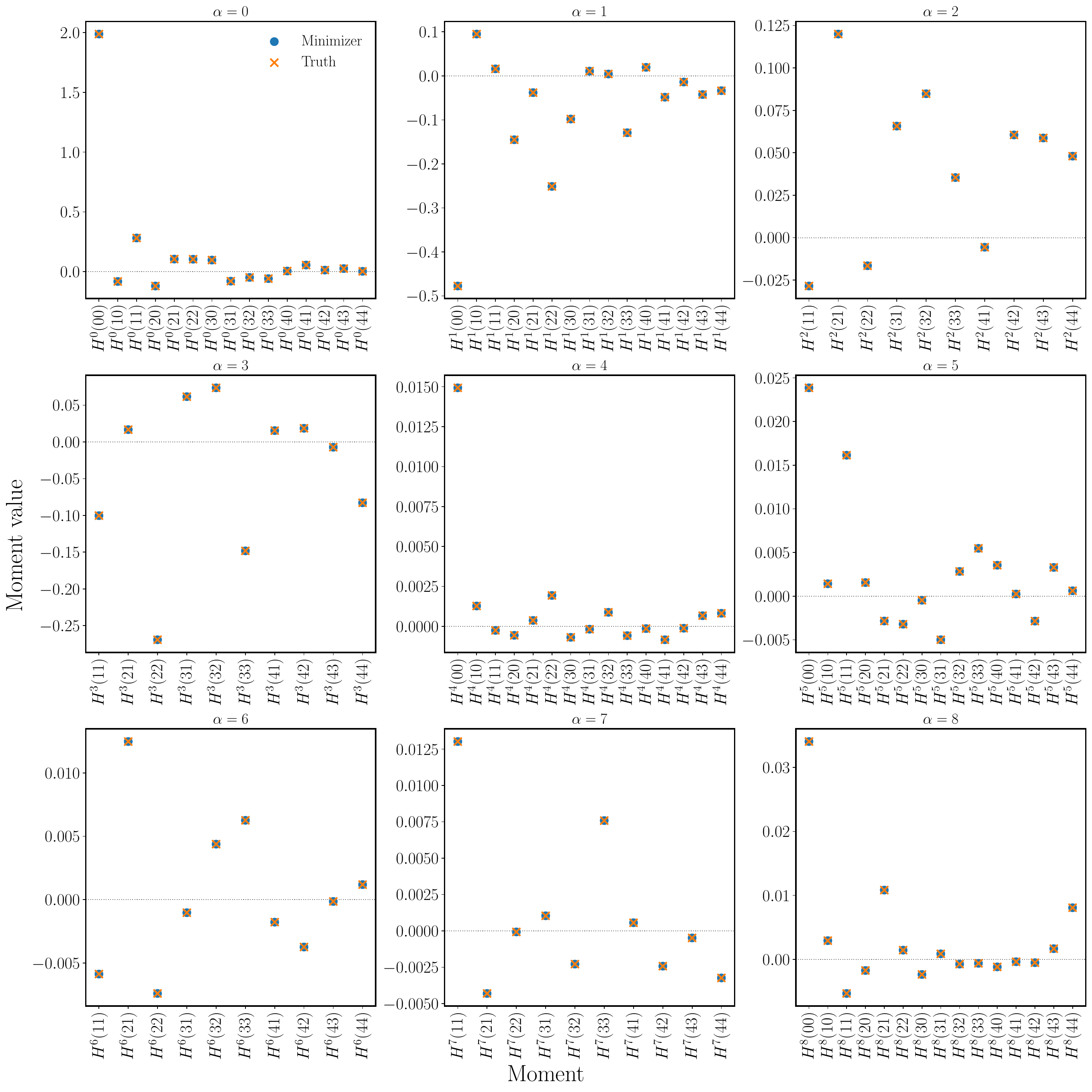}
\caption{Moment closure test for a fixed-amplitude configuration dominated by transverse amplitudes.}
\label{fig:fixedmoments_large_T}
\end{figure*}

\begin{figure*}[p]
\centering
\includegraphics[width=\textwidth]{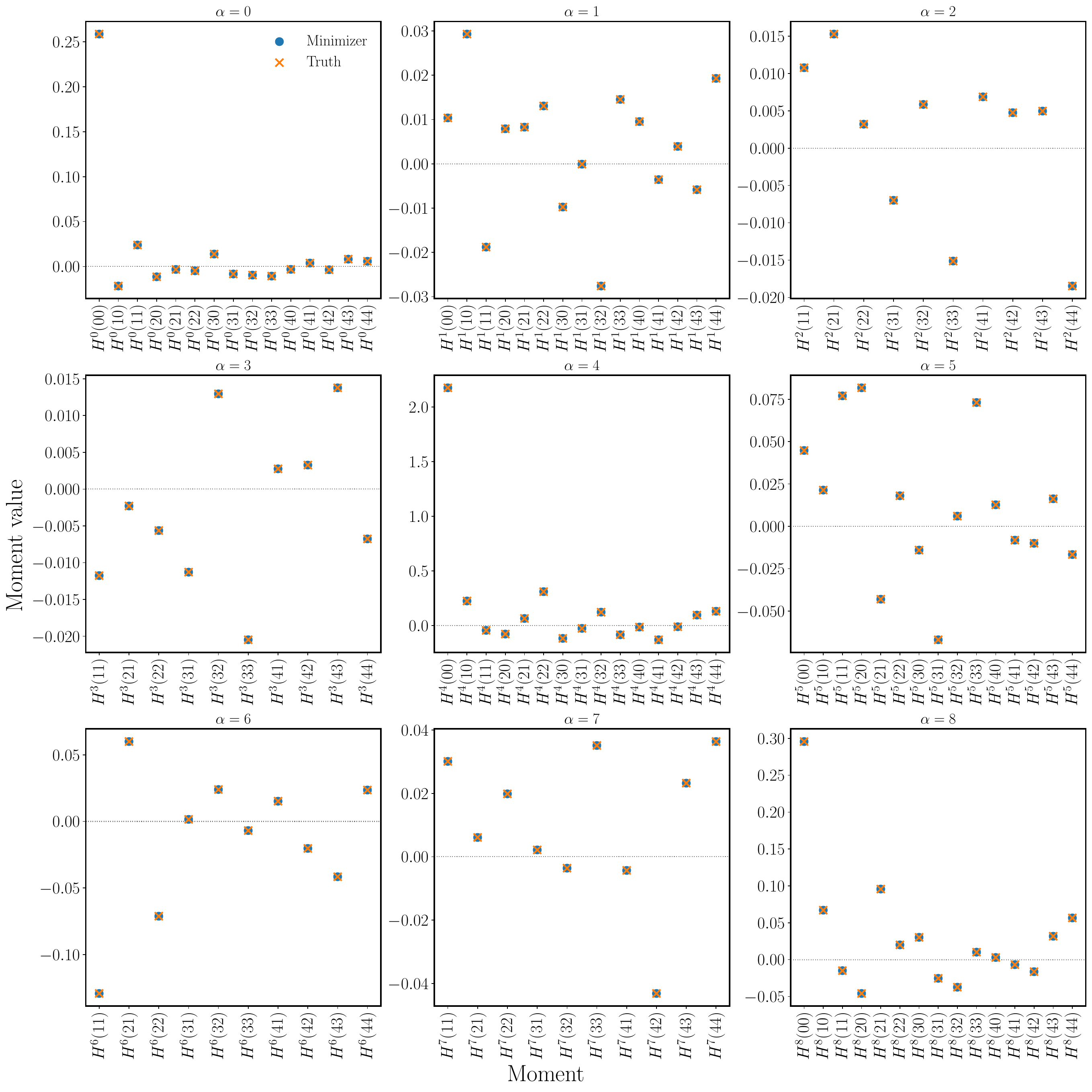}
\caption{Moment closure test for a fixed-amplitude configuration dominated by natural-reflectivity $a$ amplitudes.}
\label{fig:fixedmoments_large_a}
\end{figure*}

\begin{figure*}[p]
\centering
\includegraphics[width=\textwidth]{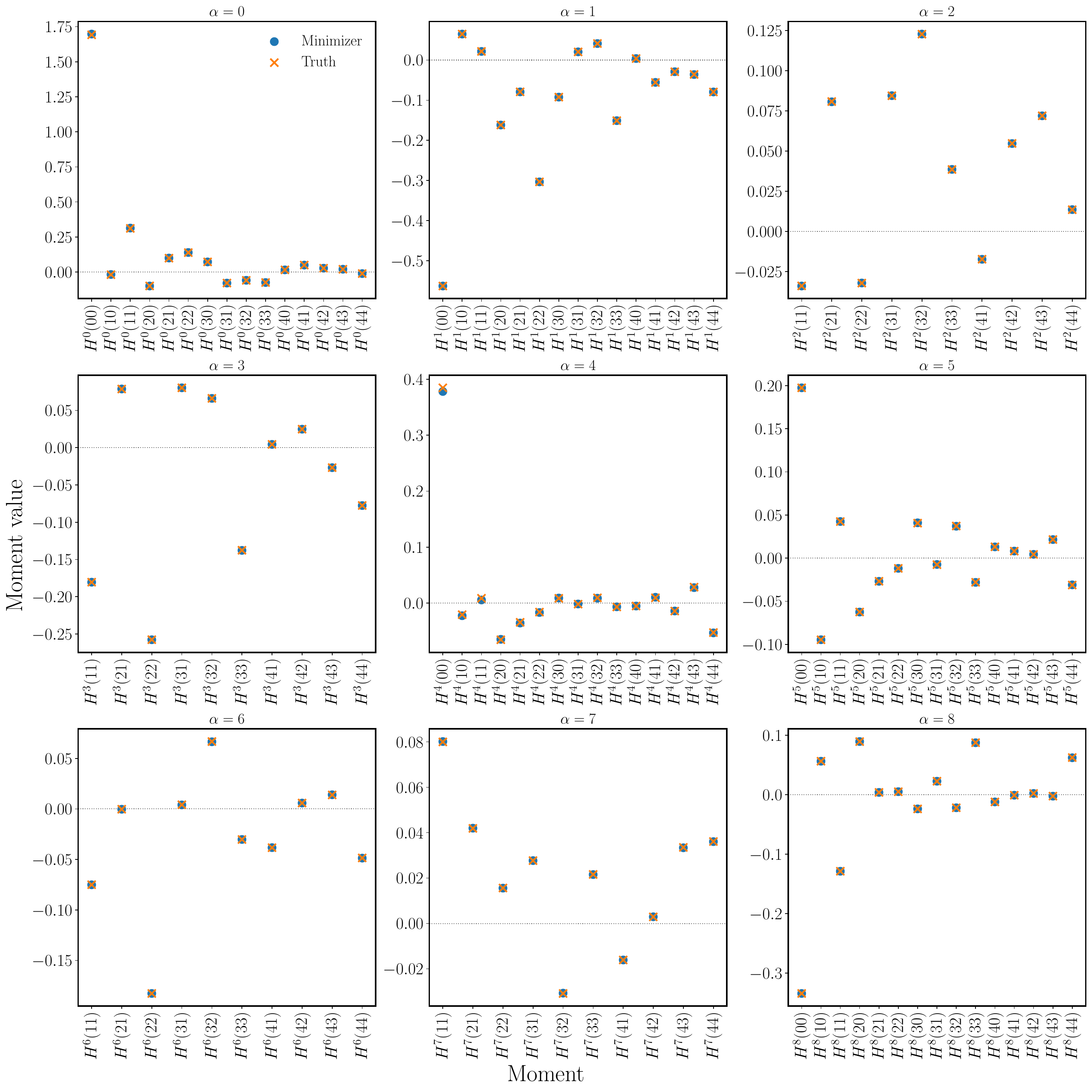}
\caption{Moment closure test for a fixed-amplitude configuration dominated by unnatural-reflectivity $b$ amplitudes.}
\label{fig:fixedmoments_large_b}
\end{figure*}

\begin{figure*}[p]
\centering
\includegraphics[width=\textwidth]{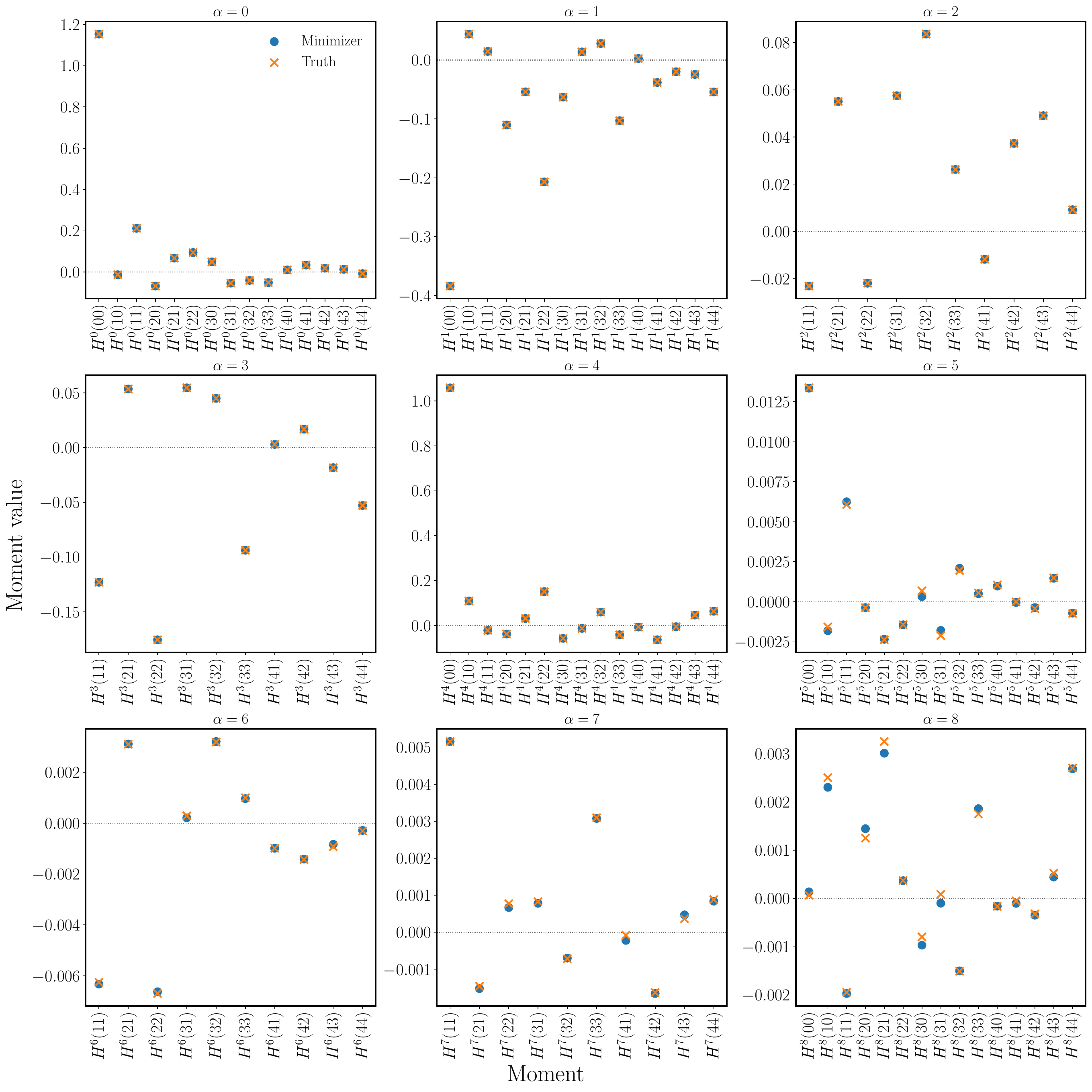}
\caption{Moment closure test for a mixed fixed-amplitude configuration with large $b_\TT$ and $a_\LL$ amplitudes.}
\label{fig:fixedmoments_large_bTaL}
\end{figure*}

\begin{figure*}[p]
\centering
\includegraphics[width=\textwidth]{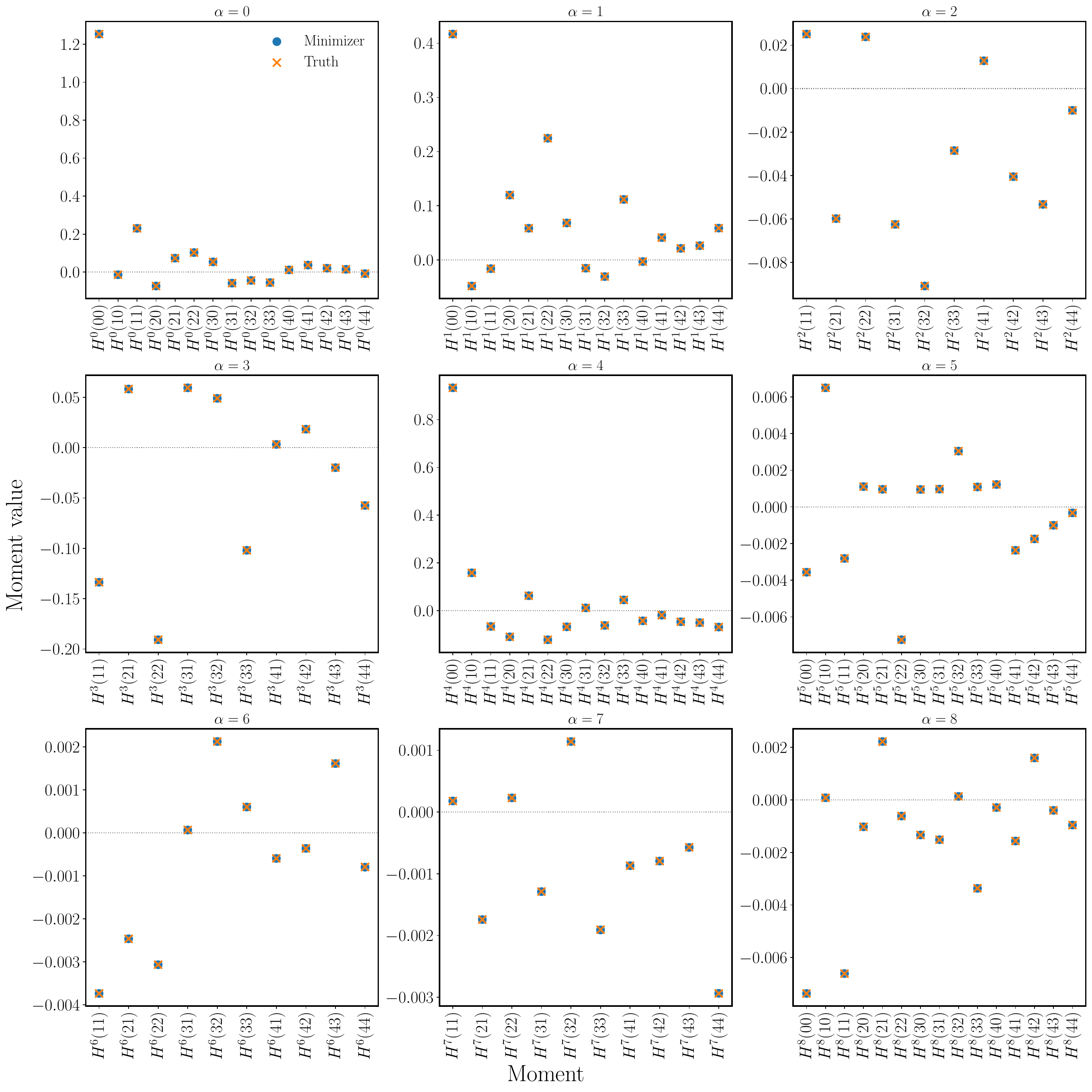}
\caption{Moment closure test for a mixed fixed-amplitude configuration with large $a_\TT$ and $b_\LL$ amplitudes.}
\label{fig:fixedmoments_large_aTbL}
\end{figure*}

The corresponding moment closure plots are also given below. The configuration dominated by longitudinal amplitudes is shown in Fig.~\ref{fig:fixedmoments_large_L}, while the transverse-dominated configuration is shown in Fig.~\ref{fig:fixedmoments_large_T}. The natural- and unnatural-reflectivity dominated cases are shown in Figs.~\ref{fig:fixedmoments_large_a} and \ref{fig:fixedmoments_large_b}, respectively. Finally, denoting by $a_c$ and $b_c$ the natural and unnatural reflectivity amplitudes with $c\in\{\TT,\LL\}$, the two mixed configurations with large $b_\TT$ and $a_\LL$ amplitudes and large $a_\TT$ and $b_\LL$ amplitudes are shown in Figs.~\ref{fig:fixedmoments_large_bTaL} and \ref{fig:fixedmoments_large_aTbL}, respectively.

\subsection{Amplitude plots}

\begin{figure*}[p]
\centering
\subfloat[$Q^2=0.82,\mathrm{GeV}^2$]{\includegraphics[width=0.31\textwidth]{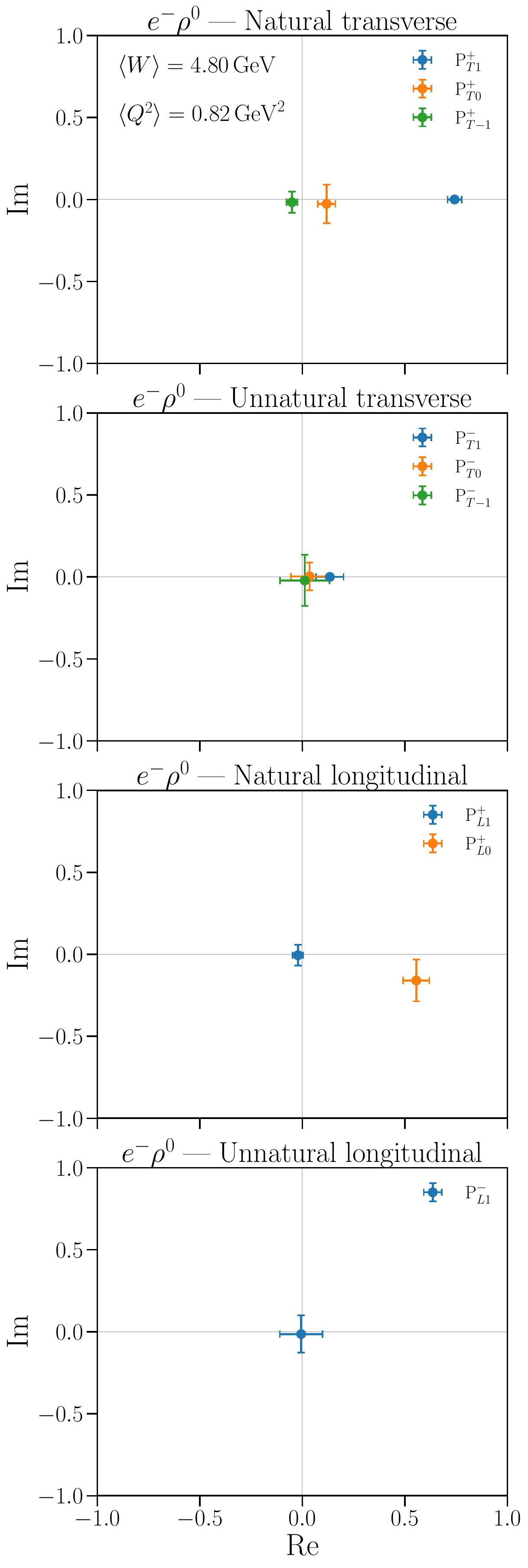}}\hfill
\subfloat[$Q^2=1.19,\mathrm{GeV}^2$]{\includegraphics[width=0.31\textwidth]{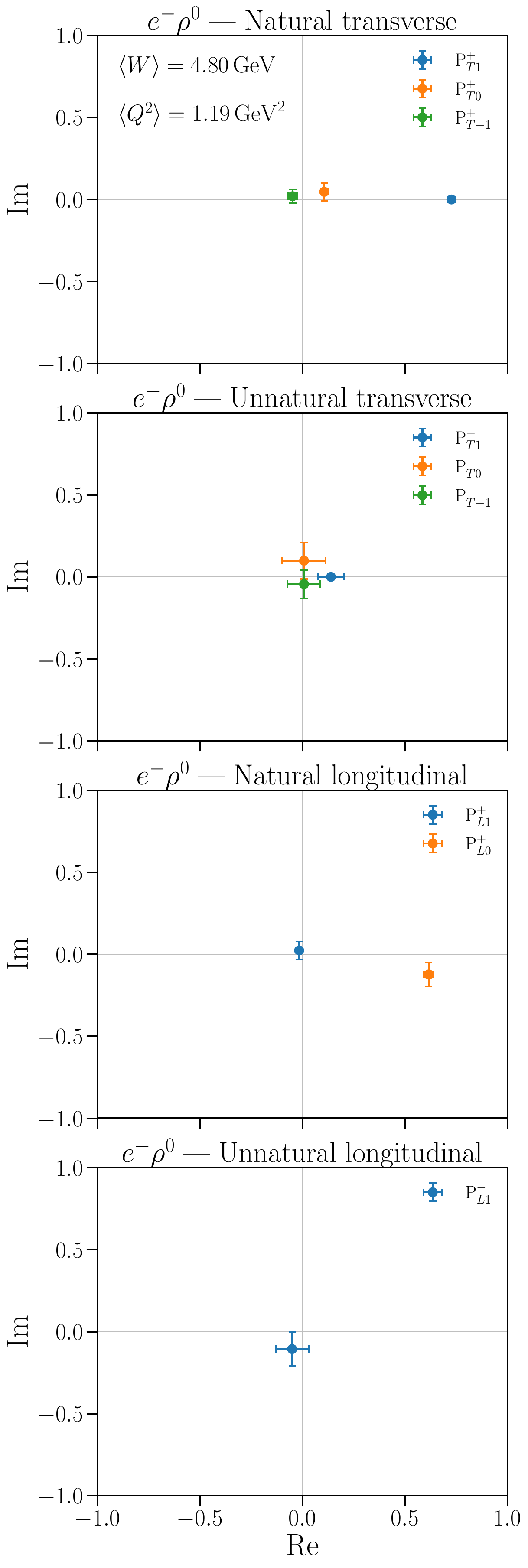}}\hfill
\subfloat[$Q^2=1.66,\mathrm{GeV}^2$]{\includegraphics[width=0.31\textwidth]{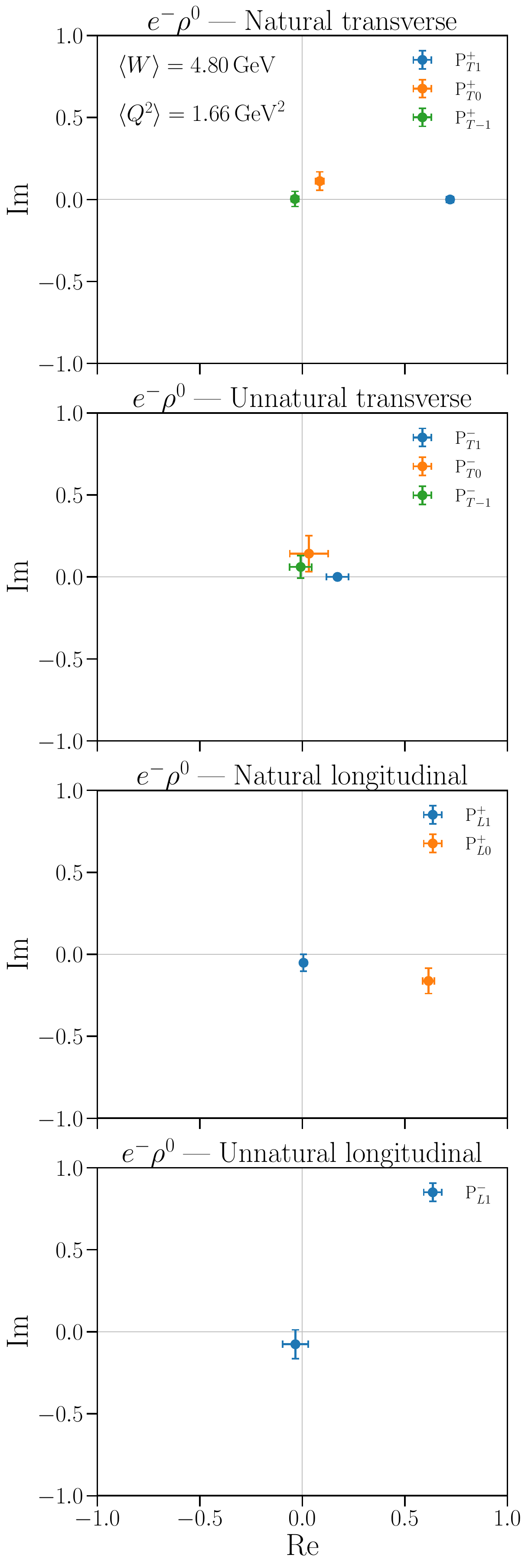}}
\caption{Supplementary $e^-\rho^0$ amplitude solutions. The highest-$Q^2$ representative result is shown in the main text in Fig.~\ref{fig:amp_erho}.}
\label{fig:app_erho}
\end{figure*}

\begin{figure*}[p]
\centering
\subfloat[$Q^2=1.14,\mathrm{GeV}^2$]{\includegraphics[width=0.31\textwidth]{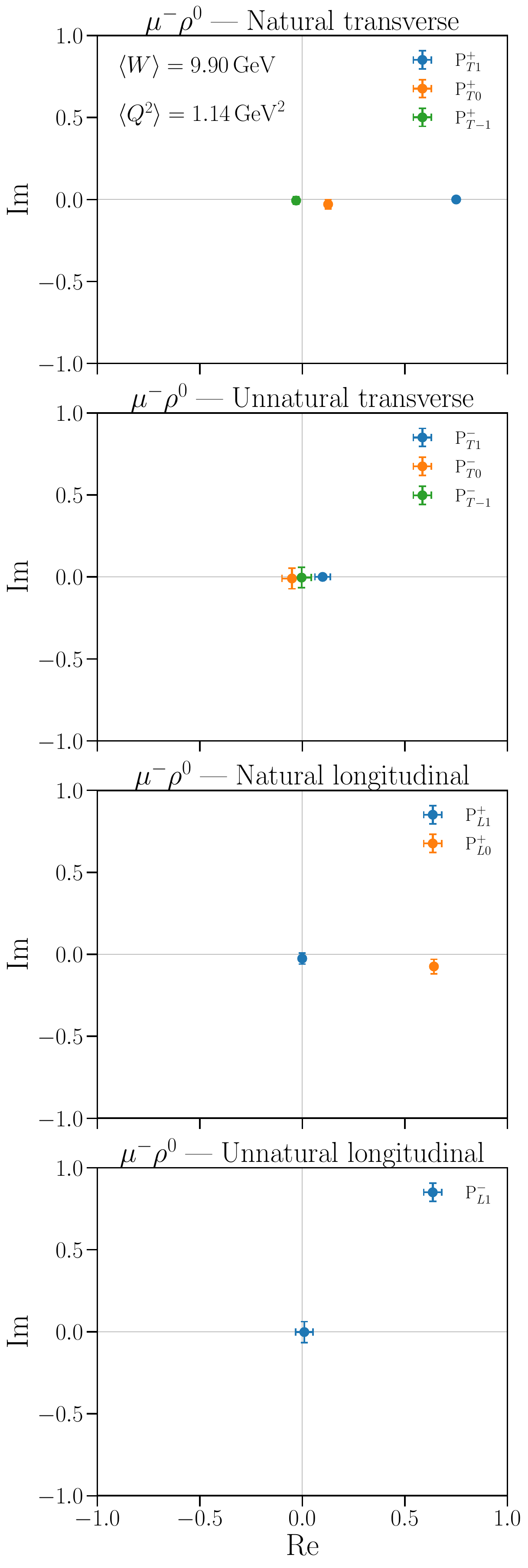}}\hfill
\subfloat[$Q^2=1.60,\mathrm{GeV}^2$]{\includegraphics[width=0.31\textwidth]{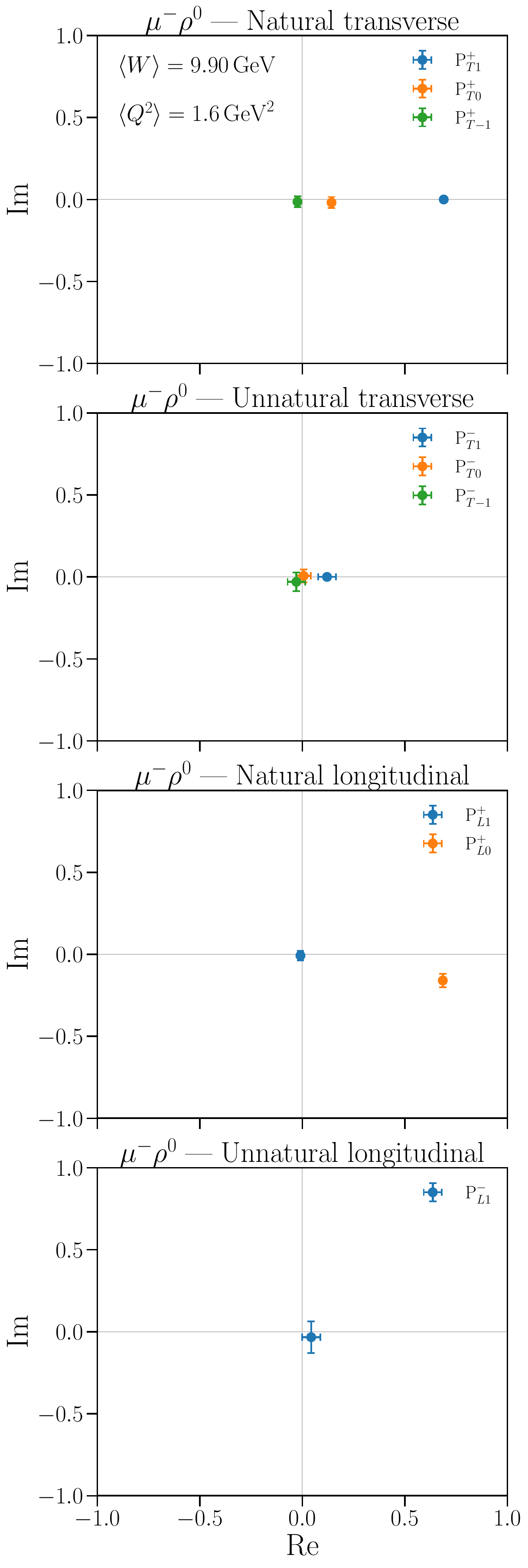}}\hfill
\subfloat[$Q^2=2.80,\mathrm{GeV}^2$]{\includegraphics[width=0.31\textwidth]{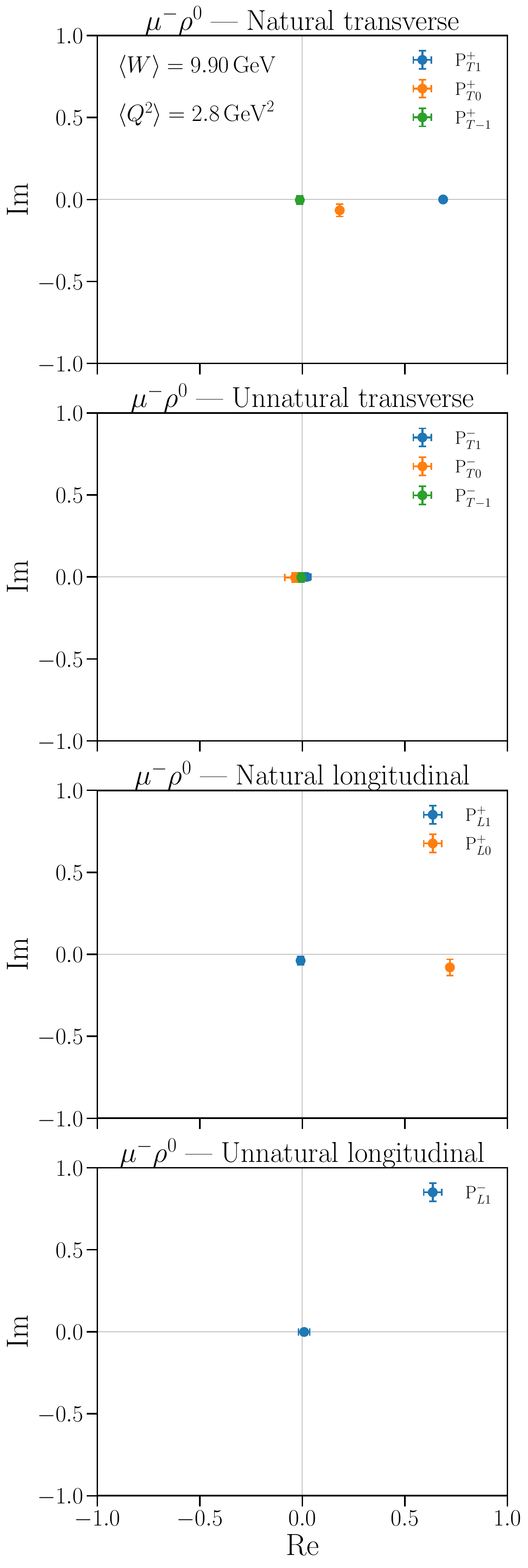}}
\caption{Supplementary $\mu^-\rho^0$ amplitude solutions. The representative highest-$Q^2$ result is shown in the main text in Fig.~\ref{fig:amp_murho}. Across the remaining bins the natural sector remains dominant, with the unnatural amplitudes staying comparatively small.}
\label{fig:app_murho}
\end{figure*}

\begin{figure*}[p]
\centering
\subfloat[]{\includegraphics[width=0.31\textwidth]{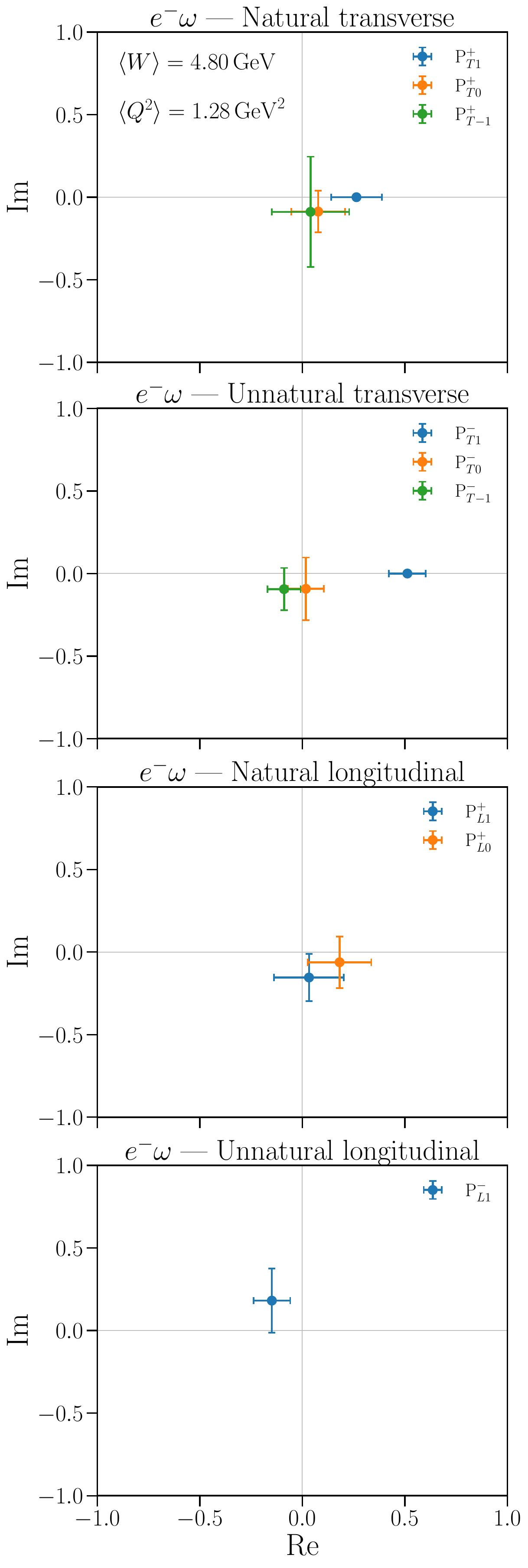}}
\subfloat[]{\includegraphics[width=0.31\textwidth]{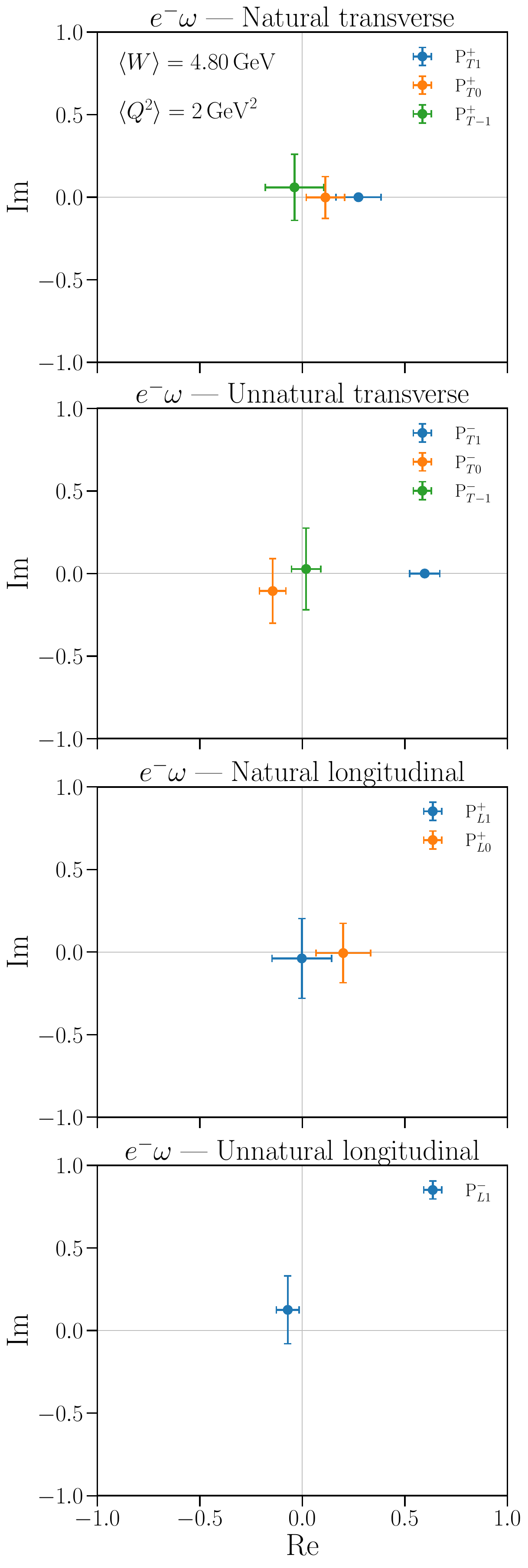}}
\caption{Supplementary $e^-\omega$ amplitude solutions. The highest-$Q^2$ representative result is shown in the main text in Fig.~\ref{fig:amp_eomega}. The enhanced unnatural-longitudinal strength remains visible already in the lower-$Q^2$ bins.}
\label{fig:app_eomega}
\end{figure*}

\begin{figure*}[p]
\centering
\subfloat[]{\includegraphics[width=0.31\textwidth]{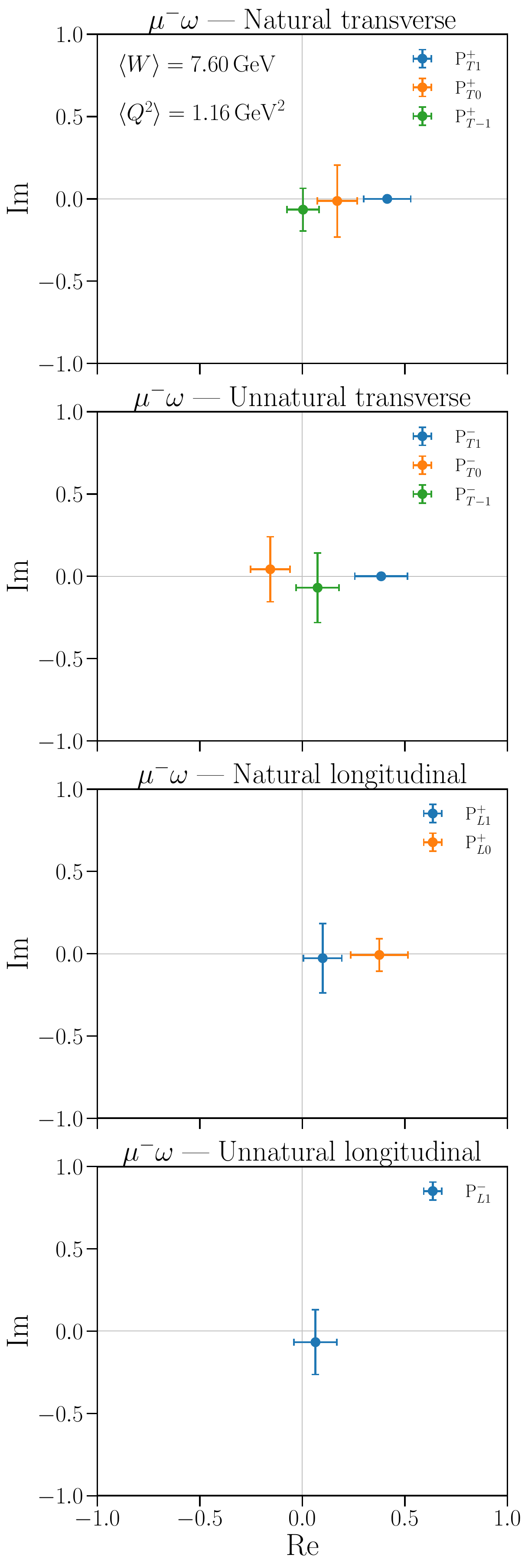}}
\subfloat[]{\includegraphics[width=0.31\textwidth]{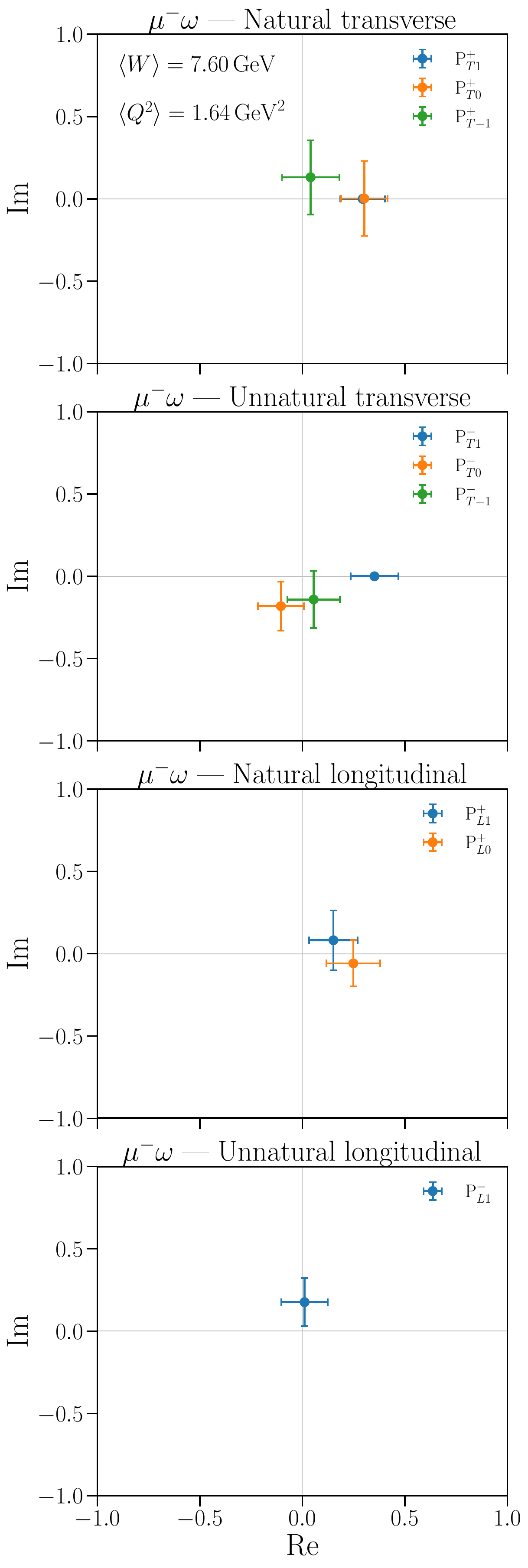}}
\caption{Supplementary $\mu^-\omega$ amplitude solutions. The representative highest-$Q^2$ result is shown in the main text in Fig.~\ref{fig:amp_muomega}. As in the electron-induced channel, the unnatural contributions are substantially larger than in $\rho^0$ production.}
\label{fig:app_muomega}
\end{figure*}

The main text shows one representative amplitude solution for each of the four reaction classes. The remaining amplitude plots are collected here for completeness. The supplementary $e^-\rho^0$ solutions are shown in Fig.~\ref{fig:app_erho}, the $\mu^-\rho^0$ solutions in Fig.~\ref{fig:app_murho}, the $e^-\omega$ solutions in Fig.~\ref{fig:app_eomega}, and the $\mu^-\omega$ solutions in Fig.~\ref{fig:app_muomega}.

\newpage 

\bibliographystyle{apsrev4-2}
\bibliography{refs}

\end{document}